\documentclass{CASthesis}

\usepackage{listings}
\usepackage{soul}

\begin{document}
\begin{CJK}{UTF8}{gbsn}
\begin{spacing}{1.3}


  \title{Time-Delay Interferometry for ASTROD-GW}

 \englishtitle{Time-Delay Interferometry for ASTROD-GW}
  \englishauthor{Gang Wang}
  \englishadvisor{Prof. Wei-Tou Ni}
  \englishinstitute{Purple Mountain Observatory \\ Chinese Academy of Sciences}
  \englishdegree{Master}
  \englishmajor{Astrophysics-Gravitational Waves Detection}

  \makeenglishtitle
\frontmatter

\begin{englishabstract}

In the detection of gravitational waves in space, the arm lengths between spacecraft are not equal due to their orbital motion. Consequently, the equal arm length Michelson interferometer used in Earth laboratories is not suitable for space. To achieve the necessary sensitivity for space gravitational wave detectors, laser frequency noise must be suppressed below secondary noise sources such as optical path noise and acceleration noise. To suppress laser frequency noise, time-delay interferometry (TDI) is employed to match the two optical paths and retain gravitational wave signals. Since planets and other solar system bodies perturb the orbits of spacecraft and affect TDI performance, we simulate the time delay numerically using the CGC2.7 ephemeris framework. To examine the feasibility of TDI for the ASTROD-GW mission, we devised a set of 10-year and a set of 20-year optimized mission orbits for the three spacecraft starting on June 21, 2028, and calculated the path mismatches in the first- and second-generation TDI channels. The results demonstrate that all second-generation TDI channels meet the ASTROD-GW requirements. A geometric approach is used in the analysis and synthesis of both first-generation and second-generation TDI to clearly illustrate the construction process.

\englishkeywords{ASTROD, Time-Delay Interferometry, Gravitational Waves Detection, Numerical Calculation, Geometric Construction}
\end{englishabstract}

\tableofcontents
  \listoftables
  \listoffigures
  
\mainmatter

\chapter{Introduction} 
\footnote{This chapter is derived from review \cite{Ni2010,Ni2010cn}.}
In 1905, Poincaré \cite{[1]} and Einstein \cite{[2]} proposed the theory of special relativity. Poincaré \cite{[1]} attempted to establish a relativistic theory of gravity, mentioned gravitational waves (GWs), and inferred their propagation speed to be the same as that of light based on Lorentz invariance. Subsequently, physicists (including Einstein himself) attempted to establish a relativistic theory of gravity \cite{[3],[4]}. It was not until 1915 that Einstein proposed the general theory of relativity, which successfully explained the anomalous precession of Mercury's perihelion \cite{[5]}. After proposing the general theory of relativity in 1915, Einstein predicted the existence of GWs and estimated their strength \cite{[6]}. To this day, Einstein's general theory of relativity has become the standard theory of gravity, widely applied in the global positioning system, planetary and lunar ephemeris calculations, solar system space navigation and exploration, astrophysics, and cosmology.

Maxwell's electromagnetic theory predicts the existence of electromagnetic waves. Einstein's general theory of relativity and other relativistic theories of gravity predict the existence of GWs. GWs propagate through spacetime, forming ripples. The status of GWs in gravitational physics is analogous to that of electromagnetic waves in electromagnetic physics. The existence of GWs is a direct consequence of general relativity and is an inevitable result of all gravitational theories with a finite propagation speed.

The importance of GW detection is twofold:
\begin{description}
\item{(i)} To explore fundamental physics and cosmology, especially black hole physics and the early universe;
\item{(ii)} To serve as a tool for astronomical and astrophysical research, studying compact celestial bodies and calculating their density distribution, complementing electromagnetic wave astronomy and cosmic ray (including neutrino) observations.
\end{description}

The evolution of the binary pulsar orbit demonstrates the existence of GW radiation \cite{[7]}. In general relativity, a binary system emits energy in the form of GWs. The energy loss leads to a shrinking orbit and a shorter orbital period. Thirty years of observations of the relativistic binary B1913+16 show a cumulative advance in periastron time by 35 s. In relativity, the orbital decay rate of a binary pulsar can be calculated from the pulsar system parameters determined by pulsar timing observations. After correcting for the relative acceleration between the solar system and the pulsar binary system, Weisberg and Taylor \cite{[7]} found that the measured orbital decay rate matches the predicted rate due to GW radiation from general relativity to within $(1.3 \pm 2.1) \times 10^{-3}$. Hulse and Taylor, who discovered this binary system, were awarded the 1993 Nobel Prize in Physics.

Similar to how electromagnetic waves are divided into radio waves, microwaves, infrared rays, light waves, ultraviolet rays, X-rays, and gamma rays based on frequency, GWs can also be categorized into different frequency bands \cite{Ni2010,Ni2010cn,[8],[9],[10],[11]}:
\begin{description}
\item{(i)} Ultra-high frequency band ($>$ 1 THz): Detection methods include terahertz resonant cavities, optical resonant cavities, and innovative methods yet to be invented.
\item{(ii)} Very high frequency band (100 kHz - 1 THz): This is the most sensitive frequency band for laboratory detection of GWs using microwave resonant systems and short-arm length laser interferometers.
\item{(iii)} High frequency band (10 Hz - 100 kHz): This is the most sensitive frequency band for ground-based detection of GWs using cryogenic resonators and laser interferometers.
\item{(iv)} Mid frequency band (0.1 Hz - 10 Hz): This is the most sensitive frequency band for space-based laser interferometers with short arm lengths ($10^3 - 10^5$ km).
\item{(v)} Low frequency band (100 nHz - 0.1 Hz): This is the most sensitive frequency band for space-based laser interferometers with long arm lengths ($10^6 - 10^9$ km).
\item{(vi)} Very low frequency band (300 pHz - 100 nHz): This is the most sensitive frequency band for pulsar timing experiments.
\item{(vii)} Ultra-low (sub-Hubble) frequency band (10 fHz - 300 pHz): This is the frequency band between the very low frequency and extremely low frequency bands, and is most sensitive for precise measurements of quasar and radio source proper motion experiments.
\item{(viii)} Hubble (extremely low) frequency band (1 aHz - 10 fHz): This is the most sensitive frequency band for cosmic background radiation anisotropy and polarization experiments.
\item{(ix)} (Infrared) Hubble frequency band ($<$ 1 aHz): Inflationary cosmological models predict GWs in this band. Confirmation of these inflationary models can indirectly provide evidence for GWs in this band.
\end{description}

The main activities for high-frequency GW detection are in ground-based long-arm laser interferometers. The TAMA 300 m arm length interferometer\cite{[13]}, GEO 600 m arm length interferometer\cite{[14]}, and kilometer-scale laser interferometer GW detectors LIGO\cite{[15]} (two 4 km arm lengths, one 2 km arm length) and Virgo\cite{[16]} have essentially achieved their original design sensitivity goals. Around 100 Hz, the sensitivity of LIGO and Virgo has reached $10^{-23}/\sqrt{\text{Hz}}$. Currently, both LIGO and Virgo are undergoing upgrade plans for the next generation detectors — AdLIGO\cite{[17]} and AdVirgo\cite{[18]}, which will improve sensitivity by ten times, increasing the number of detectable GW sources by about 1000 times. The 3 km cryogenic laser interferometer GW detector LCGT has begun construction, with the first phase being a room temperature detector, which will be converted directly to a cryogenic third-generation long-arm interferometer after completion, commissioning, and observation\cite{LCGT}, with sensitivity comparable to AdLIGO and AdVirgo. The European third-generation long-arm interferometer ET has begun planning\cite{ET}. It is expected that within about five years, humans will be able to directly detect GWs for the first time.

Space-based GW detection laser interferometers (LISA\cite{[19]}, ASTROD\cite{[20],[21]}, ASTROD-GW\cite{Ni2010,[9],[11]}, Super-ASTROD\cite{[22]}, DECIGO\cite{[23]}, and Big Bang Observer (BBO)\cite{[24],[25]}) offer the high signal-to-noise ratio, and are crucial for studying astrophysics, cosmology, and fundamental physics. We review space-based GW detectors in the next chapter. Achieving the required target sensitivity for laser interferometry space GW detectors necessitates reducing laser frequency noise. If the time delay matching of the two beams is achieved, their interference signal laser frequency noise can be subtracted, achieving the goal. The better the matching (smaller time delay difference), the better the noise reduction. In the third chapter, we discuss the principles and methods of time-delay interferometry (TDI). Designing appropriate mission orbits and establishing a suitable ephemeris is necessary. In the fourth chapter, we establish the CGC2.7 ephemeris framework. In the fifth chapter, we further discuss and select the mission orbits for ASTROD-GW based on previous work. In the sixth chapter, we introduce numerical methods for TDI and calculate the path mismatches in TDI for dynamic orbit. In the following seventh chapter, we analyze and construct various interferometry paths when the three interferometer arms work simultaneously and calculate the time delays for some of these paths for ASTROD-GW. Finally, in the eighth chapter, we provide conclusions and a brief discussion.

\chapter{Space-borne Gravitational Wave Detectors}

\footnote{This chapter is derived from references \cite{Ni2010,Ni2010cn,[11]}.}
The gravitational field of the solar system is determined by three factors: the dynamic distribution of matter within the solar system, the dynamic distribution of matter outside the solar system (Milky Way, extragalactic systems, universe, etc.), and GWs passing through the solar system. Different relativistic gravitational theories predict different gravitational fields for the solar system; thus, precise measurements of the solar system's gravitational field can test these relativistic gravitational theories. Additionally, these measurements can detect GWs, determine the distribution of matter in the solar system, and observe the measurable (testable) effects of the Milky Way and the universe. We can determine the gravitational field of the solar system by measuring/monitoring various natural and artificial celestial bodies. In the solar system, the motion of celestial bodies or spacecraft follows the following astronomical dynamics equation:
\begin{equation}
 \label{eqa2.1}
\mathbf{a} = \mathbf{a}_{\text{N}} + \mathbf{a}_{\text{1PN}} + \mathbf{a}_{\text{2PN}} + \mathbf{a}_{\text{Gal-Cosm}} + \mathbf{a}_{\text{GW}} + \mathbf{a}_{\text{non-grav}}
\end{equation}
where $\mathbf{a}$ is the acceleration of the celestial body or spacecraft, $\mathbf{a}_{\text{N}}$ is the acceleration due to the Newtonian gravitational theory from the mass distribution in the solar system, $\mathbf{a}_{\text{1PN}}$ is the first post-Newtonian correction acceleration, $\mathbf{a}_{\text{2PN}}$ is the second post-Newtonian correction acceleration, $\mathbf{a}_{\text{Gal-Cosm}}$ is the acceleration due to the mass distribution in the Milky Way and the universe, $\mathbf{a}_{\text{GW}}$ is the acceleration due to GWs, and $\mathbf{a}_{\text{non-grav}}$ is the acceleration from all non-gravitational sources. The distance between spacecraft/celestial bodies is determined by the gravitational field of the solar system (including the gravitational effects of solar oscillations), the fundamental gravitational theories, and the GWs passing through the solar system. Accurately measuring these time-varying distances can identify the causes of these variations. Some orbital combinations are better for testing relativistic gravity; some are easier for measuring solar system parameters; some are easier for detecting GWs. These are all integral parts of mission design.

Space-borne GW detectors mostly use drag-free spacecraft, so in using Equ. (\ref{eqa2.1}), $\mathbf{a}_{\text{non-grav}}$ is considered noise, and generally, local gravitational variations are also considered part of this noise.

\section{LISA}

LISA\cite{[19]} (Laser Interferometer Space Antenna) has an interferometer arm length of approximately 5 million kilometers. Its goal is to detect GWs in the $10^{-4}$ to 1 Hz frequency band, primarily in the low-frequency region, with some coverage in the mid-frequency region. Its strain detection sensitivity at 1 mHz is $4\times 10^{-21}/ \sqrt{\text{Hz}}$. LISA, ASTROD, and ASTROD-GW have a wealth of GW sources: galactic compact binaries (neutron stars, white dwarfs, etc.) and extragalactic sources. Extragalactic targets include supermassive black hole binaries, the formation of supermassive black holes, and the cosmic GW background. \st{The LISA space mission is hoped to be launched by 2020.}
\begin{figure}[htbp]
    \centering 
    \includegraphics[width=0.55\textwidth]{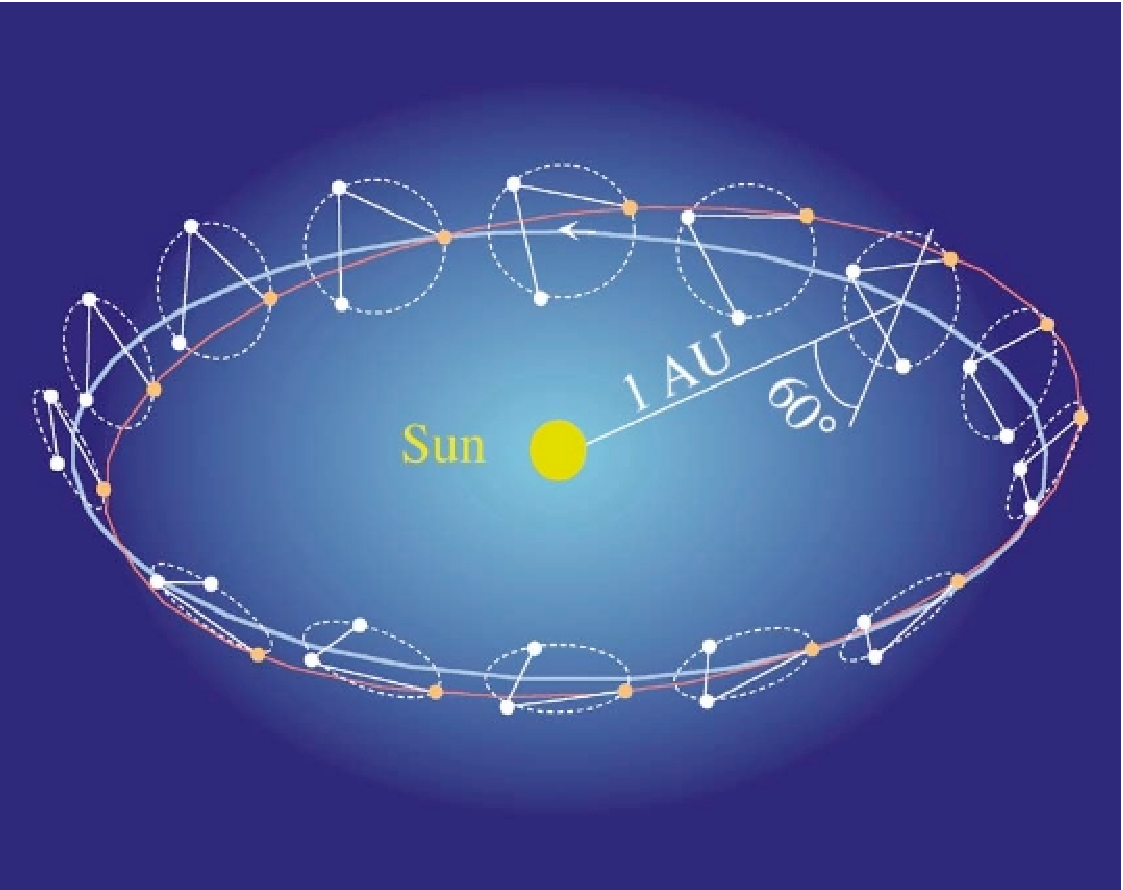} 
    \caption{\small{Orbit of the LISA mission [Credit: ESA]}} \label{fig:LISA orbit}
 \end{figure}

\section{ASTROD}

The general concept of ASTROD (Astrodynamical Space Test of Relativity Using Optical Devices) is to use drag-free spacecraft flying in formation within the solar system, using optical ranging between them, to map the gravitational field of the solar system, measure related solar system parameters, test relativistic gravity, observe solar g-mode oscillations, and detect GWs. The baseline plan for ASTROD was proposed in 1993, with conceptual and laboratory studies initiated simultaneously.

\section{ASTROD-GW Space Gravitational Wave Detection Program}

\begin{figure}[htbp]
    \centering  
    \includegraphics[width=0.6\textwidth]{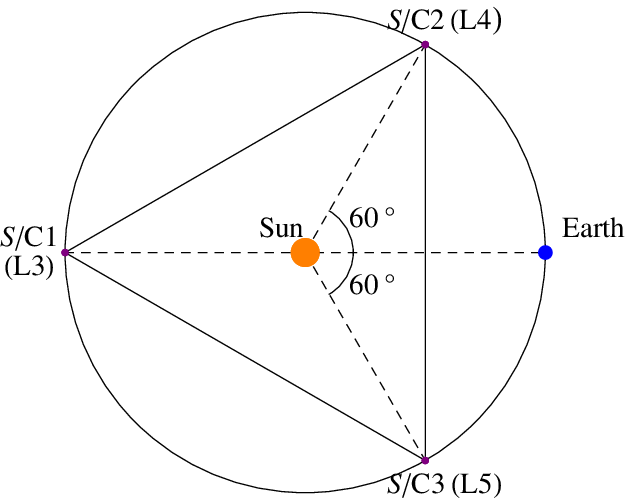} 
    \caption{\small{Orbit of the ASTROD-GW mission}} \label{fig:ASTROD-GW orbit}
 \end{figure}

Considering the need to optimize ASTROD missions for GW detection, the spacecraft forming the space interferometer implement nearly equal arm lengths. The three spacecraft are located near the Sun-Earth Lagrangian points L3, L4, and L5, forming an almost equilateral triangular array, as shown in Figure \ref{fig:ASTROD-GW orbit}, with an arm length of about 260 million kilometers (1.732 astronomical units). The three spacecraft conduct laser interferometric ranging between each other.

For ASTROD-GW, which focuses on GW detection, the orbit design scheme can be optimized. For the Earth-Sun Lagrangian points L3, L4, and L5, L4 and L5 are stable, while L3 is unstable, but with an instability timescale of about 50 years, making it effectively a stable point for missions lasting 10-20 years. With this choice, the spacecraft are about 260 million kilometers apart, 52 times the arm length of LISA, allowing detection of GWs at frequencies 52 times lower than LISA. Considering ASTROD-GW will be after LISA, the requirement for acceleration noise is assumed to be as same as LISA's requirements, and for the Doppler shift between the three spacecraft, the requirements are less stringent than for LISA, allowing the use of LISA's Doppler frequency synthesis and related technologies.

\section{Super-ASTROD, DECIGO, BBO Space Gravitational Wave Detection Programs}

The Super-ASTROD \cite{[22]} mission concept involves using 3-5 spacecraft in 5 AU orbits and one spacecraft at the Sun-Earth Lagrangian point L1/L2. These spacecraft will use optical ranging to detect primordial GWs in the 100 nHz - 1 mHz frequency band, test fundamental of  space-time, and map the mass distribution and dynamics of the outer solar system.

DECIGO \cite{[23]} (DECi-hertz Interferometer Gravitational Wave Observatory) is a future Japanese space GW detector. Its goal is to detect various GWs in the 1 mHz - 100 Hz range, opening a new observation window for GW astronomy, primarily studying primordial GWs from the Big Bang. Its concept involves using 3 drag-free spacecraft separated by 1000 km, using Fabry-Perot Michelson interferometers to measure the relative displacement between the spacecraft.

The BBO \cite{[24],[25]} (Big Bang Observer) has an orbit similar to LISA, but BBO's arm length is 2-5 $\times 10^4$ km. As a follow-up to LISA, BBO aims to detect GWs in the 0.01 - 10 Hz band, filling the gap between ground and space GW detectors. Its primary goal is to study primordial GWs from the Big Bang.

\chapter{Principle of Time-Delay Interferometry}

To achieve the required sensitivity for laser interferometer space GW detectors, it is essential to reduce laser frequency noise. The closer the time delays of the two interfering laser beams match, the better the laser frequency noise can be reduced, and the closer the sensitivity can approach the target. In this section, we discuss the principles and methods of TDI.

ASTROD employs TDI in its interferometric scheme. In space interferometers, due to long distances, the laser reaching another spacecraft must be amplified before it can be transmitted back and to other spacecraft. The method of amplification involves phase-locking the local laser with the incoming weak light before transmitting it. In the development of phase-locking with weak light, National Tsing Hua University (Hsinchu) first achieved phase-locking with 2 pW of incoming light in 2000 \cite{25a}. The Jet Propulsion Laboratory (JPL) at Caltech further improved this to 40 fW of incoming light in 2008 \cite{25b}. In 1996, Wei-Tou Ni et al. \cite{25c,25d} proposed the following two TDI schemes:
\begin{description}
\item{(i)} \ \ 
Path 1:  S/C 3 $\rightarrow$ S/C 1 $ \rightarrow$ S/C 3 $ \rightarrow$ S/C 2 $\rightarrow$ S/C 3 \\
Path 2:  S/C 3 $\rightarrow$ S/C 2 $ \rightarrow$ S/C 3 $ \rightarrow$ S/C 1 $\rightarrow$ S/C 3 \\
After phase-locking and amplification, the laser beams following Path 1 and Path 2 interfere at S/C 3. If the optical paths of Path 1 and Path 2 are equal (i.e., the distances between spacecraft are constant), then the phase of the two beams will be the same at the start and after traveling equal paths, thus canceling out the laser noise. If GWs, Lense-Thirring effects, or other factors cause the optical paths to differ, the difference is relatively small. Due to the variation in spacecraft distances over time, the corresponding noise from these changes must be considered when stabilizing the laser frequency.
\item{(ii)} \ 
Path 1:  S/C 3 $\rightarrow$ S/C 1  $ \rightarrow$ S/C 2 $\rightarrow$ S/C 3 \\
Path 2:  S/C 3 $\rightarrow$ S/C 2  $ \rightarrow$ S/C 1 $\rightarrow$ S/C 3 \\
After phase-locking and amplification, the laser beams following Path 1 and Path 2 interfere at S/C 3.
\end{description}
These two schemes were later referred to as first-generation TDI for LISA mission, with second-generation schemes providing better cancellation \cite{Tinto1}. In the following, we provide a detailed explanation of first- and second-generation TDI as discussed in \cite{Tinto1}.

In a traditional Michelson interferometer, the two interferometric arms can be made precisely equal, ensuring that laser passing through both arms have the same time delay, effectively canceling out the laser phase noise. However, in space-based GW missions, due to the orbital dynamics of the spacecraft, the distances between the three spacecraft are continuously changing, preventing the formation of an equal-arm space interferometer. Consequently, laser frequency noise cannot be canceled out at the traditional receiver, necessitating the development of methods to handle unequal-arm space interferometers and remove laser noise from the signal.

\begin{figure}[htbp]
    \centering  
    \includegraphics[width=0.8\textwidth]{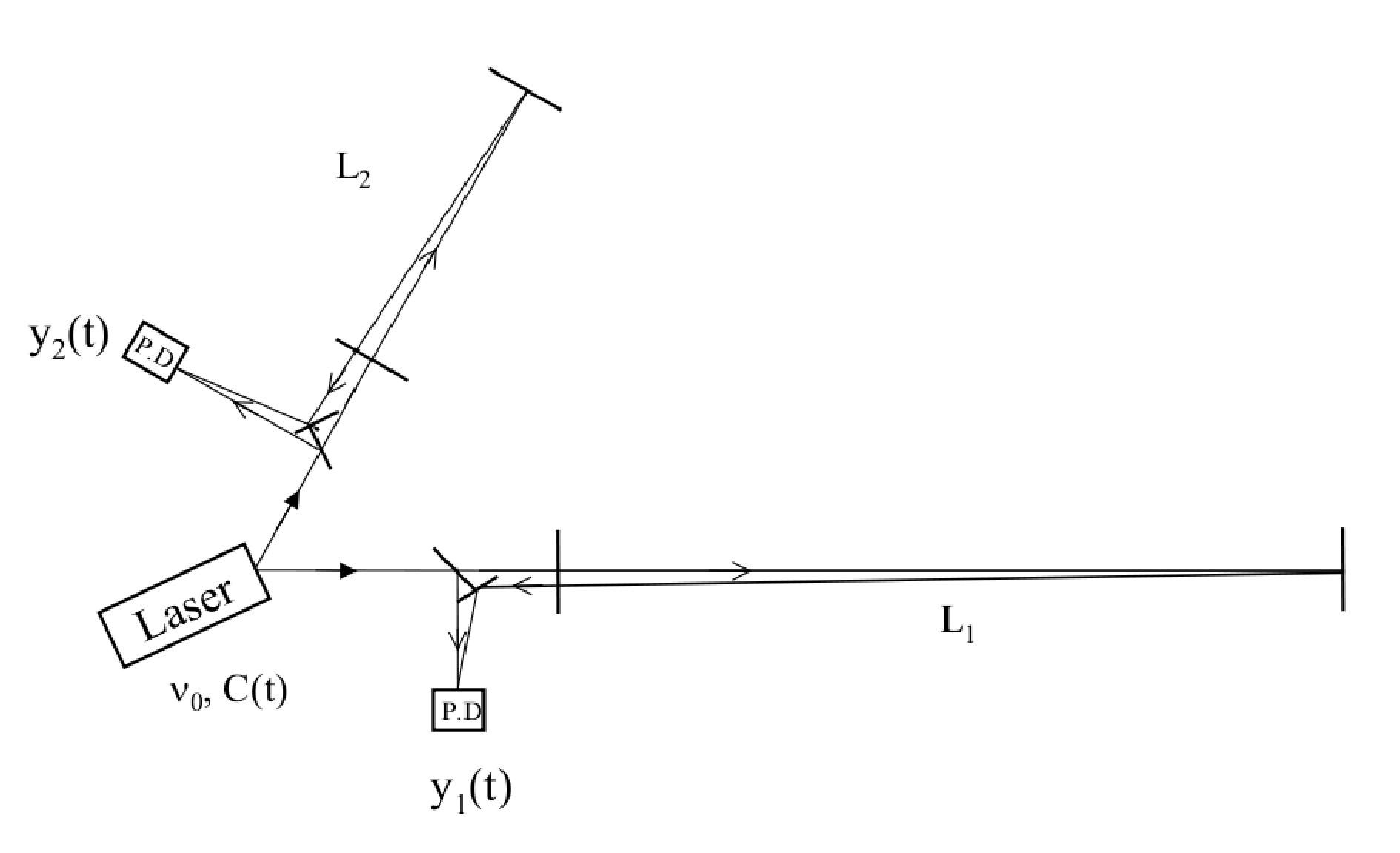} 
    \caption{\small{Schematic of unequal-arm interferometry. (figure reused from \cite{Tinto1})}}
    \label{fig:uneqarm}
 \end{figure}

For an unequal-arm Michelson interferometer, as shown in Fig. \ref{fig:uneqarm}, the two interferometric arms have different lengths, $L_1$ and $L_2$. The light beams passing through these arms do not interfere directly at a single photodetector but rather interfere with their respective outgoing beams upon return. The Doppler measurement results at their respective photodetectors are denoted as $y_1(t)$ and $y_2(t)$, with $C(t)$ representing the laser frequency noise. The GW signals entering each Doppler measurement are denoted as $h_1(t)$ and $h_2(t)$, and the remaining noise components as $n_1(t)$ and $n_2(t)$. The measurements $y_1(t)$ and $y_2(t)$ can be expressed as:
  \begin{equation}
   \label{y1t}
     y_1(t) = C(t - 2 L_1) - C(t) + h_1(t) + n_1(t),
  \end{equation}
   \begin{equation}
   \label{y2t}
     y_2(t) = C(t - 2 L_2) - C(t) + h_2(t) + n_2(t).
  \end{equation}
From equations (\ref{y1t}) and (\ref{y2t}), it is important to note the time-varying stochastic process $C(t)$ in the Doppler measurements $y_1$ and $y_2$. The laser signals entering the photodetector at arm 1 at time $t$ include noise $C(t-2L_1)$ from $2L_1$ earlier (assuming $c=1$), from which the current noise $C(t)$ must be subtracted, as shown in Eq. (\ref{y1t}). A similar result can be obtained for arm 2.

By comparing the difference between the measurements in Eqs. (\ref{y1t}) and (\ref{y2t}), we obtain:
 \begin{equation}
   \label{y1y2}
     y_1(t) - y_2(t) = C(t-2L_1) - C(t-2L_2) + h_1(t) - h_2(t) + n_1(t) - n_2(t).
  \end{equation}
Considering how laser noise enters Eq. (\ref{y1y2}) rather than Eqs. (\ref{y1t}) and (\ref{y2t}), a time-shifting $y_1(t)$ could be implemented before the laser traverses arm 2, i.e., $y_1(t-2L_2)$, and shifting $y_2(t)$ to before the laser traverses arm 1, i.e., $y_2(t-2L_1)$, we can derive:
 \begin{equation}
   \label{y1L2}
     y_1(t - 2 L_2) = C(t - 2 L_1 - 2 L_2) - C(t - 2 L_2) + h_1(t - 2 L_2) + n_1(t - 2 L_2),
  \end{equation}
 \begin{equation}
    \label{y2L1}
     y_2(t - 2 L_1) = C(t - 2 L_2 - 2 L_1) - C(t - 2 L_1) + h_2(t - 2 L_1) + n_2(t - 2 L_1).
  \end{equation}
  Comparing the difference between Eqs. (\ref{y1L2}) and (\ref{y2L1}), we obtain:
 \begin{equation}
 \begin{split}
     y_1(t - 2 L_2) - y_2(t - 2 L_1) = & C(t - 2 L_1) - C(t - 2 L_2) + h_1(t - 2 L_2) - h_2(t - 2 L_1) \\
         & + n_1(t - 2 L_2) - n_2(t - 2 L_1).
         \label{y1y2L}
  \end{split}  
  \end{equation}
  Comparing Eqs. (\ref{y1y2}) and (\ref{y1y2L}), we can see that both contain the same laser frequency noise. This implies that by subtracting Eq. (\ref{y1y2L}) from Eq. (\ref{y1y2}), we obtain new data $X$ that is free from laser frequency noise $C(t)$:
  \begin{equation}
    \begin{split}
     X  & \equiv [y_1 (t) - y_2 (t)] - [y_1 (t - 2 L_2) - y_2 (t - 2 L_1)] \\
        & = h_1(t) - h_2(t) + n_1(t) - n_2(t)  \\
         & -[h_1(t - 2 L_2) - h_2(t - 2 L_1) + n_1(t - 2 L_2) - n_2(t - 2 L_1)].
     \end{split}
  \end{equation}
  From the expression of $X$, we can see that by performing appropriate time delay operations in time domain and combining different Doppler measurements, it becomes feasible to eliminate the laser frequency noise. This is the essence of TDI.

\chapter{CGC2.7 ephemeris framework}

For the purposes of orbital design and numerical calculations for TDI, having a planetary ephemeris with sufficient accuracy is crucial. Therefore, it is necessary to introduce the ephemeris framework we used, CGC2.7 (CGC: Center for Gravitation and Cosmology). The early ephemeris framework CGC1.0 was established by Dah-Wei Chiou and Wei-Tou Ni \cite{Qiu1,Qiu2}, and the subsequent improved CGC2.0 was developed by Chien-Jen Tang and Wei-Tou Ni \cite{Tang,Tang2}. Our ephemeris framework, based on CGC2.0, is numbered as CGC2.7 ephemeris framework.
The main interactions considered in the CGC2.7 ephemeris framework include:
 \begin{itemize}
     \item Newtonian and post-Newtonian interactions among major celestial bodies (the Sun, the nine planets, the Moon, Ceres, Pallas, and Vesta);
     \item The effects of the Sun's second zonal harmonic and Earth's second to fourth zonal harmonics on other celestial bodies and on themselves;
     \item Newtonian perturbations of 349 asteroids on major celestial bodies.
   \end{itemize}
Additionally, we provide a brief introduction to the numerical integration algorithm of the ephemeris framework and compare the accuracy of the ephemeris of JPL's DE405 in this section.

\section{Newtonian and Post-Newtonian Interactions Between Celestial Bodies}
Based on the relativistic gravitational theory with two PPN (Parameterized Post-Newtonian) parameters, $\gamma$ and $\beta$, Brumberg\cite{Brumberg} derived the equations for Newtonian and post-Newtonian acceleration corrections that need to be considered for a celestial body $i$ under the influence of other celestial bodies in the adopted barycentric coordinate system, as shown in Eqs. (\ref{equ:PPN}-\ref{equ:PPNB}).
\begin{equation}
\label{equ:PPN}
\ddot{\mathbf{r}}_i=-\displaystyle\sum_{j\neq{i}}\frac{GM_{j}}{r_{ij}^3}\mathbf{r}_{ij}
  +\displaystyle\sum_{j\neq{i}} m_j (A_{ij} \mathbf{r}_{ij} + B_{ij} \dot{\mathbf{r}}_{ij})
\end{equation}
\begin{equation}
\label{equ:PPNA}
\begin{split} 
 A_{ij} =& \frac{{\dot{\mathbf{r}}_i}^2}{r^3_{ij}} -(\gamma+1)\frac{\dot{\mathbf{r}}^2_{ij}}{r_{ij}^3}+\frac{3}{2 r_{ij}^5}(\mathbf{r}_{ij}\dot{\mathbf{r}}_j)^2  + G[(2\gamma+2\beta+1)M_i+(2\gamma+2\beta)M_j]\frac{1}{r_{ij}^4} \\
 & + \displaystyle\sum_{k\neq{i,j}} GM_k [(2\gamma+2\beta)\frac{1}{r_{ij}^3 r_{ik}} + (2\beta-1)\frac{1}{r_{ij}^3 r_{jk}} + \frac{2(\gamma+1)}{r_{ij} r_{jk}^3} \\
 & -(2\gamma+\frac{3}{2})\frac{1}{r_{ik} r_{jk}^3}-\frac{1}{2r_{jk}^3}\frac{\mathbf{r}_{ij}\mathbf{r}_{ik}}{r_{ij}^3}] 
\end{split} 
\end{equation}
\begin{equation}
\label{equ:PPNB}
B_{ij} = \frac{1}{r_{ij}^3}[(2\gamma+2)(\mathbf{r}_{ij}\dot{\mathbf{r}}_{ij})+\mathbf{r}_{ij}\dot{\mathbf{r}}_{j}]
\end{equation}
In Eq. (\ref{equ:PPN}), the first term on the right-hand side represents the Newtonian interaction, while the second term accounts for the post-Newtonian interaction. Here, $\mathbf{r}_i, \dot{\mathbf{r}}_i, \ddot{\mathbf{r}}_i$ denote the position, velocity, and acceleration vectors of celestial body $i$ in the barycentric coordinate system of the solar system, respectively. $G$ is the gravitational constant, $M_i$ is the mass of celestial body $i$, and $m_i = GM_i/c^2$; the vector $\mathbf{r}_{ij} = \mathbf{r}_i - \mathbf{r}_j$ and $r_{ij} = |\mathbf{r}_i - \mathbf{r}_j|$ represent the position vector and distance between the centers of mass of celestial bodies $i$ and $j$, respectively; $c$ is the speed of light. 
For Einstein's General Relativity, the PPN parameters $\gamma = \beta = 1$. With $\alpha = 0$ in Eqs. \eqref{equ:PPNA} and \eqref{equ:PPNB}, we obtain the relativistic post-Newtonian acceleration correction.

\section{Effects from Extended Bodies}
For non-spherical extended celestial bodies, their interaction with other bodies, which are treated as point masses, is handled in the $O-\xi\eta\zeta$ coordinate system. The origin is placed at the center of mass of the non-spherical perturbing body, with the body's rotation axis as the polar axis and the equatorial plane as the coordinate plane. The $\xi$ axis points from the origin to the point mass, and the perturbing body's rotation axis lies in the $\xi\zeta$ plane. The $\eta$ axis is determined by the right-hand rule.

For any point K outside the perturbing body, its coordinates are expressed in spherical coordinates as $(r,\lambda,\theta)$. Let $\phi = \frac{\pi}{2} - \theta$, and the unit coordinate vector at point K is given by:
\begin{displaymath}
\mathbf{u}=\frac{\mathbf{r}}{r}=
\begin{pmatrix}
\cos\phi \cos\lambda \\  \cos\phi \sin\lambda \\  \sin \phi
\end{pmatrix}.
\end{displaymath}
The effect of the shape of a celestial body on the acceleration of a point mass is given by (see, e.g., Moyer, 1971),
\begin{small}
\begin{equation*}
\begin{split}
     \begin{pmatrix}
       \ddot{ \xi }_1 \\   \ddot{ \eta }_1 \\    \ddot{ \zeta }_1 
     \end{pmatrix}  
     = &  \frac{GM}{r^2}        \left\lbrace 
     \displaystyle\sum_{i=2}^\infty J_i (\frac{a}{r})^i
           \begin{pmatrix}
            (i+1) P_i (\sin \phi) \\     0 \\    -\cos \phi P^{'}_i (\sin \phi)
          \end{pmatrix}  \right\rbrace 
           \\ 
     & + \frac{GM}{r^2}    \left\lbrace 
       \sum_{i=2}^\infty (\frac{a}{r})^i \sum_{j=1}^i
       \begin{pmatrix}
         -(i+1)      P^j_i (\sin \phi) ( C_{ij} \cos{j \lambda}+S_{ij} \sin{j \lambda}) \\
        j \sec \phi  P^j_i (\sin \phi) (- C_{ij} \sin{j \lambda}+S_{ij} \cos{j \lambda}) \\
         \cos \phi   P'^{j}_i (\sin \phi) ( C_{ij} \cos{j \lambda}+S_{ij} \sin{j \lambda}) 
       \end{pmatrix}       
       \right\rbrace 
  \end{split} 
\end{equation*}
\end{small}
where $G$ is the gravitational constant, $M$ is the mass of the non-spherical perturbing body, $a$ is the radius of the perturbing body, $r$ is the distance between the center of mass of the non-spherical perturbing body and the point mass, $J_i$ are the zonal harmonic coefficients of the body, $P_i (\sin \phi)$ are the Legendre polynomials of degree $i$, $P^j_i (\sin \phi)$ are the $j$th power of $P_i (\sin \phi)$, and $C_{ij}$ and $S_{ij}$ are the spherical harmonic coefficients of the perturbing body\cite{Liu1}.

Let $ \mathbf{p}=\cos\phi \mathbf{\Phi}^0 + \sin\phi \mathbf{u} $ be the unit vector in the direction of the polar axis, where $ \mathbf{\Phi}^0$ is the unit gradient vector of $ \phi$. Thus, the expression for the effect of the 2-4 order zonal harmonics of the non-spherical perturbing body on the acceleration of the point mass can be respectively obtained:
\begin{equation}
\label{equ:2nd}
\ddot{\mathbf{r}}_{pm}=\frac{GM a^2 J_2 }{r^4} [(7.5 \sin^2 \phi -1.5) \mathbf{u} -3 \sin \phi \mathbf{p}],
\end{equation}
\begin{equation}
\ddot{\mathbf{r}}_{pm}=\frac{GM a^3 J_3 }{r^5} [(17.5 \sin^3 \phi -7.5 \sin\phi) \mathbf{u} -(7.5 \sin^2 \phi -1.5) \mathbf{p}],
\end{equation}
\begin{equation}
\label{equ:4th}
\ddot{\mathbf{r}}_{pm}=\frac{GM a^4 J_4 }{r^6} [(39.375 \sin^4 \phi -26.25 \sin^2 \phi +3) \mathbf{u} -(17.5 \sin^3 \phi -7.5 \sin\phi) \mathbf{p} ],
\end{equation}
where $J_i$ represents the $i$th zonal harmonic coefficient of the celestial body.

In this ephemeris framework, zonal harmonic terms of celestial bodies consider interactions between extended body and a point mass, including the following cases:
\begin{itemize}
\item Solar second-order zonal harmonics interacting with point masses of other celestial bodies. According to the formula for second-order zonal harmonics, the solar second-order zonal harmonic coefficient $J_2 = 2 \times 10^{-7}$, and the solar radius is $a = 696000$ km.
\item Earth's 2nd to 4th order zonal harmonics interacting with point masses of other celestial bodies. The Earth's zonal harmonic coefficients are $J_2 = 0.1082626 \times 10^{-2}$, $J_3 = -0.2533 \times 10^{-5}$, and $J_4 = -0.1616 \times 10^{-5}$, with the Earth's equatorial radius $a = 6378.137$ km. To compute Earth's polar direction, precession and nutation of the Earth need to be considered.
\end{itemize}

The unit rotation matrices are defined as:
 \begin{equation*}  
 \begin{split}
       \mathbf{R}_x (\theta)= &
      \begin{pmatrix}
       1 &  0 & 0 \\  0 & \cos \theta & \sin \theta \\ 0 & -\sin \theta & \cos \theta
      \end{pmatrix},    \\
        \mathbf{R}_y (\theta)= &
      \begin{pmatrix}
       \cos \theta & 0 & -\sin \theta \\  0 & 1 & 0 \\  \sin \theta & 0 & \cos \theta
      \end{pmatrix} ,     \\
        \mathbf{R}_z (\theta)= &
      \begin{pmatrix}
       \cos \theta  & \sin \theta  & 0 \\ -\sin \theta  & \cos \theta  & 0 \\  0 & 0 & 1
      \end{pmatrix} .
     \end{split}
   \end{equation*}
   
\subsection{the Sun's quadrupole moment harmonics} 

When calculating the solar quadrupole moment, the formula for the solar axis vector $\mathbf{p}$ in Eq. (\ref{equ:2nd}) is given by:
 \begin{equation*}
   \mathbf{p}=\mathbf{R}_x(- \varepsilon)
     \begin{pmatrix}
    \cos \Omega \sin I \\  \sin \Omega \sin I \\    \cos I
   \end{pmatrix} 
   =
     \begin{pmatrix}
      1 &  0 & 0 \\  0 & \cos \varepsilon & -\sin \varepsilon \\ 0 & \sin \varepsilon & \cos \varepsilon
     \end{pmatrix} 
     \begin{pmatrix}
     \cos \Omega \sin I \\  \sin \Omega \sin I \\    \cos I
     \end{pmatrix} 
    \end{equation*}
    where $\Omega$ and $I$ are respectively the longitude of the ascending node and inclination of the solar equatorial plane relative to the ecliptic plane; $I = 7^{\circ} 15'$,
$\Omega = 75^{\circ} 46' + 84'' \cdot T$, and $\varepsilon = 23^{\circ} 26'21''.448$ is the obliquity of the ecliptic for the J2000 epoch.
  \begin{equation}
  T=\frac{JD(t)-JD(2000)}{36525.0}=\frac{JD(t)-2451545.0}{36525.0},
  \end{equation}
 where $t$ is the dynamical time, JD represents the Julian Date, JD(t) denotes the Julian Date corresponding to the dynamical time, and JD(2000) represents the Julian Date corresponding to the J2000 epoch, which is 2451545.0.


\subsection{Earth's 2nd to 4th degree zonal harmonics}

To obtain the Earth's pole vector $\mathbf{p}$, we need to compute the Earth's precession and nutation. In CGC2.7, instead of numerical integration, we calculate the precession and nutation using the theoretical framework outlined by J. Wahr and H. Kinoshita in the IAU 1980 theory \cite{Xia1,Cal}.

\textbf{Earth's Precession}

Precession refers to the transformation between the epoch mean equatorial geocentric coordinates and the true equatorial geocentric coordinates, representing the difference between these two coordinate systems. The transformation method is outlined as follows \cite{Liu}:
\begin{equation*}
\mathbf{r}_m = (\mathbf{PR})\mathbf{R},
\end{equation*}
where $\mathbf{PR}$ is the precession matrix, composed of three rotation matrices:
\begin{equation*}
   \mathbf{PR} = \mathbf{R}_z(-z_A) \mathbf{R}_y(\theta_A) \mathbf{R}_z(-\xi_A),
\end{equation*}
where $\xi_A$, $z_A$, and $\theta_A$ are the equatorial precession angles, computed as:
\begin{equation*}
\left\{
\begin{aligned}
& \xi_A = 2306''.2181 T + 0''.30188 T^2 + 0''.017998 T^3 \\
& z_A = 2306''.2181 T + 1''.09468 T^2 + 0''.018203 T^3 \\
& \theta_A = 2004''.3109 T - 0''.42665 T^2 - 0''.041833 T^3
\end{aligned}
\right.
\end{equation*}
The corresponding right ascension precession $m_A$ and declination precession $n_A$ are:
\begin{equation*}
\left\{
\begin{aligned}
& m_A = \xi_A + z_A = 4612''.4362 T + 1''.39656 T^2 + 0''.036201 T^3 \\
& n_A = \theta_A
\end{aligned}
\right.
\end{equation*}

\textbf{Nutation of the Earth} 
 
 Instantaneous mean equatorial geocentric coordinate system and instantaneous true equatorial geocentric coordinate system conversion. The difference between these two coordinate systems is known as \textbf{nutation}. The conversion method is according to \cite{Liu}:
  \begin{equation*}
   \mathbf{r}_t = (\mathbf{NR})\mathbf{r}_m
   \end{equation*}
where $\mathbf{NR}$ is the nutation matrix, composed of three rotation matrices:
  \begin{equation*}
     \mathbf{NR}  = \mathbf{R}_x (-\Delta \varepsilon ) \mathbf{R}_y (\Delta \theta) \mathbf{R}_z (-\Delta \mu),
   \end{equation*}
where $\Delta \mu$, $\Delta \theta$, and $\Delta \varepsilon$ represent nutation in right ascension, nutation in declination, and nutation in obliquity, respectively. The nutation series adopted is from IAU(1980), and for meter precision, the first 20 terms of this series are used. The computation formulas are as follows:
  \begin{equation*}
     \left\{
     \begin{aligned}
     & \Delta \psi = \sum_{j=1}^{20} (A_{0j}+A_{1j} t) \sin (\sum_{i=1}^5 k_{ji} \alpha (t))   \\
     & \Delta \varepsilon = \sum_{j=1}^{20} (B_{0j}+B_{1j} t) \cos (\sum_{i=1}^5 k_{ji} \alpha (t)) 
     \end{aligned}
     \right.
    \Rightarrow
     \left\{
     \begin{aligned}
     & \Delta \mu = \Delta \psi \cos \varepsilon \\
     & \Delta \theta = \Delta \psi \sin \varepsilon
     \end{aligned}
     \right.
     \end{equation*}
   where $\psi$ is nutation in longitude, and $\Delta \mu$ and $\Delta \theta$ are nutation in right ascension and nutation in declination, respectively. The formula for the obliquity of the ecliptic $\varepsilon$ is:
     \begin{equation*}
     \varepsilon = 23^{\circ} 26' 21''.448 - 46''.8150 T.
     \end{equation*}
  The formulas for calculating the five angular terms in the nutation series are:
     \begin{equation*}
     \left\{
     \begin{aligned}
     & \alpha_1 = 134^{\circ} 57'46''.733 + (1325^r + 198^{\circ} 52'02''.633) T + 31''.310 T^2 \\
     & \alpha_2 = 357^{\circ} 31'39''.804 + (  99^r + 359^{\circ} 03'01''.224) T -  0''.577 T^2 \\
     & \alpha_3 =  93^{\circ} 16'18''.877 + (1342^r +  82^{\circ} 01'03''.137) T - 13''.257 T^2 \\
     & \alpha_4 = 297^{\circ} 51'01''.307 + (1236^r + 307^{\circ} 06'41''.328) T -  6''.891 T^2 \\
     & \alpha_5 = 125^{\circ} 02'40''.280 - (   5^r + 134^{\circ} 08'10''.539) T +  7''.455 T^2 
     \end{aligned}
     \right.
     \end{equation*}
 where $1^r = 360^{\circ}$, and the relevant coefficients for the first 20 terms of the nutation series are shown in Table \ref{tab:tab20}.

\begin{table}[tbp]
\small
\caption{IAU1980 coefficients for the first 20 terms of the nutation} \label{tab:tab20}
\centering
\renewcommand{\arraystretch}{1.15}
\begin{tabular}{r|l|ccccc|cc|cc}
\hline
$j$ & period (Days) & $k_{j1}$ & $k_{j2}$ & $k_{j3}$ &$k_{j4}$ &$k_{j5}$ & $A_{0j}$\footnotemark &$A_{1j}$  & $B_{0j}$ & $B_{1j}$ \\
\hline
 1  &  6798.4   &  0 &  0 &  0 &  0 &  1   & -171996 & -174.2 &  92025 &  8.9 \\
 2  &  182.6    &  0 &  0 &  2 & -2 &  2   & -13187  & -1.6   &  5736  & -3.1 \\
 3  &  13.7     &  0 &  0 &  2 &  0 &  2   & -2274   & -0.2   &  977   & -0.5 \\
 4  &  3399.2   &  0 &  0 &  0 &  0 &  2   &  2062   &  0.2   & -895   &  0.5 \\
 5  &  365.2    &  0 &  1 &  0 &  0 &  0   &  1426   & -3.4   &  54    & -0.1 \\
\hline
 6  &  27.6     &  1 &  0 &  0 &  0 &  0   &  712    &  0.1   & -7     &  0.0 \\
 7  &  121.7    &  0 &  1 &  2 & -2 &  2   & -517    &  1.2   &  224   & -0.6 \\
 8  &  13.6     &  0 &  0 &  2 &  0 &  1   & -386    & -0.4   &  200   &  0.0 \\
 9  &  9.1      &  1 &  0 &  2 &  0 &  2   & -301    &  0.0   &  129   & -0.1 \\
 10 &  365.3  &  0 & -1 &  2 & -2 &  2  & 217    & -0.5    & -95   & 0.3 \\
 \hline
 11 &  31.8     &  1 &  0 &  0 & -2 &  0   & -158    &  0.0   & -1     &  0.0 \\
 12 &  177.8    &  0 &  0 &  2 & -2 &  1   &  129    &  0.1   &  70    &  0.0 \\
 13 &  27.1     & -1 &  0 &  2 &  0 &  2   &  123    &  0.0   & -53    &  0.0 \\
 14 &  27.7     &  1 &  0 &  0 &  0 &  1   &  63     &  0.1   & -33    &  0.0 \\
 15 &  14.8     &  0 &  0 &  0 &  2 &  0   &  63     &  0.0   & -2     &  0.0 \\
 \hline
 16 &  9.6      & -1 &  0 &  2 &  2 &  2   & -59     &  0.0   &  26    &  0.0 \\
 17 &  27.4     & -1 &  0 &  0 &  0 &  1   & -58     & -0.1   &  32    &  0.0 \\
 18 &  9.1      &  1 &  0 &  2 &  0 &  1   & -51     &  0.0   &  27    &  0.0 \\
 19 &  205.9    &  2 &  0 &  0 & -2 &  0   &  48     &  0.0   &  1     &  0.0 \\
 20 &  1305.5   & -2 &  0 &  2 &  0 &  1   &  46     &  0.0   & -24    &  0.0 \\
\hline
\end{tabular} 
\end{table}
\footnotetext{In, Table \ref{tab:tab20}, $A_{0j}, A_{1j}, B_{0j},B_{1j}$ are in unit of $0''.0001$}
Based on the above discussion, the transformation from the epoch mean equatorial geocentric coordinates to the instantaneous true equatorial geocentric coordinates is given by:
 \begin{equation}
   \mathbf{r}_t = (\mathbf{NR})(\mathbf{PR})\mathbf{r}.
 \end{equation}
 
\section{Perturbations of Asteroids}

Asteroid data numbered up to April 14, 2010, were downloaded from the Lowell Asteroid Database \cite{Bowell}. Asteroids with diameter data and classified density were selected to calculate asteroid masses. Due to the large number of asteroids, the handling in the ephemeris is as follows:
\begin{itemize}
    \item Select asteroids with mass $ \mu > 10^{-15} $ for (1) Ceres, (2) Pallas, and (4) Vesta, along with the Sun, the eight major planets, Pluto, and the Moon, for numerical integration to obtain state data.
    \item For the remaining selected 349 asteroids, numerical integration is not used to compute their state at a given dynamical time. Instead, their states at specific times are calculated based on their orbital elements and Keplerian motion, using the method of undisturbed two-body motion, followed by calculation of the asteroids' perturbation on the other 11 major integrated bodies.
\end{itemize}

Table \ref{tab:astorid} presents the classification statistics of the selected 349 asteroids. Density values for C, S, and M classifications are taken from JPL DE405 (Standish, 1998). The density for E classification is based on the average value from Wasson (1974). The G classification is considered a subtype of C, therefore using the same density value as C. For U-type asteroids, the density is calculated as a weighted average of densities from the other types \cite{Tang2}.

\begin{table}[htbp]
\small
\caption{Classification and Statistics of Asteroids(Bowell, 1999)}
\label{tab:astorid}
\centering
\vspace{6pt}
\renewcommand{\arraystretch}{1.2}
\begin{tabular}{ccccc}
\hline  Type & Symbol & Density (g/$\text{cm}^2$) & Number & Mass $(\text{AU}^3 / \text{day}^2)$ \\
\hline 
Carbonaceous chondrite & C & 1.8 & 141 & $6.36 \times10^{-14} $ \\
Silicate (S-type) &  S &  2.4 & 93  & $ 2.37\times 10^{-14} $ \\
Metallic (M-type)  &  M &  5.0 & 30  &  $1.39\times 10^{-14}$  \\
Enstatite achondrite  &  E &  3.65 & 3  &  $1.88\times 10^{-16}$  \\
Extremely ultraviolet     &  G &  1.8 & 5   &  $ 3.07\times 10^{-15} $ \\
Others          &  U &  2.16 & 77 &  $2.11\times 10^{-14}$  \\
\hline
\end{tabular} 
\end{table}

 \textbf{Newtonian Perturbations of Selected Asteroids}\cite{Liu} \\
The Kepler orbit for asteroids is described as
  \begin{equation}
  \label{equ:Eorbit}
  E-e \sin E =n(t-\tau) ,
  \end{equation}
where $\tau$ is the time when the asteroid passes through perihelion, and $n=\frac{2 \pi}{T}$. By using Eq. (\ref{equ:Eorbit}), we can determine $E$, and subsequently, with known orbital elements of the asteroid, derive its position and velocity vectors.  
The transformation of asteroid orbital elements to position and velocity vectors in a Cartesian coordinate system is given by:
   \begin{equation} \label{equ:R4}
    \mathbf{r}  = a( \cos E -e) \widehat{P} + a \sqrt{1-e^2} \sin E \widehat{Q},
     \end{equation} 
   \begin{equation} \label{equ:Rdot4}
   \dot{\mathbf{r}} = \frac{\sqrt{\mu a}}{r} [-\sin E \widehat{P} +\sqrt{1-e^2} \cos E \widehat{Q}],
   \end{equation}
  in which
    \begin{equation*}
   \widehat{P}=
     \begin{pmatrix}
    \cos \Omega \cos\omega -\sin \Omega \sin\omega \cos i \\
    \sin \Omega \cos\omega +\cos \Omega \sin\omega \cos i \\ 
    \sin \omega \sin i
   \end{pmatrix} ,
   \end{equation*}
   \begin{equation*}
   \widehat{Q}=
     \begin{pmatrix}
    -\cos \Omega \sin\omega -\sin \Omega \cos\omega \cos i \\
    -\sin \Omega \sin\omega +\cos \Omega \cos\omega \cos i \\ 
     \cos \omega \sin i
   \end{pmatrix} ,
    \end{equation*}
where the orbital elements are defined as follows:
$a$ is the semi-major axis of the orbit;
$e$ is the eccentricity of the orbit;
$\omega$ is the argument of periapsis;
$\Omega$ is the longitude of the ascending node;
$i$ is the inclination of the orbital plane;
$M$ is the mean anomaly.
 
\section{Numerical Integration Methods}

For a second-order differential equation
\begin{equation}
\frac{d^2 r}{dt^2} = f(t,r,\dot{r}),
\end{equation}
the classical fourth-order Runge-Kutta numerical integration method is employed \cite{CWXu}:
\begin{equation} \label{equ:RK4}
\left\{
    \begin{aligned}
   & r_{n+1} = r_n + h \dot{r}_n + \frac{h}{6} (M_1 + M_2 +M_3)  \\
   & \dot{r}_{n+1} = \dot{r}_n + \frac{1}{6} (M_1 + 2 M_2 + 2 M_3 + M_4)  \\
   & M_1 = h f(t_n, r_n, \dot{r}_n) \\
   & M_2 = h f(t_n + \frac{h}{2} , r_n +\frac{h}{2} \dot{r}_n , \dot{r}_n + \frac{M_1}{2} ) \\
   & M_3 = h f(t_n + \frac{h}{2} , r_n +\frac{h}{2} \dot{r}_n + \frac{h}{4} M_1 , \dot{r}_n + \frac{M_2}{2} ) \\
   & M_4 = h f(t_n + h ,r_n + h \dot{r}_n + \frac{h}{2} M_2 , \dot{r}_n + M_3 ) 
   \end{aligned}   
    \right.
\end{equation}
where $h$ is a step, and $\dot{r}$ is the first-order differential with respect to $t$.
CGC2.7 ephemeris utilizes the fourth-order Runge-Kutta numerical integration method, implemented in C++ programming language.
The integration step size is 0.02 days. The starting time is June 21, 2028, 12:00 (JD 2461944.0). Initial positions for 14 celestial bodies are provided by JPL's DE405 ephemeris \cite{DE}, and the masses and harmonic parameters for each body are obtained from the DE405 ephemeris header files.
 
\section{Accuracy of CGC2.7 comparing to DE405}

\begin{figure}[htbp]
    \centering  
     \vspace{12pt} 
    \includegraphics[width=0.32\textwidth]{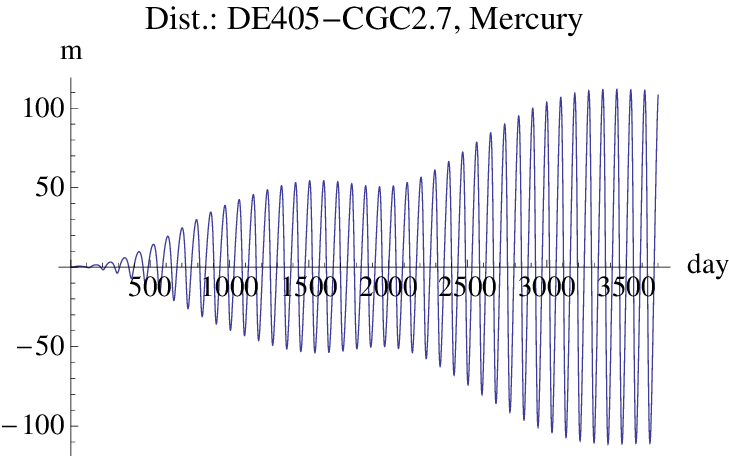} \ 
    \includegraphics[width=0.32\textwidth]{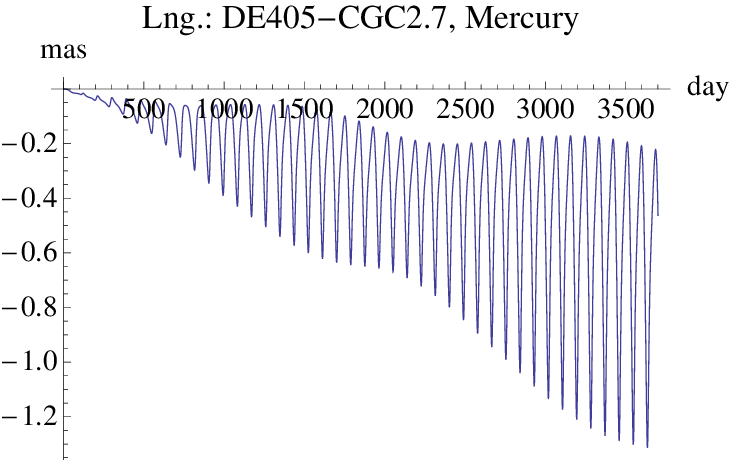} \ 
    \includegraphics[width=0.32\textwidth]{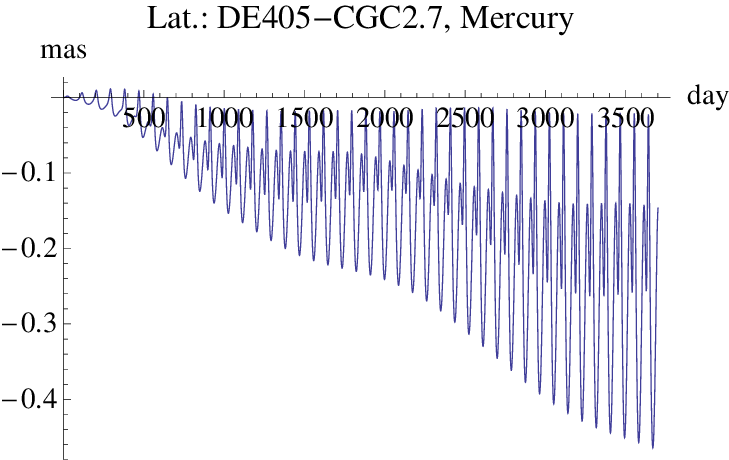}
    \vspace{12pt} 
    \includegraphics[width=0.32\textwidth]{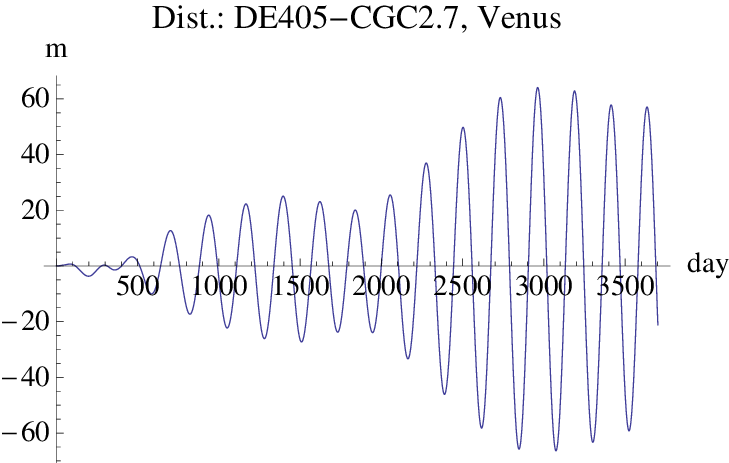} \ 
    \includegraphics[width=0.32\textwidth]{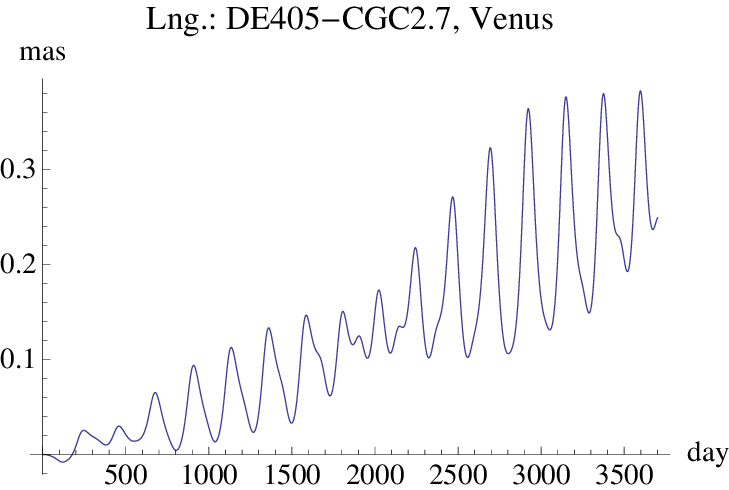} \ 
    \includegraphics[width=0.32\textwidth]{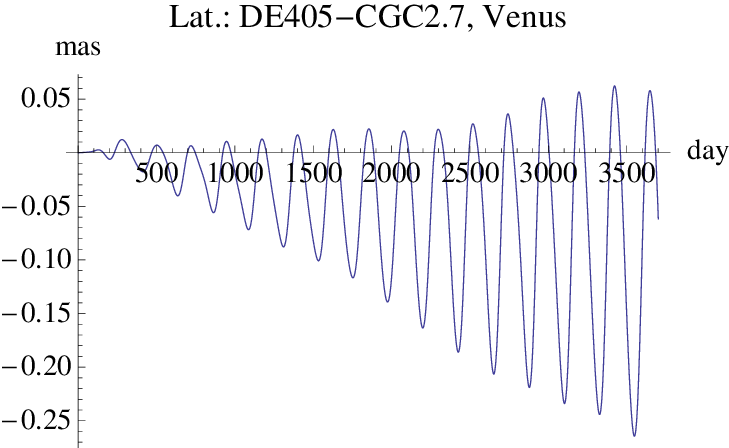}
    \vspace{12pt} 
    \includegraphics[width=0.32\textwidth]{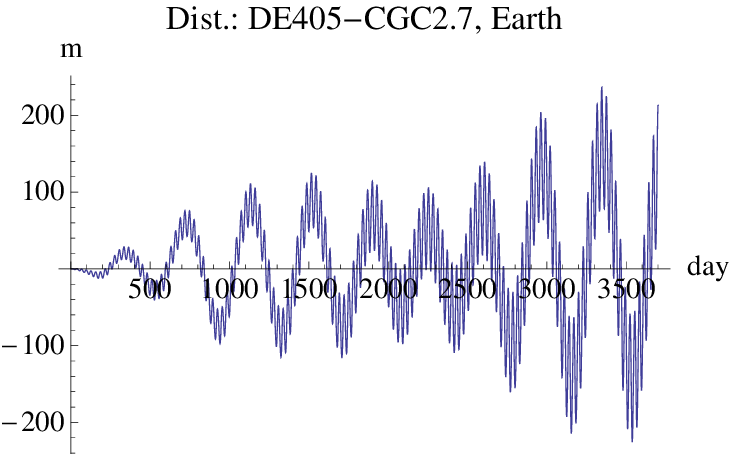} \ 
    \includegraphics[width=0.32\textwidth]{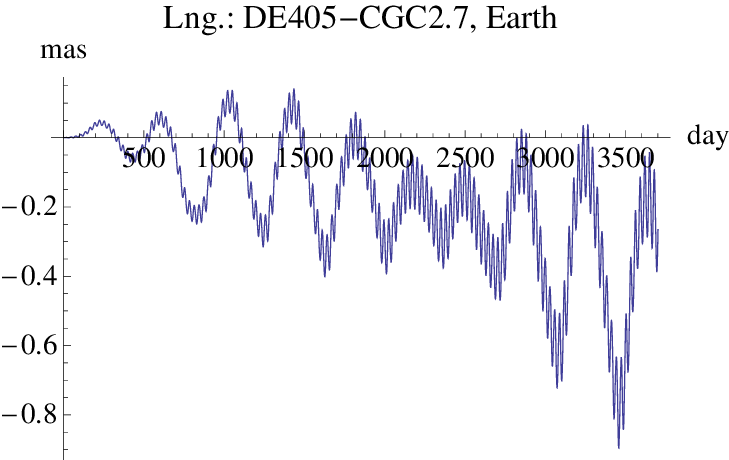} \ 
    \includegraphics[width=0.32\textwidth]{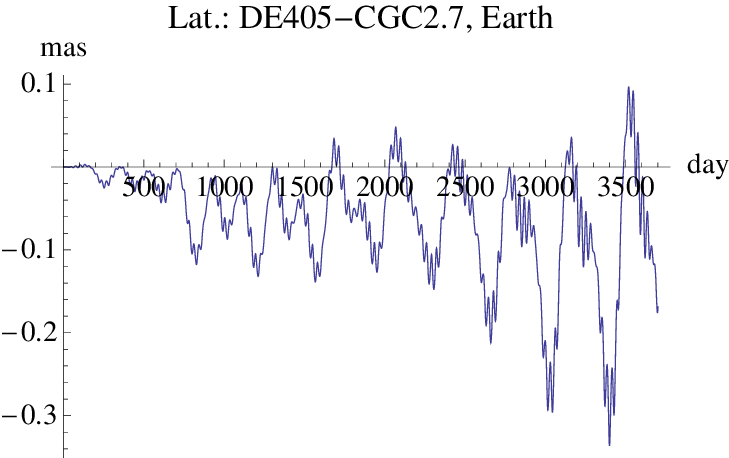}
    \includegraphics[width=0.32\textwidth]{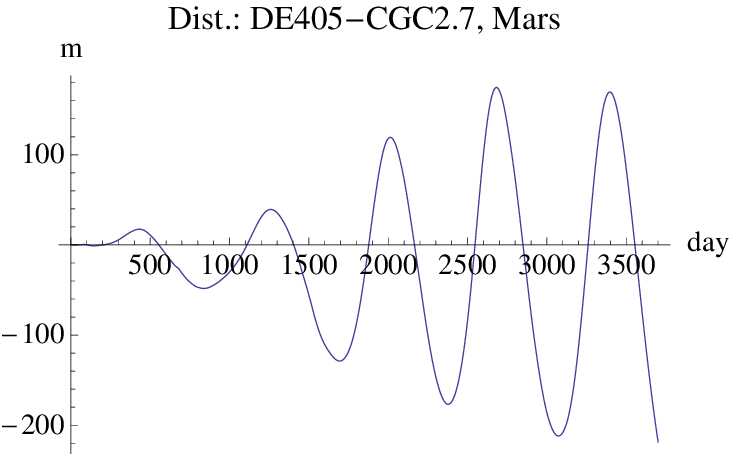} \ 
    \includegraphics[width=0.32\textwidth]{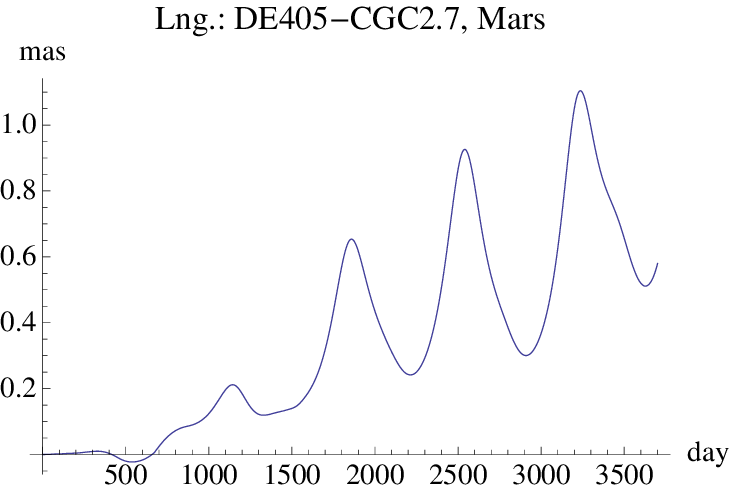} \ 
    \includegraphics[width=0.32\textwidth]{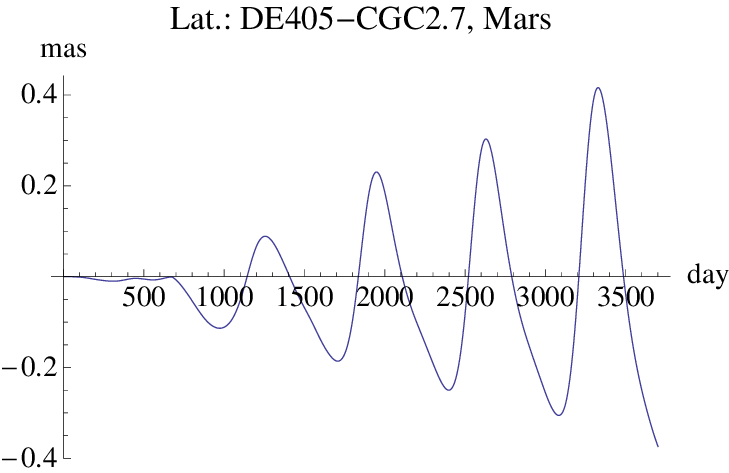}
      \caption{ Comparisons of inner planets orbit in the ecliptic coordinate heliocentric system between CGC2.7 and DE405.}
     \label{fig:DE-CGC}
\end{figure}

To assess the accuracy of CGC2.7 ephemeris framework, we compared it with the DE405 ephemeris \cite{DE}. which released by JPL.
The comparison results are shown in Figure \ref{fig:DE-CGC}. We compared the data of inner planets (Mercury, Venus, Earth, and Mars) over 10 years. Using the Sun as the origin and employing ecliptic coordinates from DE405, we plotted the differences between the two calculated orbits for each planet. 
For example, in the first row of comparisons for Mercury in Figure \ref{fig:DE-CGC}, the first column shows the variation over time of the difference in heliocentric distances between DE405 and CGC2.7. The second column displays the difference in Mercury's ecliptic longitude over time, and the third column shows the difference in ecliptic latitude.
Similarly, the second row depicts comparison for Venus's orbit, the third row shows Earth's orbit, and the fourth row shows results of Mars's orbit.
\chapter{ASTROD-GW Mission Orbit}

\begin{figure}[ht]
\centering
\includegraphics[scale=0.6]{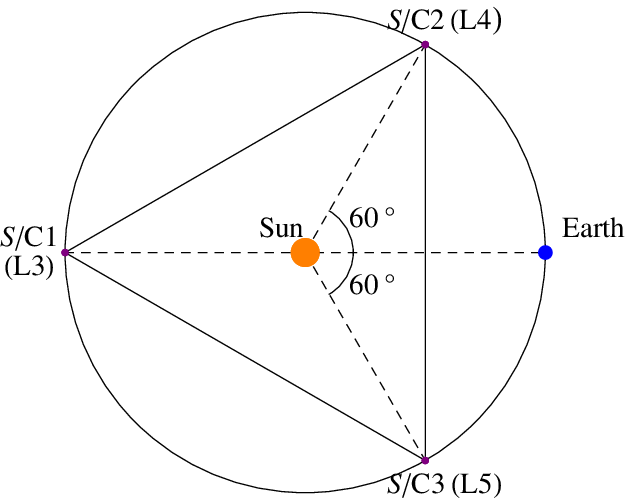}
\caption{Orbit of the ASTROD-GW Mission.} \label{fig:ASTROD-GWOrbit}
\end{figure}

The ASTROD-GW mission employs three spacecraft located near the Sun-Earth Lagrange points L3, L4, and L5. These spacecraft orbit the Sun in nearly circular orbits, forming an approximate equilateral triangle as shown in Figure \ref{fig:ASTROD-GWOrbit}. Each side of the triangle is approximately 260 million kilometers (1.732 AU). Lagrange points L4 and L5 are stable, while L3 is an unstable point with an instability timescale of about 50 years. Given that the expected mission duration of ASTROD-GW is 20 years, the spacecraft can maintain stable positions near L3, L4, and L5 throughout the mission period. Additionally, each spacecraft is equipped with micropropulsion thrusters for orbit corrections to mitigate disturbances and maintain precise positioning.

The LISA spacecraft orbits the Sun in a triangular formation with arms of approximately 5 million kilometers each. Given that ASTROD-GW has an interferometric arm length 52 times longer than LISA, if the relative arm length difference of ASTROD-GW can be controlled to be less than 1/52 of LISA's relative arm length difference and the relative velocity between spacecraft is not greater than that of LISA, then the technological requirements for laser interferometry will not be higher than those already developed for LISA. With its longer arm length, ASTROD-GW exhibits higher sensitivity to lower-frequency GWs. This chapter optimizes the mission orbit according to established methods \cite{Men,MenEn}.

\section{Initial Orbit Selection}

The initial time for ASTROD-GW mission orbit is set at 12:00 on June 21, 2028 (JD2461944). Considering the Solar System as a restricted three-body problem in an elliptical orbit on the ecliptic plane, we determine the locations of the five Lagrange points. Initially, the spacecraft are positioned near points L3, L4, and L5. The positions of the three spacecraft in the heliocentric ecliptic coordinates are as follows: L3 is located near (0, 1 AU, 0), L4 is near ($ \sqrt{3}/2 \ \text{AU} ,-1/2 \ \text{AU} ,0 $), and L5 is near ($ -\sqrt{3}/2 \ \text{AU}, -1/2 \ \text{AU}, 0$). This forms an approximately equilateral triangle with sides of about $\sqrt{3}$ AU, and each spacecraft is about 1 AU from the Sun. The initial velocities of the spacecraft are calculated based on the stellar year of the Earth's orbital period (365.26536 days). The orbital speed of the spacecraft in circular motion around the Sun is $v_0 = 0.01720209895 \ \text{AU/day}$. The initial velocity vectors of the spacecraft in the ecliptic plane are perpendicular to their initial position vectors: at L3, the spacecraft has an initial velocity of $(-v_0,0,0)$; at L4, the initial velocity is $(1/2 \ v_0 , \sqrt{3}/2 \ v_0, 0)$; and at L5, the initial velocity is $(1/2 \ v_0, -\sqrt{3}/2 \ v_0,  0 )$.

After determining the initial states of the spacecraft in the heliocentric ecliptic coordinates, the coordinates are transformed to the J2000 Solar System Barycentric (SSB) coordinates used for ephemeris calculations. The J2000.0 obliquity of the ecliptic is $\varepsilon = 23^\circ 26^\prime 21^{\prime\prime}.448$. The transformation from spacecraft positions and velocities in heliocentric ecliptic coordinates to Solar System Barycentric coordinates is given by Eq. (\ref{eq:5.1}), where $\mathbf{R}_{SC,SSB}$, $\mathbf{V}_{SC,SSB}$ represent the spacecraft positions and velocities in J2000 Solar System Barycentric coordinates, $\mathbf{R}_{SC,Sun}$, $\mathbf{V}_{SC,Sun}$ represent the spacecraft positions and velocities in heliocentric ecliptic coordinates, and $\mathbf{R}_{\odot,SSB}$, $\mathbf{V}_{\odot,SSB}$ represent the Sun's position and velocity in J2000 Solar System Barycentric coordinates.
\begin{equation} 
\label{eq:5.1}
  \begin{split}
   & \mathbf{R}_{SC,SSB} = \left[
       \begin{array}{rrr}
            1 & 0 & 0\\
            0 & \cos{\varepsilon} & -\sin{\varepsilon} \\
            0 & \sin{\varepsilon} & \cos{\varepsilon}
       \end{array}
     \right] \cdot \mathbf{R}_{SC,Sun} +\mathbf{R}_{\odot,SSB} ,
\\
   & \mathbf{V}_{SC,SSB}  = \left[
        \begin{array}{rrr}
        1 & 0 & 0\\
        0 & \cos{\varepsilon} & -\sin{\varepsilon} \\
        0 & \sin{\varepsilon} & \cos{\varepsilon}
        \end{array}
      \right] \cdot \mathbf{V}_{SC,Sun} +\mathbf{V}_{\odot,SSB} .
   \end{split}
\end{equation}

\section{Orbital Optimization}

Through the initial orbit selection, we determine the initial conditions of the spacecraft. By computing the spacecraft's orbit over a 20-year mission period and analyzing the spacecraft orbit data, we obtain graphs showing the orbital periods and heliocentric distances of the three spacecraft over time. On one hand, we adjust the average 20-year period of the spacecraft to align it as closely as possible with the Earth's orbital period. On the other hand, we adjust the spacecraft's eccentricity to approach a nearly circular orbit. However, the optimal period may not necessarily be exactly 1 stellar year. In necessary cases, we may slightly deviate the spacecraft's 20-year average period from 1 stellar year to achieve better optimization results.

\subsection{Orbital Optimization Methodology}

In this optimization process \cite{Men,MenEn,Men2}, we utilize the heliocentric coordinate system. 
The total energy of a planet's motion around the Sun is given by:
\begin{equation}
\label{eq:5.2}
E = \frac{mv^2}{2} - \frac{G(M+m)m}{r} = -\frac{G(M+m)m}{2a},
\end{equation}
where $M$ is the mass of the Sun, $m$ is the mass of the planet, $G$ is the gravitational constant, $v$ is the relative velocity between the planet and the Sun, $r$ is the heliocentric distance of the planet, and $a$ is the semi-major axis of the planet's elliptical orbit around the Sun.\\
The orbital period of a planet around the Sun is given by:
\begin{equation}
\label{eq:5.3}
T = \frac{2\pi a^{3/2}}{\sqrt{G(M+m)}}.
\end{equation}
To derive Eq. \eqref{eq:5.2}, take the total differential:
\begin{equation}
\label{eq:5.4}
mv^2 \frac{dv}{v} + \frac{G(M+m)m}{r} \frac{dr}{r} = \frac{G(M+m)m}{2a} \frac{da}{a}.
\end{equation}
Assuming a nearly circular orbit for the spacecraft, where $r \approx a$ in Eq. \eqref{eq:5.2}, we get:
\begin{equation}
\label{eq:5.5}
mv^2 \approx \frac{G(M+m)m}{a} \approx \frac{G(M+m)m}{r}.
\end{equation}
Substituting Eq. (\ref{eq:5.5}) into Eq. (\ref{eq:5.4}), we obtain the relationship:
\begin{equation}
\label{eq:5.6}
\frac{dv}{v} + \frac{dr}{r} = \frac{1}{2} \frac{da}{a}.
\end{equation}
Taking the logarithm of both sides of Eq. (\ref{eq:5.3}) and differentiating gives:
\begin{equation}
\label{eq:5.7}
\frac{dT}{T} = \frac{3}{2} \frac{da}{a} \quad \Rightarrow \quad da = \frac{2}{3} \frac{dT}{T} \cdot a.
\end{equation}

\subsubsection{Optimization of Mission Orbital Period}

Optimize the spacecraft's orbital period by adjusting the velocity and modifying the heliocentric distance. Taking the logarithm and differentiating Eq. \eqref{eq:5.5}, we obtain:
\begin{equation}
\label{eq:5.8}
2 \frac{dv}{v} \approx -\frac{da}{a} \approx -\frac{dr}{r}.
\end{equation}
From Eqs. \eqref{eq:5.7} and \eqref{eq:5.8}, combining these two equations yields:
\begin{equation}
dv \approx -\frac{1}{3} \frac{dT}{T} v, \quad dr \approx \frac{2}{3} \frac{dT}{T} r .
\end{equation}
In practical adjustment processes, the above equations should be written in vector form:
\begin{equation}
\delta \mathbf{V} \approx -\frac{1}{3} \frac{dT}{T} \mathbf{V}, \quad \delta \mathbf{R} \approx \frac{2}{3} \frac{dT}{T} \mathbf{R} .
\end{equation}
The formulas for calculating the adjusted velocity and position after adjusting the period are:
\begin{equation}
\label{eq:5.11}
\begin{split}
& \mathbf{V}_{new} = \mathbf{V} + \delta \mathbf{V} \approx \left(1 - \frac{1}{3} \frac{dT}{T}\right) \mathbf{V}, \\
& \mathbf{R}_{new} = \mathbf{R} + \delta \mathbf{R} \approx \left(1 + \frac{2}{3} \frac{dT}{T}\right) \mathbf{R}.
\end{split}
\end{equation}

\subsubsection{Optimization of Mission Orbital Eccentricity}

As following, we attempt to adjust the spacecraft's orbital eccentricity to make it closer to a circular orbit. From Eq. \eqref{eq:5.3}, it is evident that as long as the semi-major axis of the orbit remains unchanged, the orbital period will not change. Therefore, in the subsequent optimization steps, we initially set \( da = 0 \). From Eq. (\ref{eq:5.6}), we have:
\begin{equation}
\frac{dv}{v} = -\frac{dr}{r}
\end{equation}
Typically, based on the variation of heliocentric distance over time, we determine the adjustment \( dr \), and then adjust the velocity and position accordingly. The formulas used in the actual adjustment process are:
\begin{equation}
\delta \mathbf{V} = -\frac{dR}{R} \mathbf{V}, \quad \delta \mathbf{R} = \frac{dR}{R} \mathbf{R},
\end{equation}
where \textit{R} denotes the initial heliocentric distance. The formulas for calculating the adjusted velocity and position after adjusting the eccentricity are:
\begin{equation}
\label{eq:5.14}
\begin{split}
& \mathbf{V}_{new} = \mathbf{V} + \delta \mathbf{V} \approx \left(1 - \frac{dR}{R}\right) \mathbf{V}, \\
& \mathbf{R}_{new} = \mathbf{R} + \delta \mathbf{R} \approx \left(1 + \frac{dR}{R}\right) \mathbf{R}.
\end{split}
\end{equation}

\subsection{Process of Orbital Optimization}

In the process of optimizing the spacecraft's orbit, we use in the heliocentric ecliptic coordinate system. And the orbit is computed using the CGC2.7 ephemeris framework, followed by plotting and analysis. The graphs include variations over time in the arm lengths between spacecraft, variations in arm length differentials over time, changes in heliocentric distances of spacecraft over time, and changes in Doppler relative velocities between spacecraft over time.

\begin{figure}[ht]
    \centering
    \includegraphics[width=0.48\textwidth]{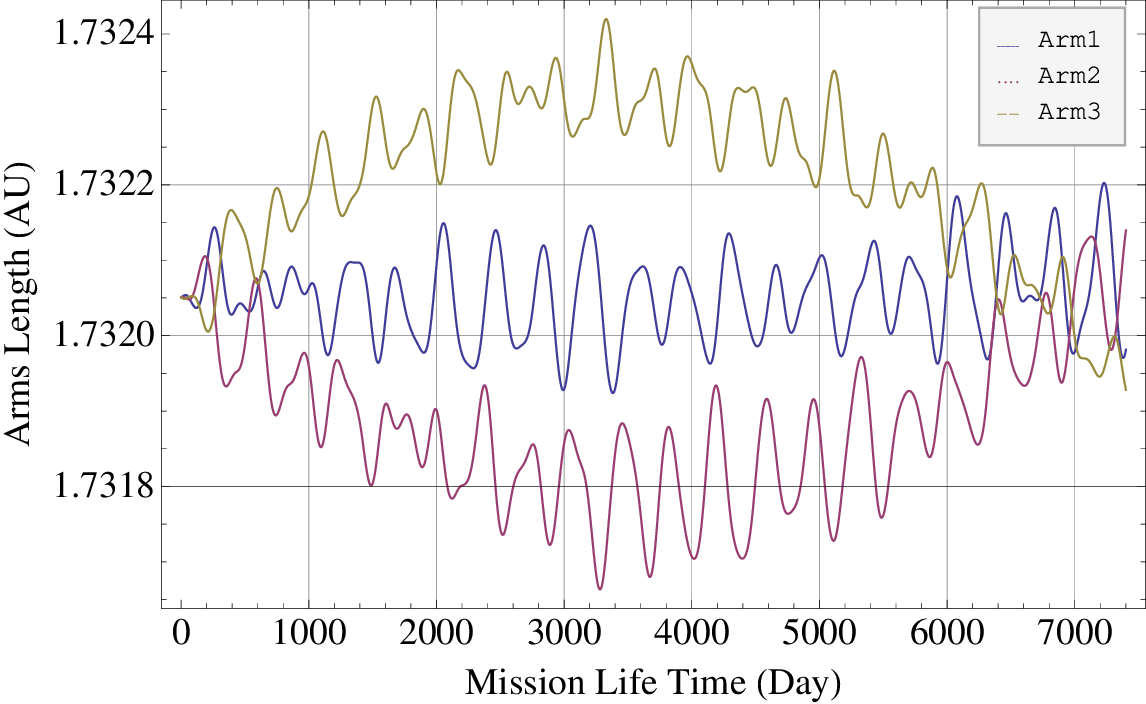} \
    \includegraphics[width=0.48\textwidth]{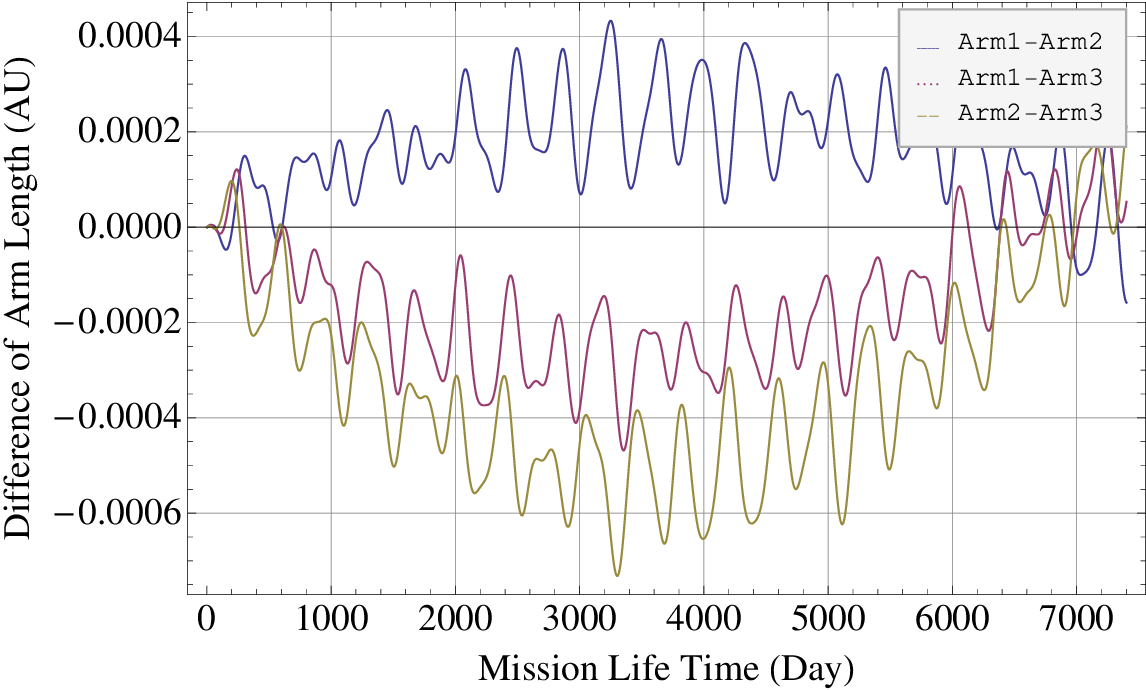} \
    \includegraphics[width=0.48\textwidth]{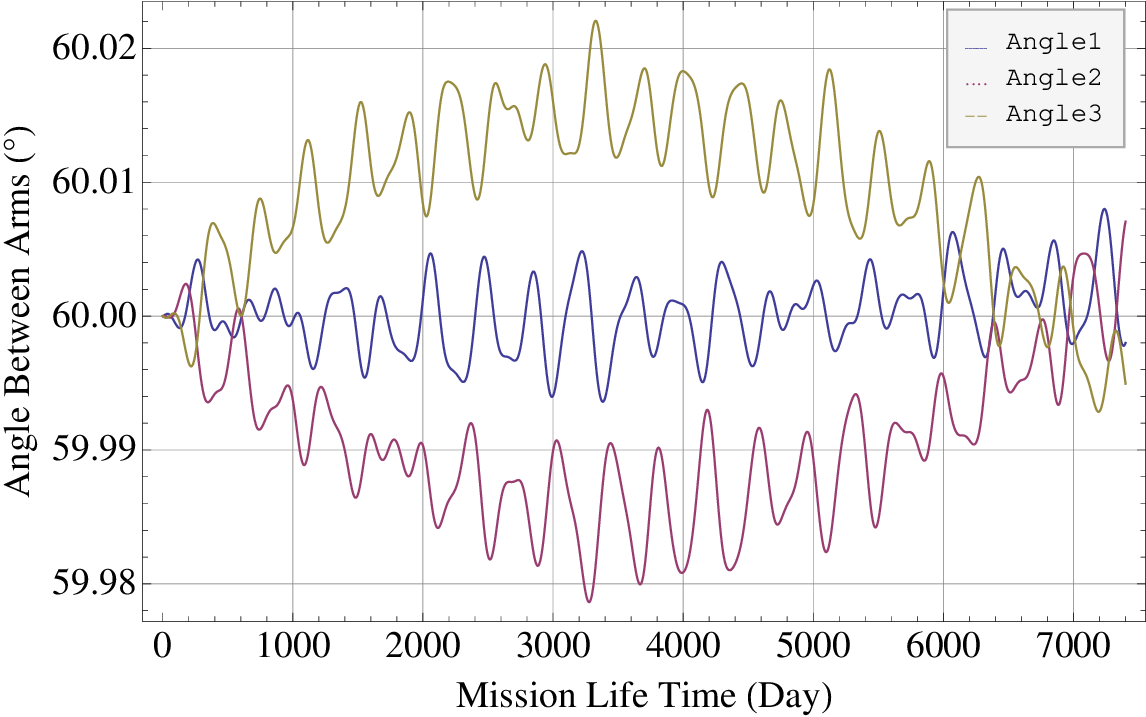} \
    \includegraphics[width=0.48\textwidth]{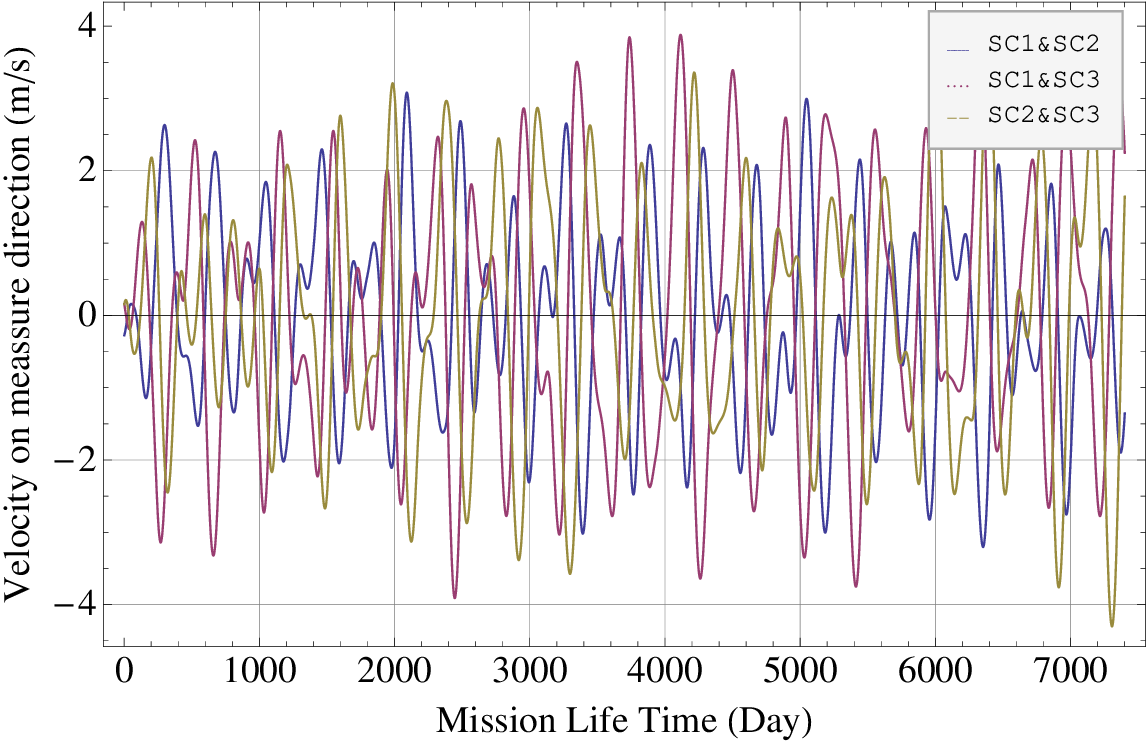}
    \caption{The arm lengths, difference between arm lengths, velocities in the measurement direction, and the angles between arms calculated using initial conditions from analytical equations. }
    \label{fig:optim1}
\end{figure}

\begin{figure}[ht]
    \centering
    \includegraphics[width=0.48\textwidth]{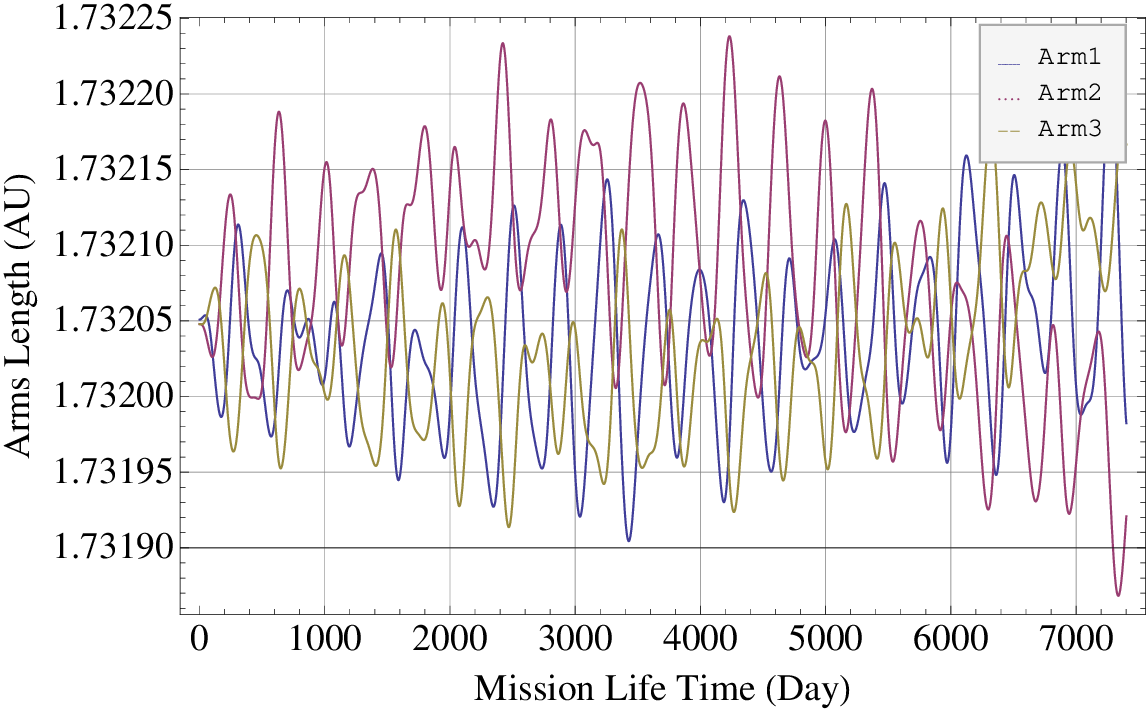} \ 
    \includegraphics[width=0.48\textwidth]{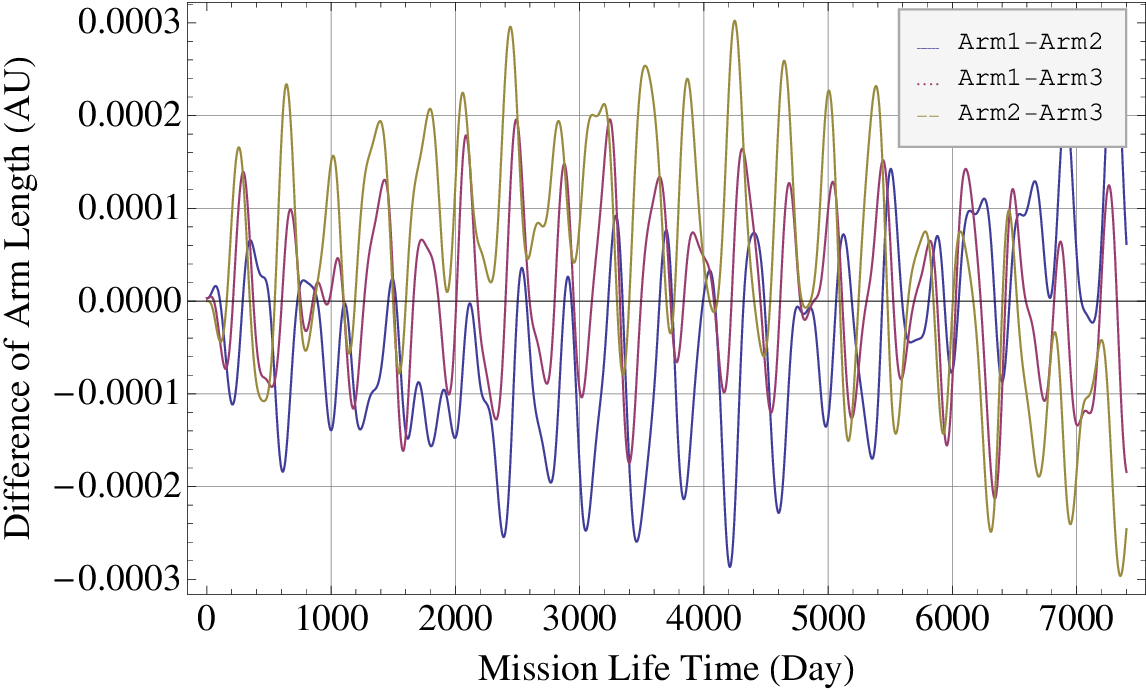} \ 
    \includegraphics[width=0.48\textwidth]{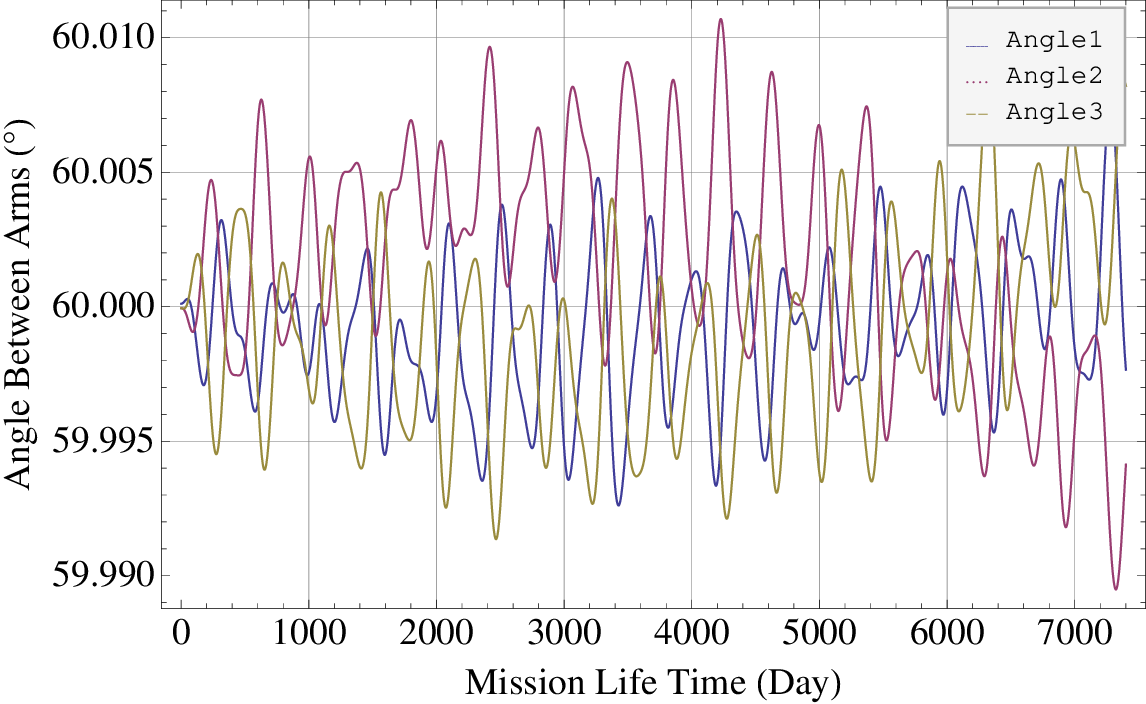} \ 
    \includegraphics[width=0.48\textwidth]{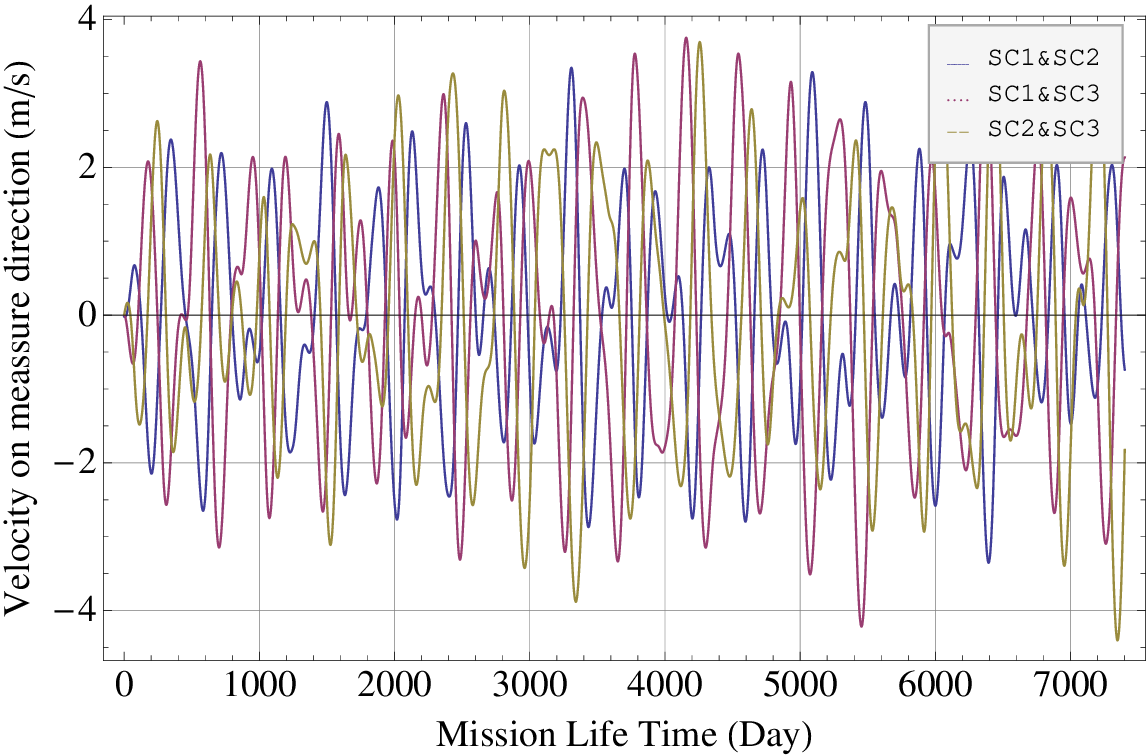}
    \caption{The changes in arm length, arm length differences, angles between arms, and spacecraft Doppler velocities over time after adjusting the orbit periods.}
   \label{fig:5.3}
\end{figure}

The optimization method for spacecraft orbit periods involves adjusting the spacecraft periods based on the variation of arm length differentials over time to minimize these differences. Due to various influencing factors, the average periods of the spacecraft change continuously over time. The period at L3 tends to decrease over time, while those at L4 and L5 tend to increase. Arm lengths Arm12 initially decrease and then increase, Arm13 increase initially and then decrease, while Arm23 shows relatively smaller changes, as shown in Figure \ref{fig:optim1}. According to Eq. (\ref{eq:5.11}), adjusting the periods involves reducing the initial velocity of S/C1 to lengthen its initial period, and increasing the initial velocities of S/C2 and S/C3 to shorten their initial periods. This approach compensates for deviations in the spacecraft periods during operation, thereby controlling the differences in periods among the three spacecraft within a certain range and appropriately managing the arm length differences. The initial conditions of the spacecraft after adjusting the mission orbit periods are summarized in Table \ref{tab:5.3}, with corresponding time-varying graphs shown in Figure \ref{fig:5.3}.

\begin{table}[ht] 
\footnotesize 
\caption{Initial position and velocities after optimizing the orbital periods at epoch JD2461944.0.}
{( J2000 solar-system-barycentric equatorial, in AU and AU/day )}
\label{tab:5.3}
\centering
\renewcommand{\arraystretch}{1.5}
\begin{tabular}{|c|r|r|r|}
\hline                  & x(AU)/vx(AU/day) & y(AU)/vy(AU/day) & z(AU)/vz(AU/day)  \\
\hline S/C1 position  &  1.15400625657242E-3   &   9.15261701184544E-1   &    3.96854368692135E-1 \\ 
\hline S/C1 velocity  & -1.72008300146903E-2   &   4.88112077380618E-6   &    2.07014410548162E-6 \\ 
\hline S/C2 position  &  8.67179419751799E-1   &  -4.60958432557285E-1   &   -1.99809748470332E-1 \\ 
\hline S/C2 velocity  &  8.60233950755836E-3   &   1.36730847422802E-2   &    5.92795834111164E-3 \\ 
\hline S/C3 position  & -8.64871407921666E-1   &  -4.60958431471891E-1   &   -1.99809747999756E-1 \\
\hline S/C3 velocity  &  8.60233303252539E-3   &  -1.36633074545593E-2   &   -5.92381152958984E-3 \\ 
\hline 
\end{tabular} 
\end{table}

The method of optimizing the orbital period can reduce the arm length difference to a certain extent, but usually such orbits cannot fully meet mission requirements, which requires us to adopt other methods for further optimization. According to the spacecraft's variation in heliocentric distance, we see that the spacecraft's orbital heliocentric distance changes within a range of approximately $(1 \pm 6 \times 10^{-5})$ AU. According to Eq. (\ref{eq:5.14}), we attempt to reduce the orbit eccentricity by adjusting the initial heliocentric distance with a variation of less than $6 \times 10^{-5}$ AU. As for whether to increase or decrease the initial heliocentric distance, both cases are calculated and compared to find the optimized results. When the optimized initial heliocentric distance approaches the periapsis or apoapsis distance of the spacecraft's orbit, further optimization of the orbit eccentricity is not feasible.

\begin{figure}[ht]
    \centering
    \includegraphics[width=0.48\textwidth]{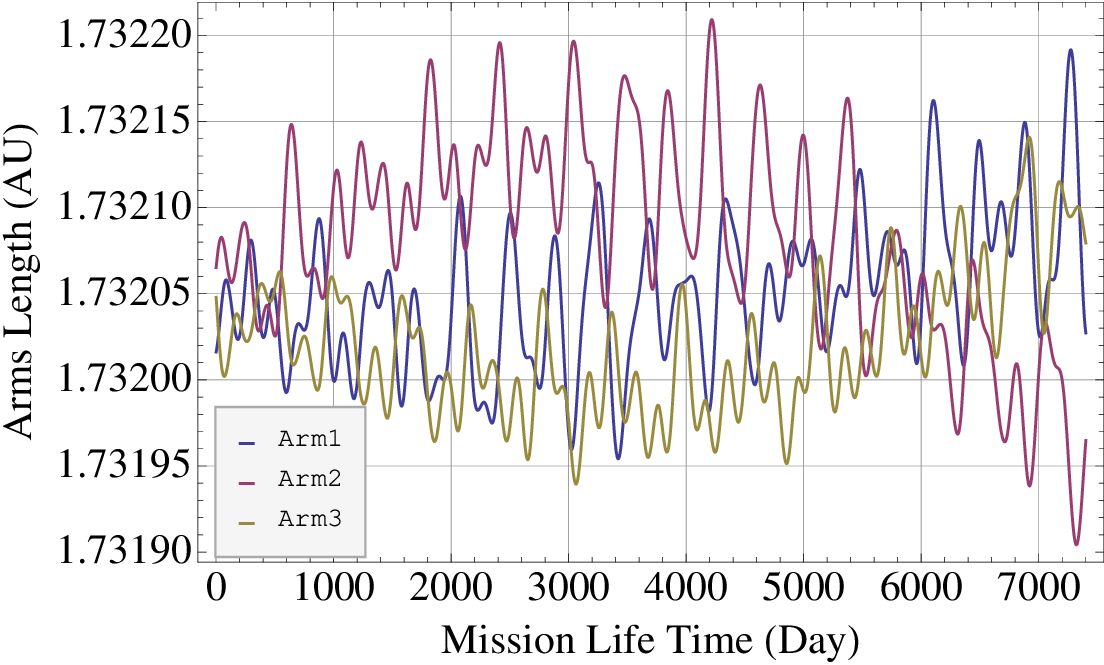} \  
    \includegraphics[width=0.48\textwidth]{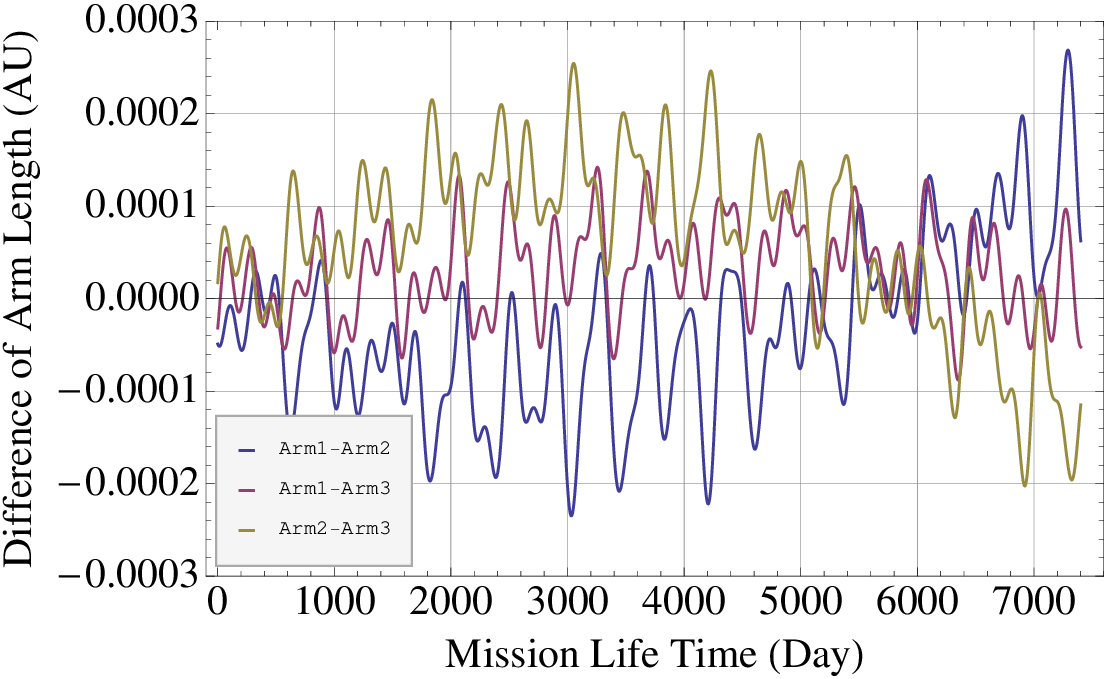} \  
    \includegraphics[width=0.48\textwidth]{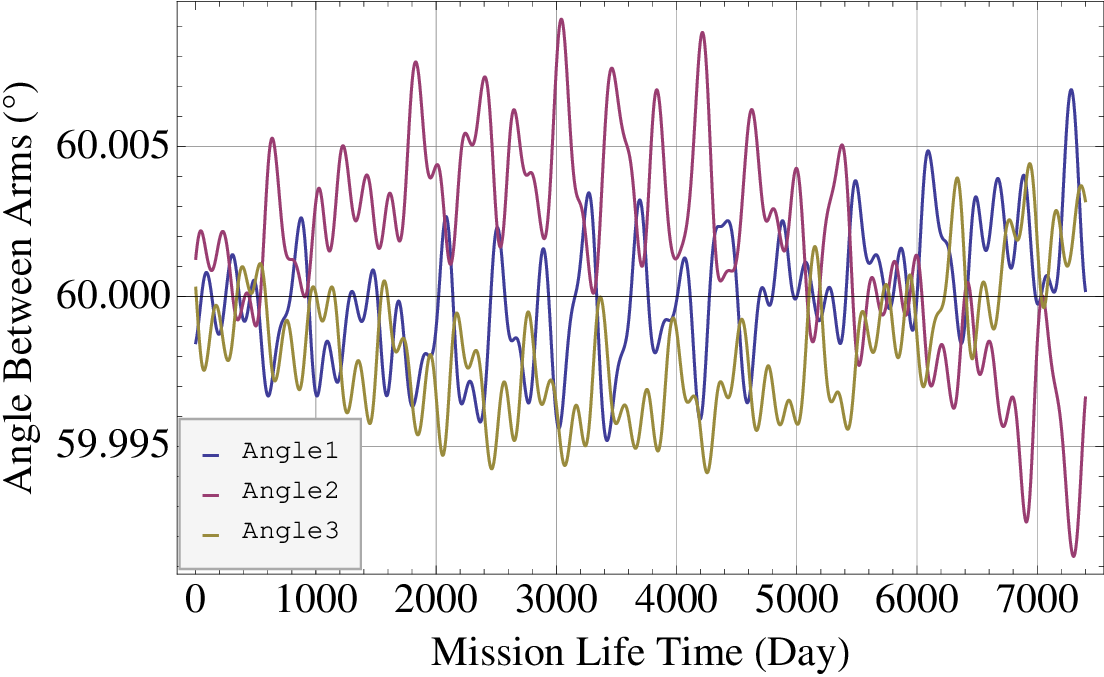} \  
    \includegraphics[width=0.48\textwidth]{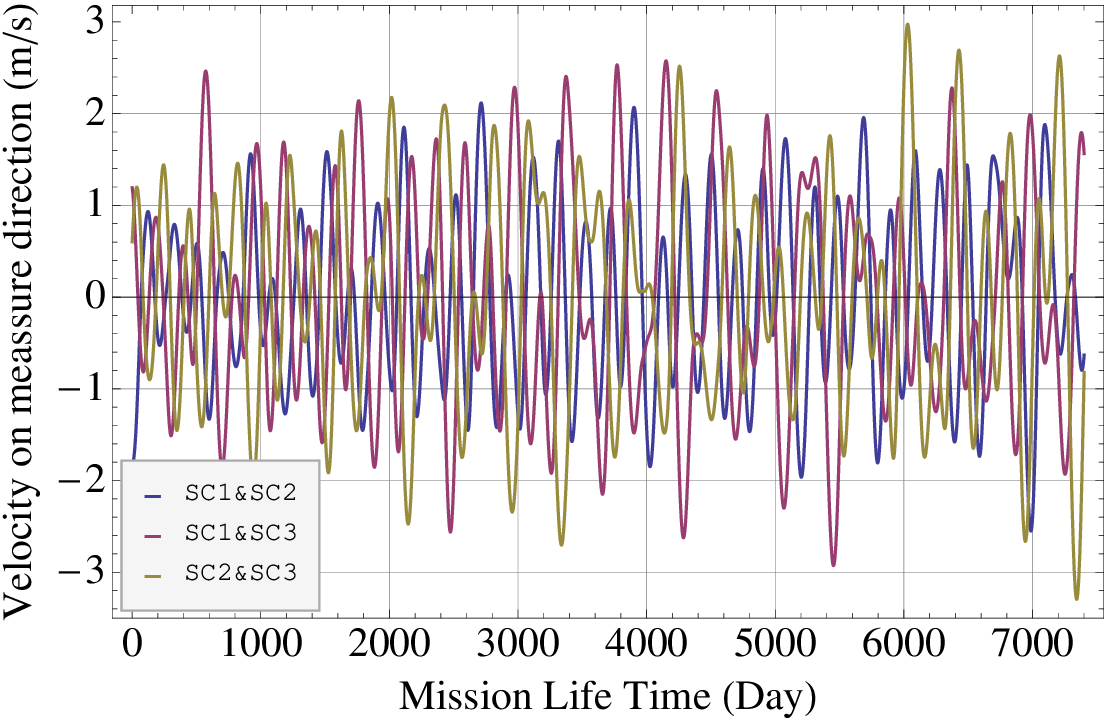}
    \caption{The final optimized results for the arm lengths, arm length differences, arm angles, and spacecraft Doppler velocities over time.}
   \label{fig:5.4}
\end{figure}

In most cases, the two orbit optimization methods cannot achieve satisfactory results in once optimization. Therefore, multiple iterations of orbit period and orbit eccentricity optimization are needed to obtain the optimized results that meet the requirements of the ASTROD-GW mission. The final optimized results for the ASTROD-GW mission orbit are shown in Table \ref{tab:5.4}, the average changes in spacecraft orbit periods are shown in Table \ref{tab:5.5}, and the variations in interference arm lengths, arm length differences, interference arm angles, and Doppler velocities over time are shown in Figure \ref{fig:5.4}.

\begin{table}[ht]
\footnotesize
\caption{Initial conditions after final optimization at epoch JD2461944.0} 
{( J2000 solar-system-barycentric equatorial)}
\label{tab:5.4}
\centering
\renewcommand{\arraystretch}{1.5}
\begin{tabular}{|c|r|r|r|}
\hline           & x(AU)/vx(AU/day) & y(AU)/vy(AU/day) & z(AU)/vz(AU/day)  \\ 
\hline S/C1 position  & 1.15400625657242E-3  &  9.15289225648841E-1  &  3.96866302001196E-1 \\ 
\hline S/C1 velocity  & -1.72003163872199E-2 & 4.88112077380618E-6   &   2.07014410548162E-6 \\ 
\hline S/C2 position  &  8.67153438989685E-1 &  -4.60944670325136E-1  &  -1.99803781815802E-1 \\ 
\hline S/C2 velocity  &  8.60259754371050E-3 &   1.36734947883888E-2  &   5.92813611775755E-3 \\ 
\hline S/C3 position  & -8.64862747667628E-1 &  -4.60953844061175E-1  &  -1.99807759114913E-1 \\
\hline S/C3 velocity  &  8.60242121981651E-3 &  -1.36634452887228E-2  &  -5.92387128797985E-3 \\ 
\hline 
\end{tabular} 
\end{table}
\begin{table}[ht]
\footnotesize
\caption{Periods of three S/C}
\label{tab:5.5}
\centering
\renewcommand{\arraystretch}{1.5}
\begin{tabular}{|c|c|c|c|c|} 
\hline            &  5-yr average  & 10-yr average  & 15-yr average & 20-yr average  \\ 
\hline  S/C1(day)   &  365.25662  &  365.25767  & 365.25636  & 365.25636  \\ 
\hline  S/C2(day)   &  365.25591  &  365.25564  & 365.25620  & 365.25646   \\ 
\hline  S/C3(day)   &  365.25624  &  365.25420  & 365.25656  & 365.25721   \\ 
\hline 
\end{tabular} 
\end{table}

\chapter{Numerical Simulation for Time-Delay Interferometry}

In Chapter 3, we briefly introduced the principles of unequal arm-length TDI. However, in a realistic space mission, due to the dynamics of spacecraft orbits, the lengths of interferometric arms vary over time. This variability makes the laser noise cannot be canceled using traditional methods. In this chapter, we will further analyze TDI and numerically calculate the optical paths of TDI to eliminate laser noise.

\section{Algorithm of Calculation}

For ASTROD-GW, the distance between spacecraft remains approximately $\sqrt{3}$ AU, and the light signal takes about 15 minutes to travel from one spacecraft to another. Therefore, the signal received by a spacecraft is emitted by the other spacecraft approximately 15 minutes earlier, necessitating time delay calculation. During laser interferometry between spacecraft, due to the continuous motion of the transmitting and receiving spacecraft in the coordinate system, and the time delay between emission and reception, it is crucial to accurately determine the position of the receiving spacecraft at the time of signal reception. The CGC2.7 ephemeris framework can calculate a series of discrete states of each spacecraft during the mission period, but it does not include all states at the moment of signal reception. To obtain the state of a spacecraft at any arbitrary time, we employ the Chebyshev polynomial interpolation method based on the spacecraft's existing state data \cite{Li,Newhall}.

In computing TDI, the instantaneous state of the spacecraft emitting the laser is known, but the state of the receiving spacecraft at the reception time is unknown in advance. It requires an iterative approach to approximate the state at reception time. The specific calculation method is as follows:
Since the propagation of light in spacetime is not perfectly straight, the accuracy of the calculation requires consideration of post-Newtonian corrections in the solar system. Taking the post-Newtonian effect of the Sun as an example, considering the solar center as the origin of the coordinate system, point P1 stationary in the reference frame, with position vector $\mathbf{R_1}$, and point P2 with position vector $\mathbf{R_2}$, from which a light signal is emitted from P1 to P2. The propagation time of the light signal $T_{Travel}$ is divided into two parts: one part is the calculation under flat spacetime, and the other part is the correction due to the post-Newtonian effect.
\begin{equation}
T_{Travel}=T_{Newton}+\Delta T_{PN},
\end{equation}
where $T_{Newton}$ is the propagation time in Newtonian flat spacetime, and $\Delta T_{PN}$ is the corresponding time correction due to the post-Newtonian effect.
(1) Calculation of propagation time in Newtonian gravity approximation \cite{Kopeikin}:
\begin{equation}
  T_{Newton}=\frac{R}{c};
\end{equation}
(2) Calculation of time correction for PN (Post-Newtonian Light Propagation) additional time delay:
\begin{equation}
\begin{split}
  \Delta T_{PN} = & \frac{2GM}{c^3} \ln \left( \frac{R_1 + R_2 + R}{R_1 + R_2 - R} \right) \\
  & + \frac{G^2 M^2}{c^5} \frac{R}{R_1 R_2} \left[ \frac{15}{4} \frac{\arccos(\mathbf{N}_1 \cdot \mathbf{N}_2 ) } {|\mathbf{N}_1 \times \mathbf{N}_2|} - \frac{4}{1+ \mathbf{N}_1 \cdot \mathbf{N}_2 } \right],
\end{split}
\end{equation}
where $GM$ is the product of the mass of the celestial body for which PN effect is to be calculated and the gravitational constant, $c$ denotes the speed of light, $R_1 = |\mathbf{R}_1|, R_2 = |\mathbf{R}_2|$ are the distances of P1 and P2 from the origin, $\mathbf{N}_1 = \mathbf{R}_1 / R_1, \mathbf{N}_2 = \mathbf{R}_2 / R_2$ are unit vectors pointing from the origin to P1 and P2 respectively, and $R = |\mathbf{R}_2 - \mathbf{R}_1|$ represents the distance between P1 and P2. Here we consider only the post-Newtonian effect of the Sun.

For calculating the propagation time between spacecraft, the method is as follows. Without loss of generality, assume at time $T_0$, S/C1 emits a laser signal towards S/C2, and S/C2 receives the signal at time $T^r_0$. The propagation time in the coordinate system is $T^r_0 - T_0$. The iterative method to compute the time when S/C2 receives the signal is as follows:
 \begin{equation}
 \begin{aligned}
      T^r_0 &= T_0 +T_1 + T_2 + T_3 +... \\
 (i)  \       T_1      &= \frac{| r_2 (T_0)-r_1(T_0) |}{c} + \Delta T_{1,PN} \\
(ii)  \    T_1 + T_2  &= \frac{| r_2(T_0 +T_1)-r_1(T_0) |}{c} + \Delta T_{2,PN} \\
(iii) \  T_1 + T_2 +T_3 &=\frac{| r_2(T_0+T_1+T_2)-r_0(T_0) |}{c} + \Delta T_{3,PN} \\
                    & ......,
\end{aligned}
\end{equation}
where $\Delta T_{i,PN}$ represents the correction to the light propagation time due to PN effect computed in the $i$-th iteration. Perform multiple iterations until $|T_{i}| < 10^{-11}$ s (3 mm). The value of $r_2(T)$ is obtained using Chebyshev polynomial interpolation \cite{Newhall}. We employ a 14th-order Chebyshev polynomial interpolation over an interval of 8 days, achieving interpolation accuracy on the order of $10^{-14}$ AU.

 \begin{figure}[htbp]
   \centering   
    \includegraphics[scale=0.6]{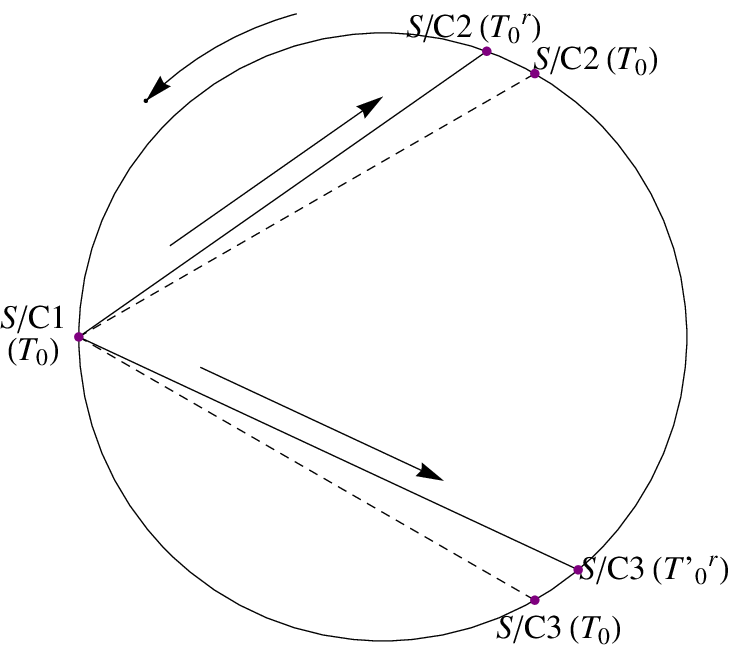} 
  \caption{\small{Illustration for time delay calculation during iteration.}}
\end{figure}

\section{Preliminary Analysis of Michelson-type TDI}

Due to the mutual laser interferometric links between the three spacecraft, each spacecraft has two optical platforms, resulting in six links when all six optical platforms are functional. However, there may be cases where optical platforms do not function properly. In the following discussion, we consider two scenarios: (1) when interference is performed using 2 out of the 3 interferometric arms in this section; (2) when all 3 interferometric arms are simultaneously operational in next chapter.

\subsection{Michelson-type TDI}

In space missions, it is possible that one of the three interferometric arms may not function properly. In such cases, we can refer to the unequal arm Michelson interference and optimize the interference method to partially meet the requirements of interferometry and achieve the corresponding sensitivity goals.

\begin{figure}[htbp]
\centering
\includegraphics[width=0.4\textwidth]{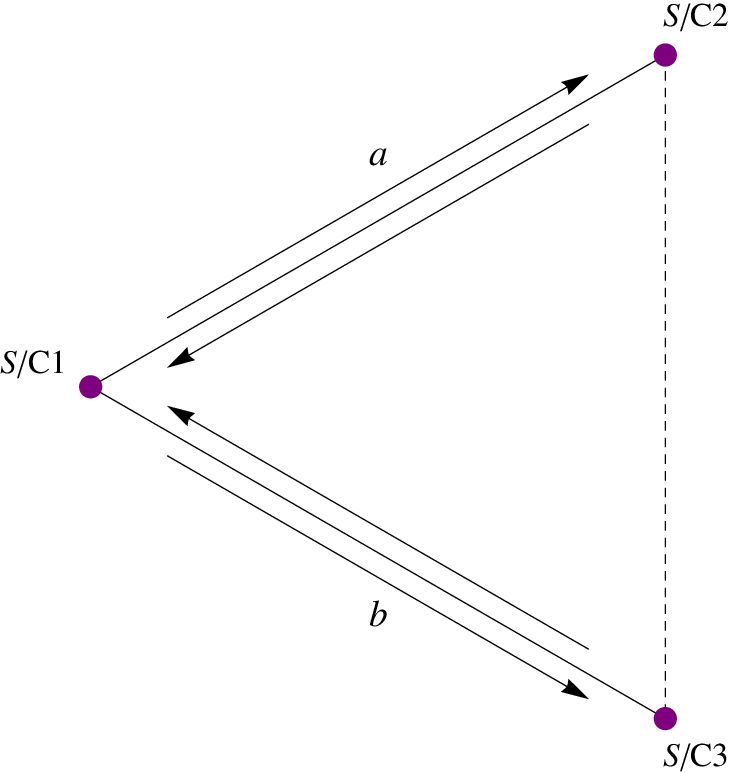} 
\caption{\small{Schematic diagram of TDI using two interferometric arms}}
\end{figure}

The path of light propagation involves path \(a\), where spacecraft S/C1 emits a laser to S/C2 and then returns to S/C1; and path \(b\), where S/C1 emits a laser to S/C3 and then returns to S/C1 via S/C3. When the two laser follow their respective paths and return to S/C1, the propagation times of the two laser are examined. When two paths exhibit symmetry, the combination of paths (will described by \(a\) and \(b\) forms the paths) traveled by the two laser, effectively minimizing the mismatch of two paths and thus reducing the laser noise.
 \begin{description}
   \centering
       \item{Path a} : $ \text{SC1} \rightarrow \text{SC2} \rightarrow \text{SC1} $; 
       \item{Path b} : $ \text{SC1} \rightarrow \text{SC3} \rightarrow \text{SC1} $.      
  \end{description}

According to theoretical calculations, for the 1st-generation TDI paths, static arm length differences can be eliminated, and for the 2nd-generation TDI paths, differences in velocities with the same relative velocities can be eliminated. There are several groups of TDI paths \cite{Dhurandhar}:
\begin{displaymath}
\begin{aligned}
\text{1st  generation}: &1\rightarrow 2 \rightarrow 1 \rightarrow 3 \rightarrow 1 (ba-ab) \\
\text{2nd  generation}: &1 \rightarrow 2 \rightarrow 1 \rightarrow 3 \rightarrow 1 \rightarrow 3 \rightarrow 1 \rightarrow 2 \rightarrow 1 (baab - abba) \\
\end{aligned}
\end{displaymath}
\begin{displaymath}
\begin{aligned}
      (i) \qquad  n  =1, &  [ab,ba] \equiv abba - baab   \\
     (ii) \qquad  n  =2, &  [a^2 b^2, b^2 a^2], [abab, baba], [ab^2a, ba^2b]   \\
    (iii) \qquad  n  =3 ,&  [a^3 b^3, b^3 a^3], [a^2bab^2, b^2aba^2], [a^2b^2ab, b^2a^2ba], [a^2 b^3 a, b^2 a^3 b], \\
        & [a b a^2 b^2,b a b^2 a^2], [ababab, bababa], [abab^2a, baba^2b], [ab^2a^2b, ba^2b^2a],  \\ 
        & [ab^2aba, ba^2bab], [ab^3a^2, ba^3b^2], \text{lexicographic (binary) order} 
\end{aligned}
\end{displaymath}

\section{Results of Numerical Simulation}

\begin{figure}[!ht]
    \centering
    \includegraphics[width=0.46\textwidth]{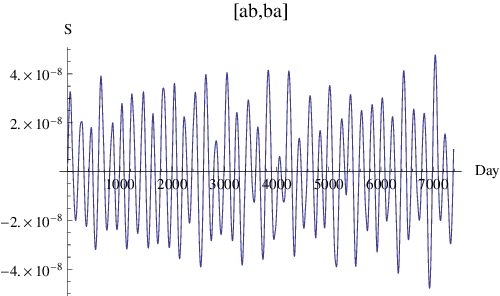} \ 
    \includegraphics[width=0.46\textwidth]{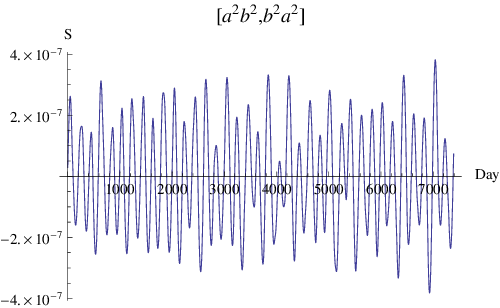} \ 
    \includegraphics[width=0.46\textwidth]{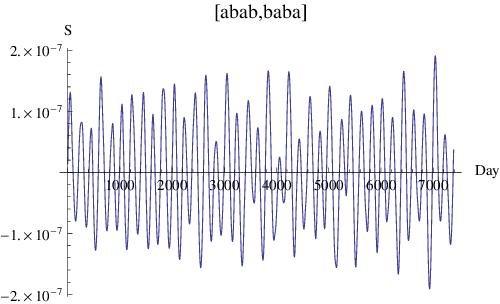} \ 
    \includegraphics[width=0.46\textwidth]{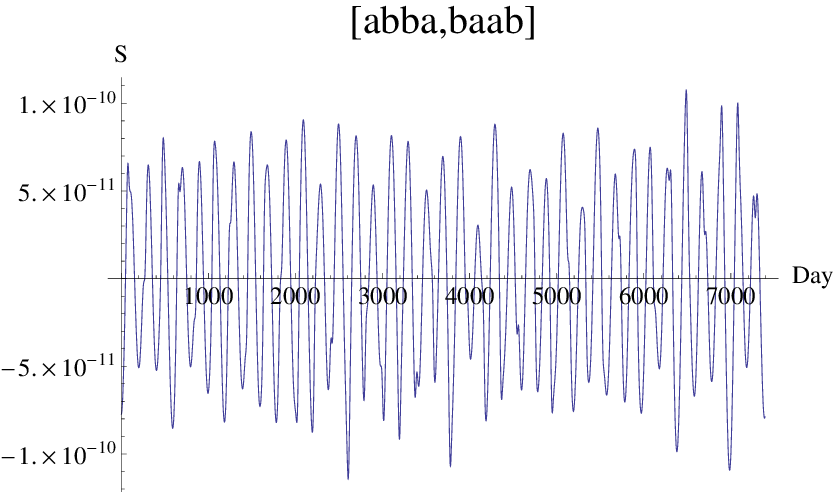} \
    \includegraphics[width=0.46\textwidth]{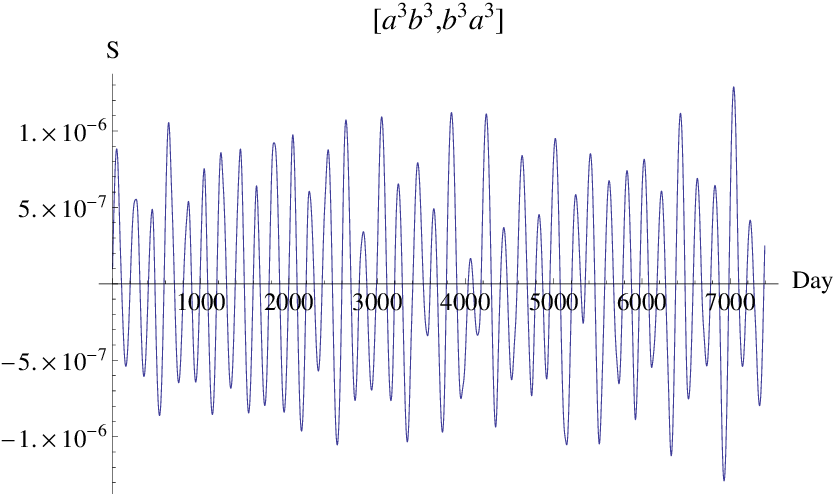} \ 
    \includegraphics[width=0.46\textwidth]{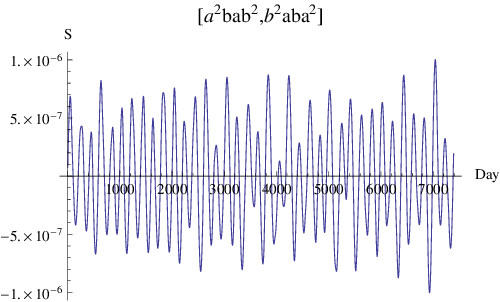} \ 
    \includegraphics[width=0.48\textwidth]{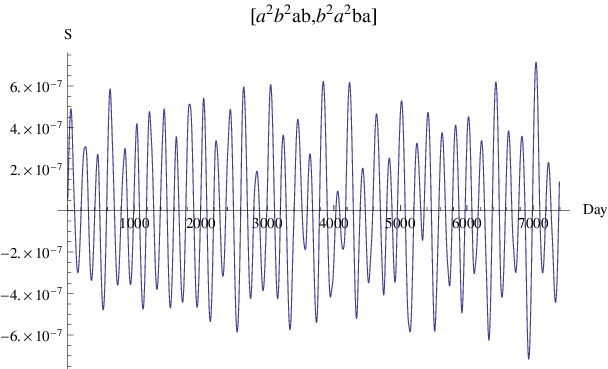} \ 
    \includegraphics[width=0.46\textwidth]{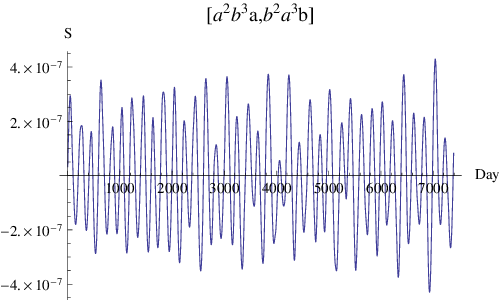} \ 
    \caption{The path mismatches for various Michelson-type TDI channels (I).}
    \label{fig:ab1}
\end{figure}

\begin{figure}[!ht]
    \centering
    \includegraphics[width=0.47\textwidth]{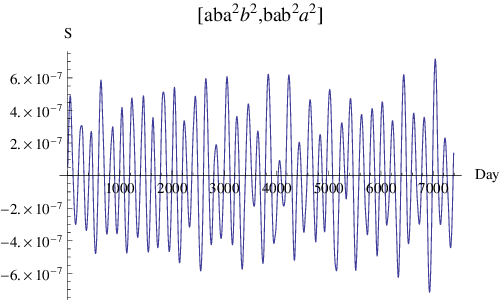} \ 
    \includegraphics[width=0.47\textwidth]{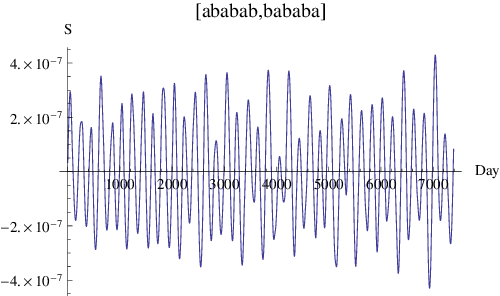} \ 
    \includegraphics[width=0.47\textwidth]{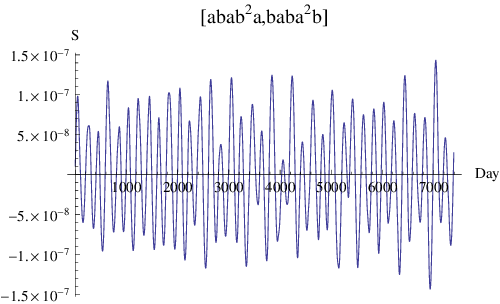} \ 
    \includegraphics[width=0.47\textwidth]{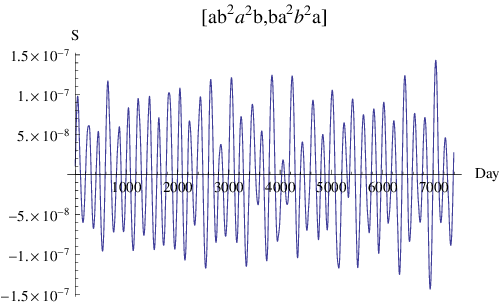} \ 
    \includegraphics[width=0.47\textwidth]{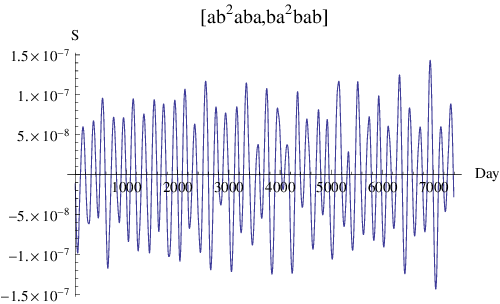} \ 
    \includegraphics[width=0.47\textwidth]{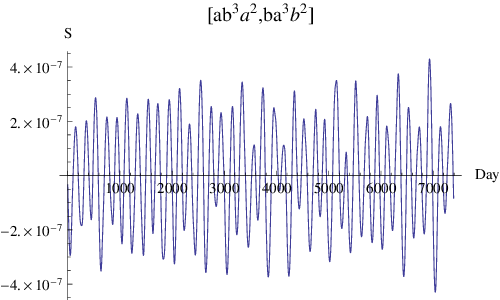}
    \caption{The path mismatches for various Michelson-type TDI channels (II).}
    \label{fig:ab2}
\end{figure}

According to the numerical calculation method, the results of mismatches in TDI channels in Section 6.2 are shown in Figures \ref{fig:ab1} and \ref{fig:ab2}. The horizontal axis represents mission time (in days), and the vertical axis represents path mismatch (time difference in seconds).

ASTROD-GW requires that the path difference between two paths be within 150 ns (50 m). Both [ab, ba] and [abba, baab] satisfy this requirement. Experimentally, after demonstrating compliance with the path difference requirement, ASTROD-GW also meets the noise requirements. In 2010, de Vine et al \cite{Vine2010} from the JPL demonstrated the implementation of LISA's TDI experiment in the laboratory. Through TDI, laser frequency noise was reduced by approximately $10^9$, and clock phase noise was reduced by $6 \times 10^4$, restoring the system's intrinsic displacement noise on the laboratory test bench. ASTROD-GW should also undergo relevant laboratory experiments, with the feasibility of the principle being similar.

\chapter{Geometry Analysis and Numerical Simulation for TDI}
 
\begin{figure}[htbp]
    \centering  
    \includegraphics[width=0.5\textwidth]{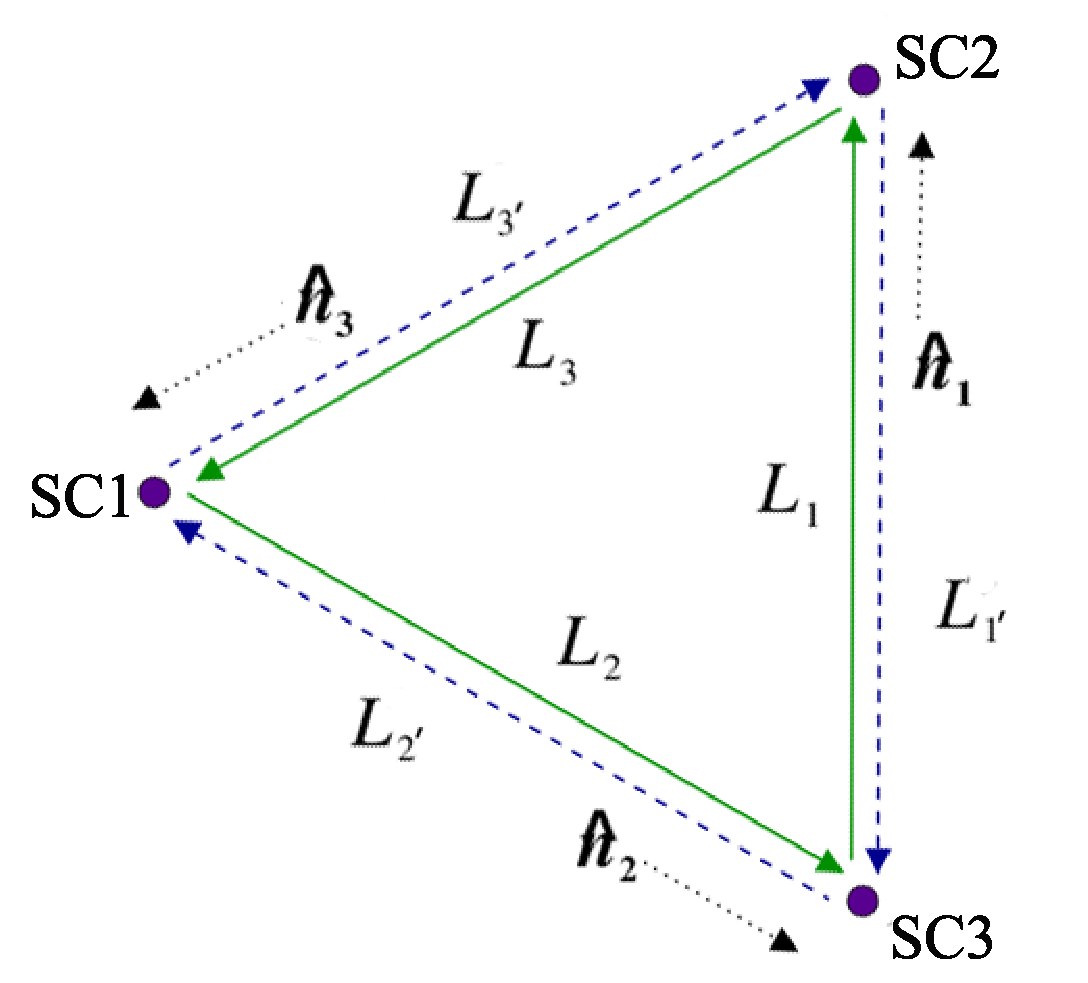}
    \caption{Labeling of spacecraft and their arm lengths.}
    \label{fig:SC-arm}
\end{figure}

To facilitate discussion, we label the interferometric arms of the three spacecraft as follows: spacecraft $i$'s interferometric arms are denoted as $L_i, L_{i'} $; clockwise, they are labeled as $ L_{1'}, L_{2'}, L_{3'} $; counterclockwise, they are $ L_{1}, L_{2}, L_{3} $. This is illustrated in Figure \ref{fig:SC-arm}. Using $ y_{ij} $ to denote interferometric measurement from spacecraft $ j $ originating from spacecraft $ i $.

\section{First-Generation TDI}

The first-generation TDI eliminates static differences in arm lengths through specific paths. Therefore, in this analysis, we initially consider the arm lengths as static. Commas denote time delays, where $y_{ij,k} = y_{ij}(t - L_k)$ and $y_{ij,kr} = y_{ij}(t - L_k - L_r)$ \cite{Vallis}. Because the arm lengths are static, the time delays following the commas can be exchanged. We proceed with a detailed analysis for first-generation TDI as the first case.

\subsection{TDI with Six Links}

 \subsubsection{Sagnac}
 
\begin{figure}[ht]
\begin{minipage}[t]{0.45\textwidth}
  \centering
  \includegraphics[width=0.8\textwidth]{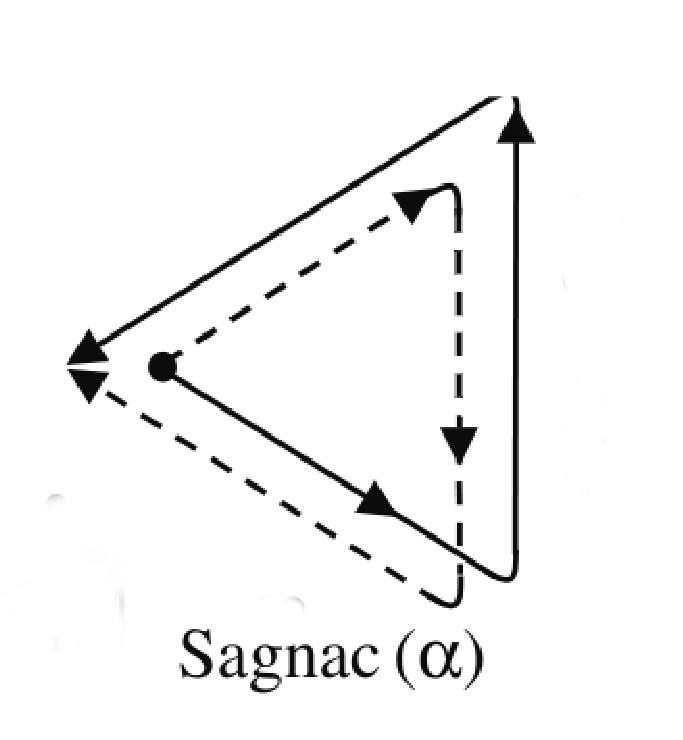}
  \caption{\small{1st-generation TDI Sagnac path}. (reused from \cite{Vallis}.) } \label{fig:Sagnac}
  \end{minipage} 
  \begin{minipage}[t]{0.45\textwidth}
  \centering
  \includegraphics[width=0.65\textwidth]{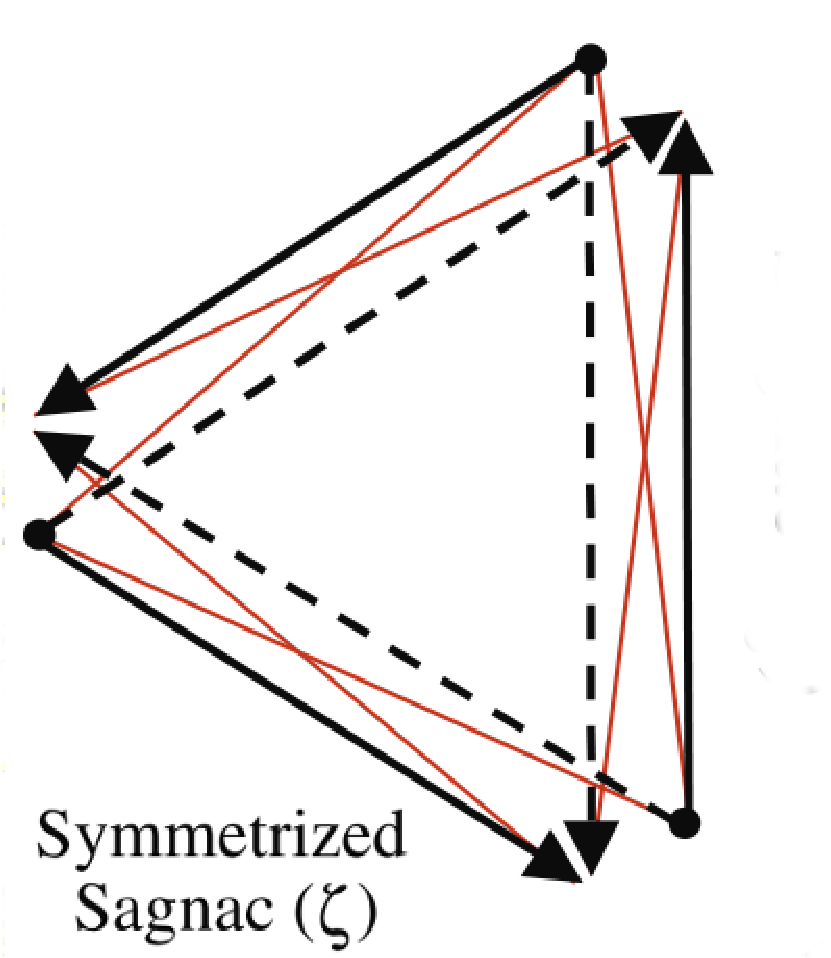}
  \caption{\small{Symmetrized Sagnac path. (reused from \cite{Vallis}.)}} \label{fig:SySagnac}
  \end{minipage}
\end{figure}

\begin{figure}[ht]
\centering
\includegraphics[width=0.7\textwidth]{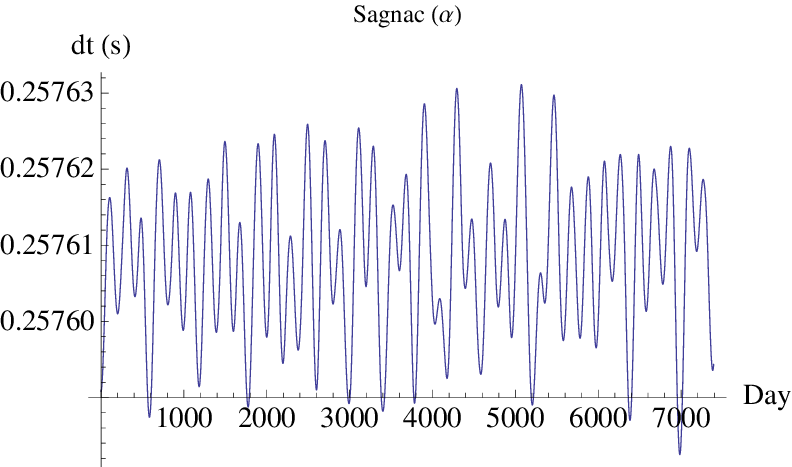}
\caption{The path mismatch in Sagnac ($\alpha$) from numerical calculation. }  \label{fig:Resulta}
\end{figure}

The Sagnac effect includes three channels, designated as $\alpha, \beta, \gamma$ depending on which spacecraft serves as the starting point \cite{Tinto2004}.
Sagnac ($\alpha$) employs six interferometric arms, illustrated in Figure \ref{fig:Sagnac}. Interference involves two laser beams: $\alpha$-Beam1 departs from spacecraft SC1, passes through SC3 and SC2, and returns to SC1; $\alpha$-Beam2 departs from SC1, passes through SC2 and SC3, and returns to SC1. Virtual interference occurs at SC1 where these two laser beams meet. The formula expressions for the $\alpha$ is given by Eq. (\ref{equ:Sagnac}),
\begin{equation} \label{equ:Sagnac}
\begin{split}
\alpha = y_{13,13} + y_{32,3} + y_{21} - y_{31} - y_{23,2'} - y_{12,1'2'}.
\end{split} \end{equation}
Through analysis, we can express the specific process as Eq. (\ref{equ:aBeam}). We need to calculate the time difference $dt$ between when these two laser beams reach SC1 after traveling their respective paths. The results for Sagnac ($\alpha$) are depicted in Figure \ref{fig:Resulta}.
Due to the influence of the Sagnac effect, the computed results at this stage are not meet the requirement. In subsequent analyses, we will refine the Sagnac interference type to achieve more satisfactory results.

\begin{equation} \label{equ:aBeam}
\begin{split}
\alpha - Beam 1:  \hspace{20pt}
  \overbrace{   1 \overset{L_{2}}{\longrightarrow} 3 \overset{L_{1}}{\longrightarrow} 2 \overset{L_{3}}{\longrightarrow} 1 
          }^{y_{13,13}+y_{32,3} +y_{21}} \  || \ t& \\ 
\alpha - Beam 2: \hspace{20pt} 
  \underbrace{  1 \overset{L_{3'}}{\longrightarrow} 2 \overset{L_{1'}}{\longrightarrow} 3 \overset{L_{2'}}{\longrightarrow} 1 
         }_{  y_{12,1'2'}+y_{23,2'}+ y_{31} } \  || \ t& 
\end{split}  \end{equation}

During the calculation, the time difference is obtained by subtracting the time taken by $\alpha$-Beam2, which travels through $L_{3',1'2'} + L_{1',2'} + L_{2'}$, from the time taken by $\alpha$-Beam1, which travels through $L_{2,12} + L_{1,3} + L_3$. After the light beam along $\alpha$-Beam1 arrives at SC1 at time $t$, we calculate backward along the negative time direction starting from time $t$ for the process of $\alpha$-Beam2. This negative time is then added to the time taken by $\alpha$-Beam1, resulting in the time difference $dt$ as given by Eq. (\ref{equ:dt}).
If we denote the positive direction of time with "$\rightarrow$" and the negative direction with "$\leftarrow$", we can represent the interference process of Sagnac ($\alpha$) as $\overrightarrow{213} \ \overleftarrow{2'1'3'}$, where the numbers under the arrows denote the arm labels.

\begin{equation} \label{equ:dt}
\begin{split}
dt & = L_{2,12} + L_{1,3} + L_3 - (L_{2'}+L_{1',2}+L_{3',1'2'}) \\
    & = L_{2,12} + L_{1,3} + L_3 + (- L_{2'}) + (- L_{1',2'}) + ( - L_{3',1'2'}) 
\end{split}  \end{equation}


\subsection{TDI with Eight Links}

\subsubsection{Unequal-Arm Michelson}

Regarding the 1st-generation Michelson TDI, detailed discussions have been previously conducted, and thus will not be further elaborated upon here.

\subsubsection{Relay}
 
\begin{figure}[ht]
\begin{minipage}[t]{0.4\linewidth}
\centering
\includegraphics[width=0.9\textwidth]{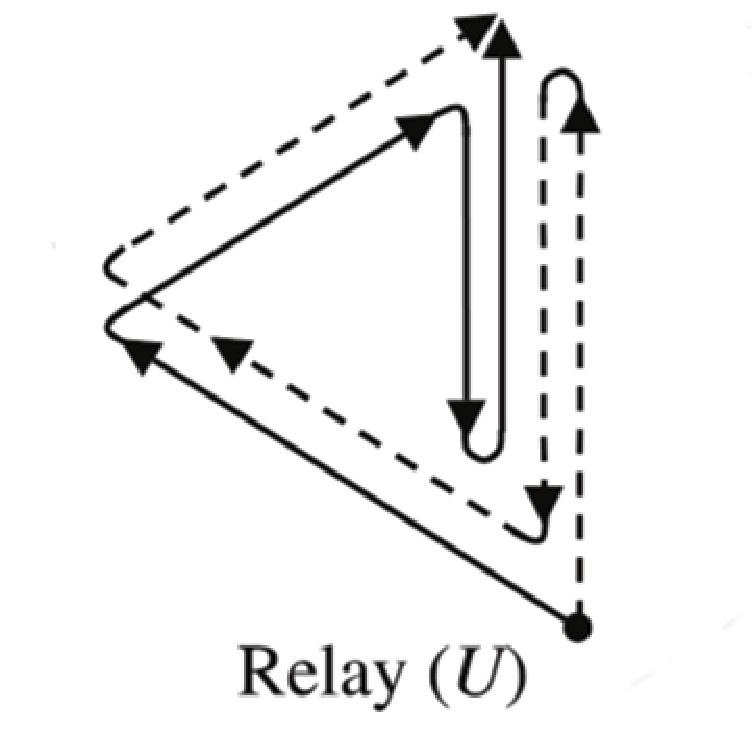}
\caption{\small{Diagram of Relay paths. (reused from \cite{Vallis}.)}} \label{fig:Relay}
\end{minipage}
\hfill
\begin{minipage}[t]{0.57\linewidth}
\centering
\includegraphics[width=0.7\textwidth]{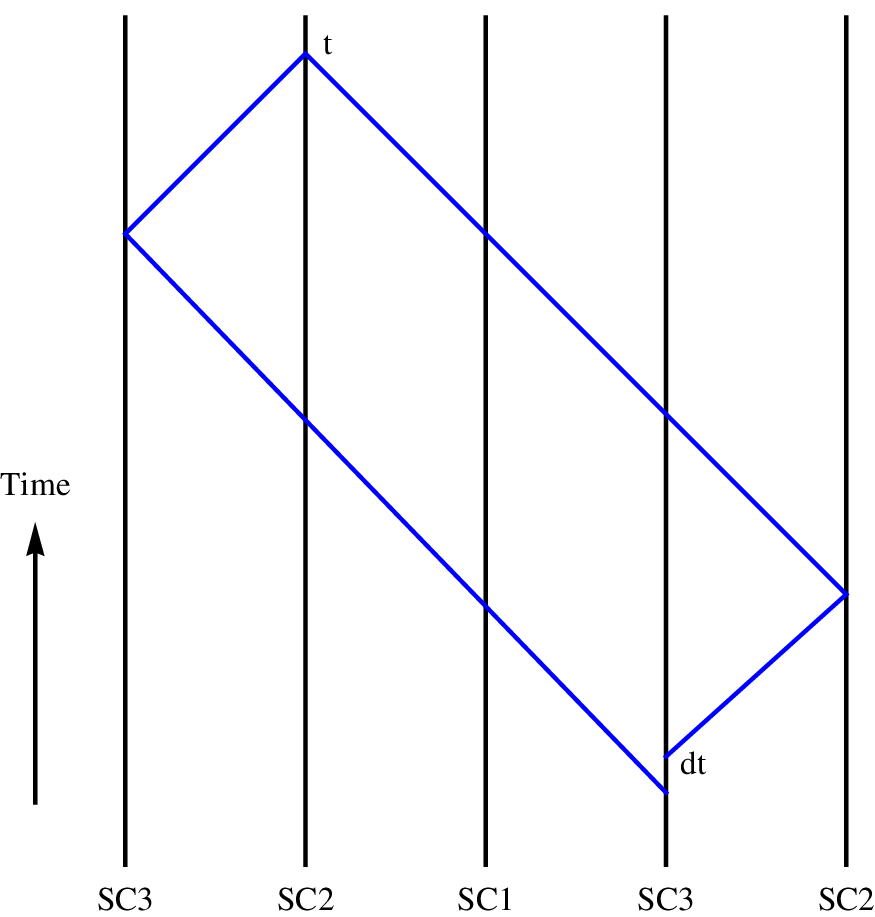}
\caption{\small{S/C layout-time delay diagrams for Relay (U).}}
 \label{fig:RelayTime}
\end{minipage}
\end{figure}

Relay involves three observables based on different spacecraft starting points, named U, V, and W, respectively. In Relay (U), four interferometric arms $ L_1, L_{1'}, L_{2'}$ and $L_{3'} $ are used, with interferometry employing two laser paths. The interferometric path of Relay (U) between spacecraft is illustrated in Figure \ref{fig:Relay}. The formula expression of Relay-U is given by Eq. (\ref{equ:U}) \cite{Tinto2004},
\begin{equation}
\label{equ:U}
 U = y_{31,3'1'1} + y_{12,1'1} +  y_{23,1} + y_{32} -  y_{12} - y_{31,3'} - y_{23,2'3'}  - y_{32,1'2'3'}.
\end{equation}
Beam U-Beam1 starts from SC3, passes through SC1 and SC2 back to SC3, then through $L_1$ to SC2; Beam U-Beam2 starts from SC3, first reaches SC2, then sequentially passes through SC3 and SC1, finally returning to SC2. The interference of two beams occurs at SC2, and the specific process is represented as Eq. \eqref{equ:UBeam}. If the relative positions of the three spacecraft are represented by worldlines, and the interference paths of TDI are reflected across spacecraft, as shown in Figure \ref{fig:RelayTime}, where the blue lines indicate the interference paths. 
\begin{equation}  \label{equ:UBeam}
\begin{split}
U-Beam 1:  \hspace{20pt} 
  \overbrace{
   3 \overset{L_{2'}}{\longrightarrow} 1 \overset{L_{3'}}{\longrightarrow} 2 \overset{L_{1'}}{\longrightarrow} 3 \overset{L_1}{\longrightarrow} 2           }^{y_{31,3'1'1}+y_{12,1'1} +y_{23,1}+y_{32}} \  || \ t& \\ 
U-Beam 2: \hspace{20pt} 
  \underbrace{  
  3 \overset{L_{1}}{\longrightarrow} 2 \overset{L_{1'}}{\longrightarrow} 3 \overset{L_{2'}}{\longrightarrow} 1 \overset{L_{3'}}{\longrightarrow} 2   }_{  y_{32,1'2'3'}+y_{23,2'3'}+ y_{31,3'} + y_{12} } \  || \ t& 
\end{split} \end{equation}

\begin{figure}
  \centering
  \includegraphics[scale=0.7]{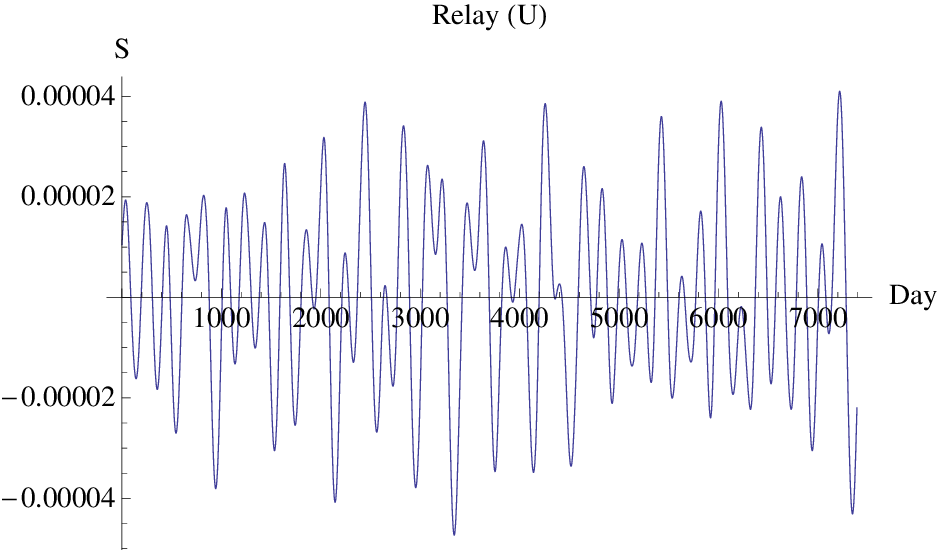}
  \caption{The path mismatch in Relay (U) from numerical calculation. } \label{fig:RelayUR}
\end{figure}

From the path diagram of Relay (U) on the spacecraft worldlines, it can be seen that the virtual TDI path meet at SC2 at time $t$, indicating that the two laser paths are continuous at $t$, thereby considering the two paths of the entire interference process as continuous, enclosing an almost closed path. In the calculation process, U-Beam1 first reaches SC2 in the positive time direction at time $t$, then returns from SC2 along U-Beam2 in the negative time direction to SC3. The resulting time difference $dt$ for Relay (U) is calculated through this process, expressed as Eq. (\ref{equ:TimeU}). 
\begin{equation} \label{equ:TimeU}
\overrightarrow{_3 2' _1 3' _2 1' _3 1 } _2  \overleftarrow{3' _1 2' _3 1' _2 1 _3}
\end{equation}
The interferometric arms used are denoted by $ i, i' (i=1,2,3)$, with subscripts indicating the spacecraft through which they pass. We select different times within the entire mission duration to simulate TDI and obtain the variation of $dt$ with time as shown in Figure \ref{fig:RelayUR}.

Further analysis reveals that if we first calculate along the path of U-Beam2, moving from SC3 in the positive time direction to SC2, and then return along the path of U-Beam1 in the negative time direction back to SC3, the result obtained will be opposite in sign to the calculation where we first move along U-Beam1 and then return along U-Beam2 in reverse. This reversal in results is easy to understand, and it precisely facilitates the subsequent self-splicing construction of second-generation TDI paths.
\begin{equation} \label{equ:Udt}
dt[\overrightarrow{2' 3' 1' 1} \ \overleftarrow{3' 2' 1' 1}] = -dt [ \overrightarrow{1 1' 2' 3'} \ \overleftarrow{1 1' 3' 2'}]
\end{equation}

\subsubsection{Beacon}
 
The three observables include Beacon, labeled P, Q, and R respectively depending on the starting spacecraft, involve different interference paths. The expression for the interference is given by Eq. (\ref{equ:P}) \cite{Tinto2004}, 
\begin{equation}  \label{equ:P}
P =  y_{13,11'3'} + y_{32,1'3'} + y_{23,3'} + y_{12,2} -  y_{13, 3'} - y_{32,2} - y_{23,12} - y_{12,1'12}.
\end{equation}
And the analysis based on different beams is detailed in Eq. (\ref{equ:PBeam}). 
\begin{equation}  \label{equ:PBeam}
\begin{split}
P-Beam 1-1:  \hspace{20pt} 
  \overbrace{1 \overset{L_{3'}}{\longrightarrow} 2 \overset{L_{1'}}{\longrightarrow} 3 \overset{L_{1}}{\longrightarrow} 2 }^{y_{12,1'12}+y_{23,12}+y_{32,2}} \  | \ 1  \overset{L_{2}}{\dashrightarrow} 3 \ || \ t& \\ 
P-Beam 1-2:  \hspace{80pt}
\underbrace{ 1 \overset{L_{2}}{\longrightarrow} 3 }_{ y_{13, 3'}} \ | \ 1  \overset{L_{3'}}{\dashrightarrow} 2 \ || \ t& \\    
P-Beam 2-1: \hspace{20pt} 
  \overbrace{1 \overset{L_{2}}{\longrightarrow} 3 \overset{L_{1}}{\longrightarrow} 2 \overset{L_{1'}}{\longrightarrow} 3 
   }^{  y_{13,11'3'}+y_{32,1'3'}+y_{23,3'}} \ | \ 1  \overset{L_{3'}}{\dashrightarrow} 2 \ || \ t& \\ 
P-Beam 2-2:  \hspace{80pt}
 \underbrace{  1 \overset{L_{3'}}{\longrightarrow} 2 }_{ y_{12,2}} \ | \ 1  \overset{L_{2}}{\dashrightarrow} 3 \ || \ t&                                  
\end{split}  \end{equation}
In Beacon (P), four interferometric arms are used: $ L_1, L_{1'}, L_{2}, $ and $ L_{3'} $, with four laser beams employed for interference. 
Beam P-Beam 1-1 starts from SC1, passes through $ L_{3'} $ to SC2, then through $ L_{1'} $ to SC3, and finally through $ L_{1} $ to SC2 at time $ t $, accounting for the delay $ L_{2} $. Beam P-Beam 1-2 starts from SC1 to SC3, where it arrives with a delay $ L_{3'} $ at time $ t $. Similar analysis applies to Beams P-Beam 2-1 and 2-2 as per Eqs. (\ref{equ:PBeam}).
From Eq. (\ref{equ:PBeam}), it's evident that Beam P-Beam 1-1 and Beam P-Beam 2-2 reach SC2 simultaneously at time $ t - L_{2} $, allowing their virtual interference at SC2. Similarly, Beam P-Beam 1-2 and Beam P-Beam 2-1 reach SC3 simultaneously at time $ t - L_{3} $, enabling their interference at SC3. Their combined results are then simulated.

\begin{figure}[!ht]
\begin{minipage}[t]{0.4\textwidth}
\centering
\includegraphics[width=0.9\textwidth]{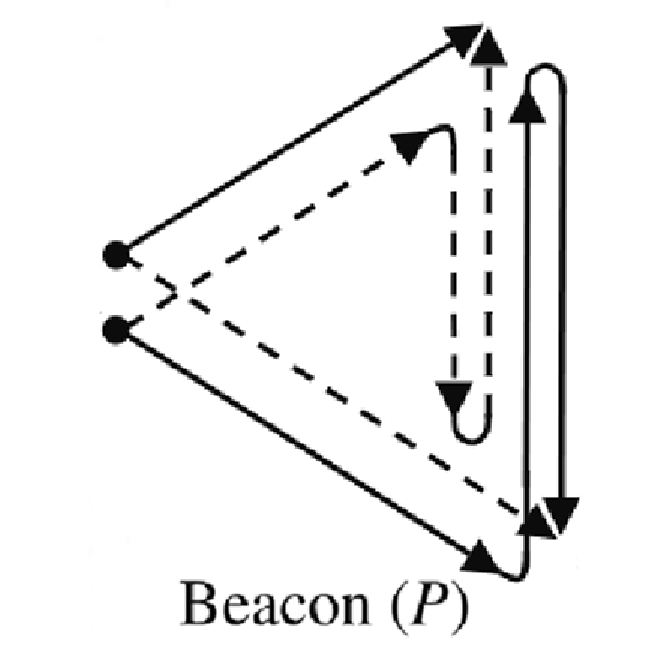}
\caption{\small{Diagram of Beacon (P) path (reused from \cite{Vallis}.)}} \label{fig:Beacon}
\end{minipage}
\hfill
\begin{minipage}[t]{0.58\textwidth}
\centering
\includegraphics[width=0.7\textwidth]{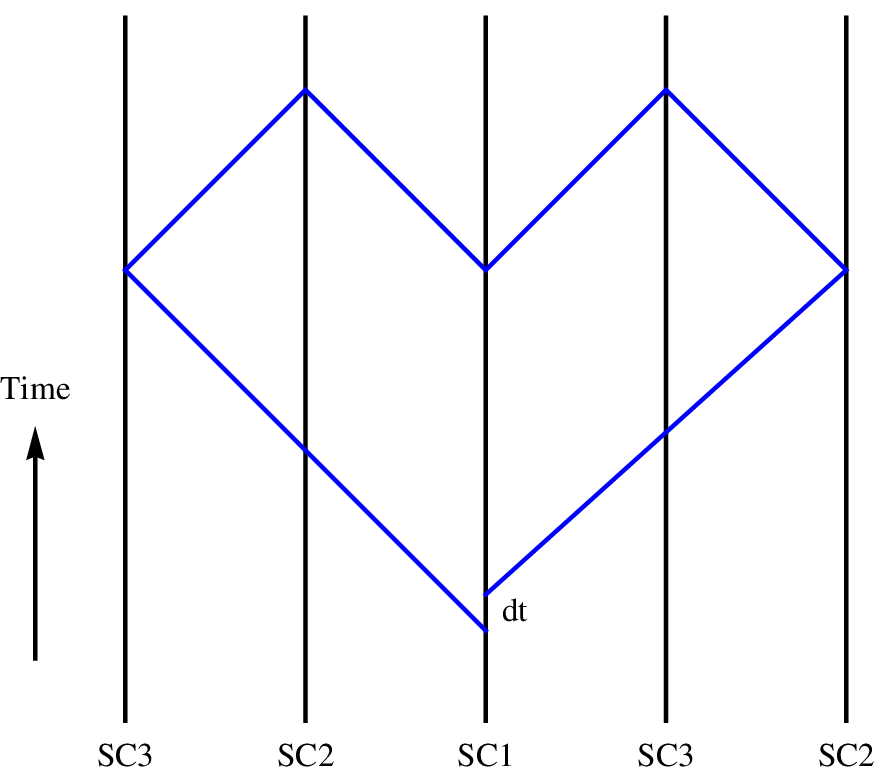}
\caption{\small{S/C layout-time delay diagrams for Beacon (P). }} \label{fig:BeaconTime}
\end{minipage}
\end{figure}

\begin{figure}[ht]
    \centering  
    \includegraphics[width=0.7\textwidth]{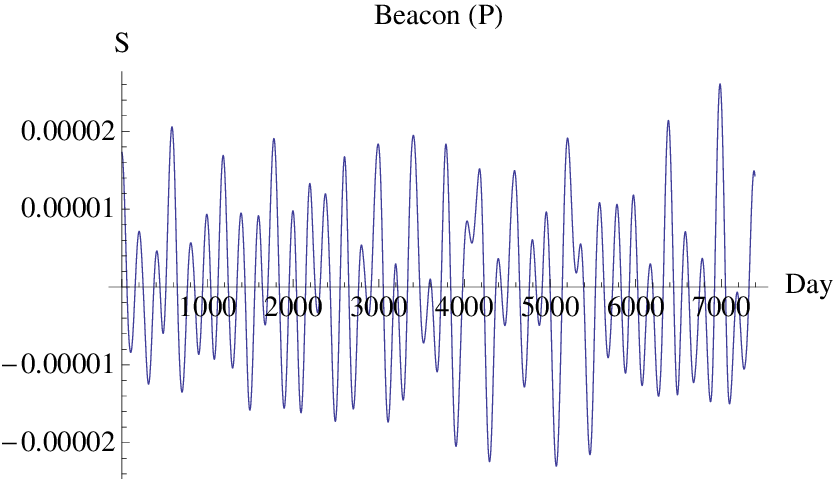} 
    \caption{\small{The path mismatch in Beacon (P) from numerical calculation. } }
    \label{fig:BeaconPR}
 \end{figure}

The interference paths of Beacon (P) are represented on the spacecraft's relatively separations, as depicted in Figure \ref{fig:BeaconTime}. When calculating the paths of TDI, for ease of processing, all beam paths can be viewed as continuous. We use " $ \rightarrow $ " to indicate the positive direction of time and "$ \leftarrow $ " to indicate the negative direction of time. Thus, Beam P-Beam 1-1 and 2-2 interference can be represented as $ \overrightarrow{ 3' 1' 1 } \ \overleftarrow{ 3' } $, and Beam P-Beam 1-2 and 2-1 interference as $ \overrightarrow{ 2 } \ \overleftarrow{ 1' 1 2 } $, where the numbers below the arrows denote the interferometric arm labels. This approach effectively addresses simultaneous interference arrival at the spacecraft during computation. These are then combined to form Eq. (\ref{equ:TimeP}). For clarity, spacecraft labels are subscripted under the numbers. The computed results are shown in Figure \ref{fig:BeaconPR}.
\begin{equation}  \label{equ:TimeP}
\overrightarrow{_1 3' _2 1' _3 1 } _2 \overleftarrow{3' } _1 \overrightarrow{2 } _3 \overleftarrow{1' _2 1 _3 2 _1}
\end{equation}
 
\subsubsection{Monitor}

In Monitor, three observables are labeled E, F, and G, respectively \cite{Tinto2004}. In first channel (E), four interferometric arms $L_1, L_{1'}, L_{2'}$, and $L_3$ are used, with four laser beams employed for interference, and the expression is given by Eq. (\ref{equ:E}) \cite{Tinto2004}. 
 \begin{equation}  \label{equ:E}
E =  y_{32,1'2'} + y_{23,2'} + y_{31} + y_{21,11'} -  y_{31,1'1} - y_{21} - y_{32,3} - y_{23,13}.
\end{equation}
According to the analysis with different beams as shown in Eq. (\ref{equ:EBeam}), Beam E-Beam 1-1 departs from SC2, passes through $L_{1'}$ to reach SC3, then passes through $L_{1}$ to return to SC2, and finally arrives at SC1 via $L_{3}$ at time $t$. Beam E-Beam 1-2 departs from SC3, passes through $L_{2}$ to reach SC1, and upon reaching SC3, it experiences a time delay of $L_{1'} + L_{1}$ until virtual interference time $t$. Similar analysis applies to Beams E-Beam 2-1 and 2-2 based on the following two Eqs. from (\ref{equ:EBeam}).
\begin{equation}  \label{equ:EBeam}
\begin{split}
E-Beam \ 1-1:  \hspace{40pt} 
  \overbrace{  2 \overset{L_{1'}}{\longrightarrow} 3 \overset{L_{1}}{\longrightarrow} 2 \overset{L_{3}}{\longrightarrow} 1 }^{y_{23,13}+y_{32,3}+y_{21}} \ || \ t& \\ 
E-Beam \ 1-2:  \hspace{30pt}
\underbrace{ 3 \overset{L_{2'}}{\longrightarrow} 1 }_{ y_{31, 1'1}} \ | \ 2  \overset{L_{1'}}{\dashrightarrow} \ 3  \overset{L_{1}}{\dashrightarrow} 2 \ || \ t& \\    
E-Beam \ 2-1: \hspace{40pt} 
  \overbrace{  3 \overset{L_{1}}{\longrightarrow} 2 \overset{L_{1'}}{\longrightarrow} 3 \overset{L_{2'}}{\longrightarrow} 1 
   }^{  y_{32,1'2'} + y_{23,2'} + y_{31}} \ || \ t& \\ 
E-Beam \ 2-2:  \hspace{30pt}
 \underbrace{  2 \overset{L_{3}}{\longrightarrow} 1 }_{ y_{21,11'}} \ | \ 3  \overset{L_{1}}{\dashrightarrow} \ 2 \  \overset{L_{1'}}{\dashrightarrow} 3 \ || \ t&                                  
\end{split}  \end{equation} 

 \begin{figure}[ht]
\begin{minipage}[t]{0.4\textwidth}
\centering
\includegraphics[width=0.9\textwidth]{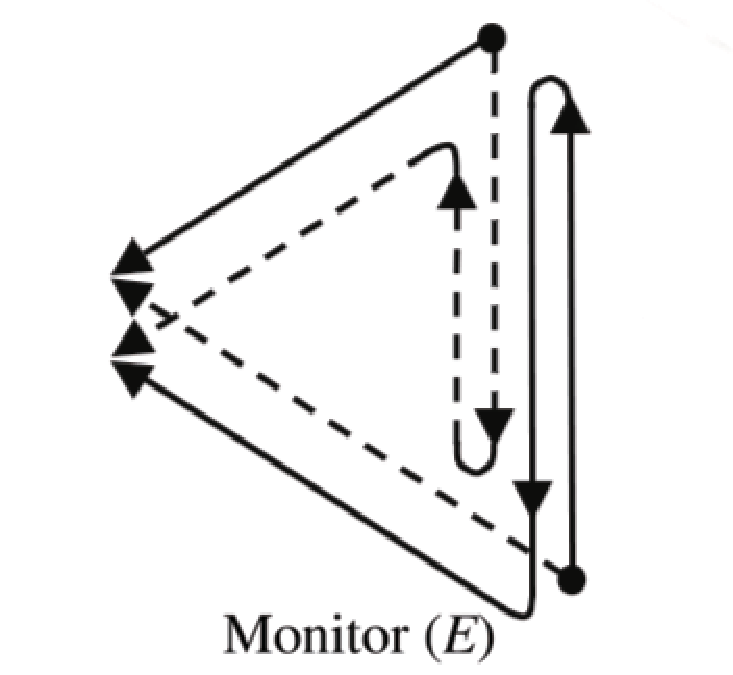}
\caption{\small{Diagram of Monitor path  (reused from \cite{Vallis}.) }}
 \label{fig:Monitor}
\end{minipage}
\hfill
\begin{minipage}[t]{0.55\textwidth}
\centering
\includegraphics[width=0.7\textwidth]{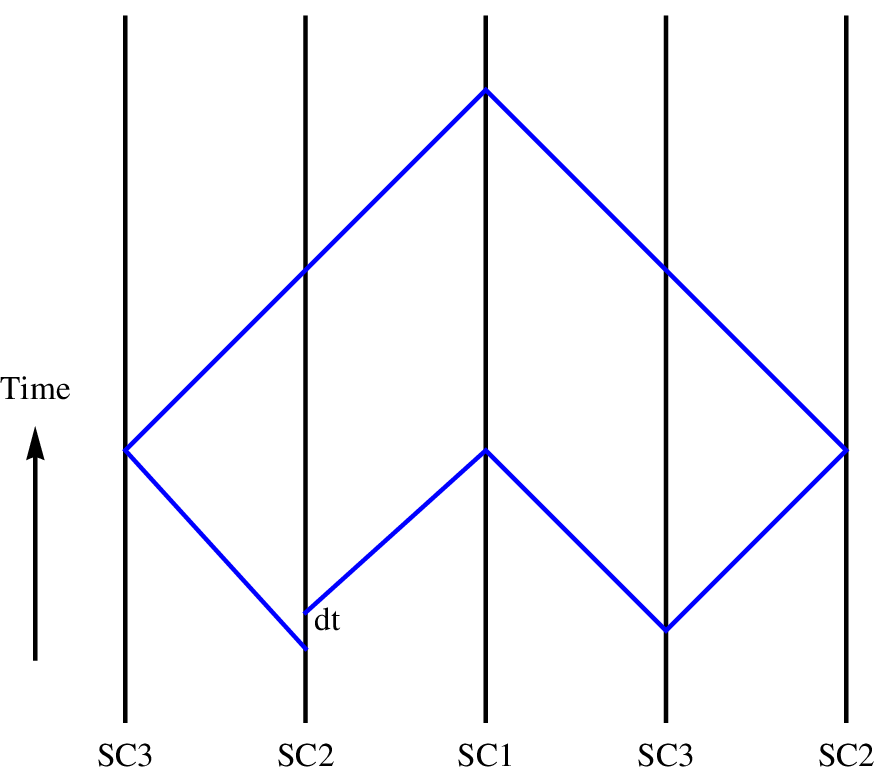}
\caption{\small{S/C layout-time delay diagrams for Monitor (E). }}
\label{fig:MonitorTime}
\end{minipage}
\end{figure}

\begin{figure}[htbp]
    \centering  
    \includegraphics[width=0.7\textwidth]{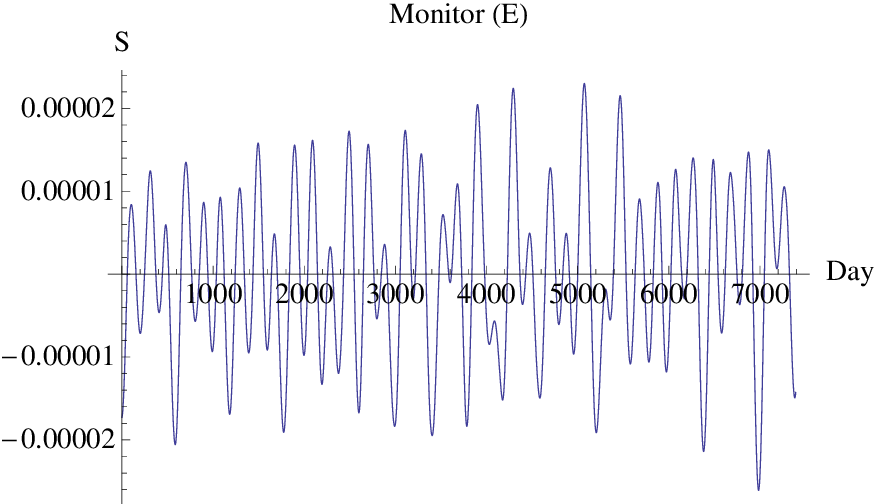} 
    \caption{\small{The path mismatch in Monitor (E) from numerical calculation.}}
    \label{fig:MonitorER}
 \end{figure}

In Eq. (\ref{equ:EBeam}), we observe that Beams E-Beam 1-1 and E-Beam 2-1 arrive at SC1 simultaneously at time $t$. Therefore, we can arrange for both to interfere at SC1 at time $t$, and Beams E-Beam 1-2 and E-Beam 2-2 to interfere at SC1 at time $t - L_1 - L_{1'}$. The results of both interferences are then combined. The TDI paths are depicted in Fig. \ref{fig:MonitorTime}. When calculating the TDI, we use a method similar to Beacon (P), where we consider all beams as continuous for ease of calculating, progressing forward or backward in time. We use "$\rightarrow$" to denote the positive direction of time and "$\leftarrow$" to denote the negative direction. Thus, we can represent the interference of E-Beam 1-1 and 2-1 as $\overrightarrow{1' 1 3}\ \overleftarrow{2' 1' 1}$, and the interference of E-Beam 1-2 and 2-2 as $\overrightarrow{2'}\ \overleftarrow{3}$, connecting them to obtain Eq. (\ref{equ:TimeE}). The numerical results are shown in Fig. \ref{fig:MonitorER}.
\begin{equation}  \label{equ:TimeE}
 \overrightarrow{_2 1' _3 1 _2 3 } _1 \overleftarrow{2' _3 1' _2 1 } _3 \overrightarrow{ 2' } _1 \overleftarrow{3 _2}
\end{equation}
 
 \newpage
 
\section{Second-Generation TDI} 

For the second-generation TDI, the goal is to eliminate spacecraft with the same relative velocity. We use the conventions and starting approach in \cite{Vallis} in this section. The semicolons is used to define the delay symbols. In this case, we consider the variation of arm lengths with time ($L_l (t) \neq L_{l'} (t)$), and after the semicolon, delay factors can no longer be exchanged. From Eqs. (\ref{equ:DF1})-(\ref{equ:DF3}), it can be seen that each delay factor $k$ generates a first-order differential term for each delay factor to its right. For example, in Eq. (\ref{equ:DF3}), $k$ generates a first-order differential term $\dot{L}_k (L_m + L_n)$ for the two delay factors $m$ and $n$ to its right, and $m$ generates a first-order differential term $\dot{L}_m L_n$ for the delay factor $n$ to its right. For the 2nd-generation, it is possible to find suitable paths to cancel out all these first-order differential terms.
 \begin{equation}
 \label{equ:DF1}
  y_{ij;k}(t) \equiv y_{ij} ( t - L_k (t) )
\end{equation} 
 \begin{equation}
 \begin{split}
  y_{ij;km}(t) & \equiv y_{ij} ( t - L_m (t)  - L_k (t - L_m (t)  ) ) \\
                   &  \simeq y_{ij} ( t - L_m (t) - L_k (t) + \dot{L}_k  L_m ) \\
                   & \simeq y_{ij,km} + \dot{y}_{ij,km} \dot{L}_k L_m 
 \end{split}
\end{equation} 
 \begin{equation}
 \label{equ:DF3}
 \begin{split}
  y_{ij;kmn}(t) & \equiv y_{ij} ( t - L_n (t)  - L_m (t - L_n(t) ) - L_k [t - L_n (t) - L_m (t - L_n (t) ) ] ) \\
                      & \simeq  y_{ij} ( t - L_n - L_m - L_k + \dot{L}_m  L_n + \dot{L}_k ( L_n + L_m ) ) \\
                      & \simeq y_{ij,kmn} + \dot{y}_{ij,kmn}[ \dot{L}_k ( L_m +L_n) + \dot{L}_m L_n ]
  \end{split}
\end{equation} 
 
We continue to use the combination of " $\rightarrow$ ", " $ \leftarrow $ ", and arm length labels to represent the paths of laser interference. For 1st-generation TDI observables, when satisfying $\textbf{N}[ \overrightarrow{l}] = \textbf{N}[\overleftarrow{l}]$ and returning to the starting spacecraft after completing the entire path, we can eliminate fixed unequal arm length differences \cite{Vallis}. We refer to this case as $L$ closed. For 2nd-generation TDI paths, based on satisfying $L$ closed, further satisfying the conditions in Eq. (\ref{equ:L-dot-close}) (the path should at least simultaneously include $\overrightarrow{l} \overrightarrow{\dot{m}},\overleftarrow{l} \overleftarrow{\dot{m}},\overrightarrow{l} \overleftarrow{\dot{m}},\overleftarrow{l} \overrightarrow{\dot{m}}$) can eliminate the first-generation differentials of arm lengths. We refer to this case as $\dot{L}$ closed \cite{Vallis}. Here, $\textbf{N}[i]$ indicates the number of occurrences of $i$ in the string. The count of $l \dot{m}$ is statistically calculated as follows: each $\overrightarrow{l}$ with itself, and all $\overrightarrow{m}$ and $\overleftarrow{m}$ to its right; each $\overleftarrow{l}$ with all $\overrightarrow{m}$ and $\overleftarrow{m}$ to its right. For example, for $\overrightarrow{3'3} \ \overleftarrow{2'2}$, all counted double-letter combinations include $\overrightarrow{3'3'}, \overrightarrow{3'3}, \overrightarrow{3'} \overleftarrow{2'}, \overrightarrow{3'}\overleftarrow{2},\ \overrightarrow{33},\ \overrightarrow{3}\overleftarrow{2'},\ \overrightarrow{3}\overleftarrow{2}$, and $\overleftarrow{2'2}$.
\begin{equation} 
 \label{equ:L-dot-close}
 \textbf{N}[\overrightarrow{l} \overrightarrow{\dot{m}}, \overleftarrow{l} \overleftarrow{\dot{m}} ]
= \textbf{N}[\overrightarrow{l} \overleftarrow{\dot{m}}, \overleftarrow{l} \overrightarrow{\dot{m}}] \hspace{20pt} (l,m=1,1',2,2',3,3')
\end{equation}

For $L$ closed interference paths, when we appropriately connect another closed interference path at a suitable point, the newly formed TDI path is at least still $L$ closed. The term "appropriate point" refers to the spacecraft at the junction point, which, after following the newly connected path, returns to the original junction spacecraft. During the joining process, encountering pairs like $\overrightarrow{l} \overleftarrow{l}$ or $\overleftarrow{l} \overrightarrow{l}$ ($l=1,1',2,2',3,3'$) that have no physical significance can be eliminated.
For example, in the case of an unequal arm length Michelson TDI path ($L$ closed):
\begin{equation}
\overrightarrow{_{1} 3' _{2} 3 _{1} 2 _{3} 2' }_{\textbf{1}} \overleftarrow{3 _{2} 3' _{1} 2' _{3} 2 _{1} }  
\end{equation}
We connect an interference path at SC1 as follows: $\overrightarrow{_1 2 _3 2' _1 3' _2 3 }_{\textbf{1}} \overleftarrow{ 2' _3 2 _1 3 _2 3' _1}$, resulting in:
\begin{equation}
\label{equ:BeamXX1}
\overrightarrow{_1 3' _2 3 _1 2 _3 2' }_{\textbf{1}} \overrightarrow{[2 _3 2' _1 3' _2 3 }_{\textbf{1}} \overleftarrow{ 2' _3 2 _1 3 _2 3' ]}_{\textbf{1}} \overleftarrow{3 _{2} 3' _{1} 2' _{3} 2 _{1}}
\end{equation}
The resulting Eq. (\ref{equ:BeamXX1}) is the previously computed second-generation unequal arm length Michelson interference, and according to (\ref{equ:L-dot-close}), it is also $\dot{L}$ closed.

\subsection{Self-Splicing to Construct Second-generation TDI Observables}

Further analysis reveals that for a first-generation TDI, reversing the order of all arm lengths also reverses the direction of time, for example $\overrightarrow{ 2 1 3 } \ \overleftarrow{2'  1'  3'} \Rightarrow \overrightarrow{ 3'  1'  2'} \ \overleftarrow{ 3  1  2}$. The TDI paths before and after transformation are essentially the same, with their computed results differing by approximately a negative sign ($dt[\overrightarrow{ 2 1 3 } \ \overleftarrow{2'  1'  3'}] \simeq -dt[\overrightarrow{ 3'  1'  2'} \ \overleftarrow{ 3  1  2}]$). By appropriately splicing these two interference modes before and after transformation, a new TDI observables is constructed that is $\dot{L}$ closed, constituting a second-generation TDI observable. 
For first-generation TDI, while maintaining $L$ closure, we can flexibly transform them to facilitate splicing construction.

Taking Sagnac ($\alpha$) type as an example, let's explain the splicing construction process. The original Sagnac ($\alpha$) interference path is $\overrightarrow{ 2 1 3 } \ \overleftarrow{2'  1'  3'}$. After the reversal transformation, we obtain $\overrightarrow{ 3'  1'  2'} \ \overleftarrow{ 3  1  2}$. Then, we transform $\overrightarrow{ 3'  1'  2'} \ \overleftarrow{ 3  1  2}$ to have SC1 as both the starting and ending points, resulting in paths starting and ending at SC2: $\overrightarrow{1'  2'} \ \overleftarrow{ 3  1  2} \ \overrightarrow{3'}$, and starting and ending at SC3: $\overrightarrow{2'} \ \overleftarrow{ 3  1  2} \ \overrightarrow{3' 1'}$. These results are then spliced into the original interference path corresponding to the spacecraft, resulting in three independent second-generation TDI paths named $\alpha$12-1, $\alpha$12-2, and $\alpha$12-3, as shown in Eq. (\ref{equ:BeamSS1}).
Sagnac ($\alpha$)-type:
\begin{equation}
\label{equ:BeamSS1}
\begin{split}
 \overrightarrow{ 2 1 3 } \ \overleftarrow{2'  1'  3'} \overset{\text{\tiny{inversed}}}{\Longleftrightarrow} 
   &\ \overrightarrow{ 3'  1'  2'} \ \overleftarrow{ 3  1  2} \\
  \overrightarrow{213} _1 \overleftarrow{2'1'3'} + \overrightarrow{_1 3'1'2'}\ \overleftarrow{312 _1} \Longrightarrow
  &\ \overrightarrow{2 1  3 [ 3'  1' 2'} \ \overleftarrow{3 1 2 ] 2' 1' 3'} \ (\alpha \text{12-1}) \\
  \overrightarrow{21 _2 3}\ \overleftarrow{2'1'3'}+ \overrightarrow{_2 1'2'} \ \overleftarrow{312} \ \overrightarrow{3' _2} \Longrightarrow
  &\ \overrightarrow{21[1'2'} \ \overleftarrow{312} \ \overrightarrow{3']3} \ \overleftarrow{2'1'3'} \ (\alpha \text{12-2}) \\
 \overrightarrow{2 _3 13} \ \overleftarrow{2'1'3'} + \overrightarrow{_3 2'} \ \overleftarrow{312} \ \overrightarrow{3'1' _3} \Longrightarrow
 &\ \overrightarrow{2 [ 2'} \ \overleftarrow{3 1 2} \ \overrightarrow{3' 1' ] 1 3} \ \overleftarrow{2' 1' 3'} \ ( \alpha \text{12-3})
\end{split}
\end{equation}
\begin{figure}[!ht]
\centering
\includegraphics[width=0.45\textwidth]{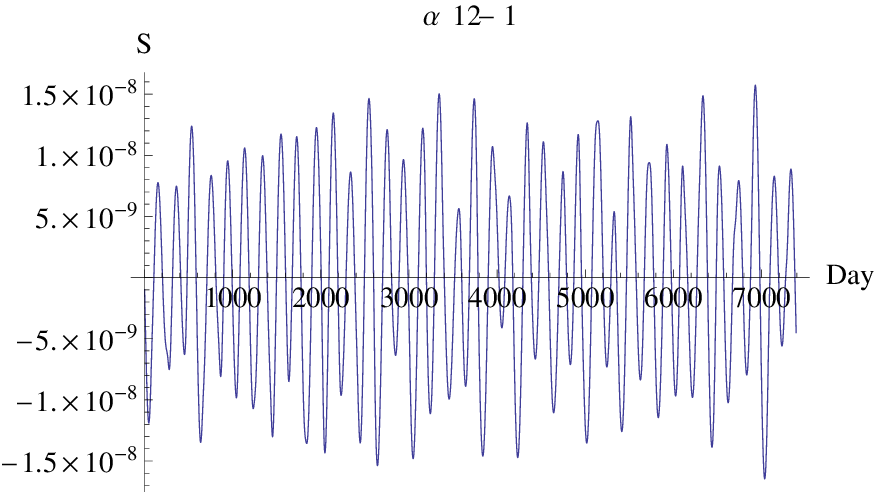}
\includegraphics[width=0.45\textwidth]{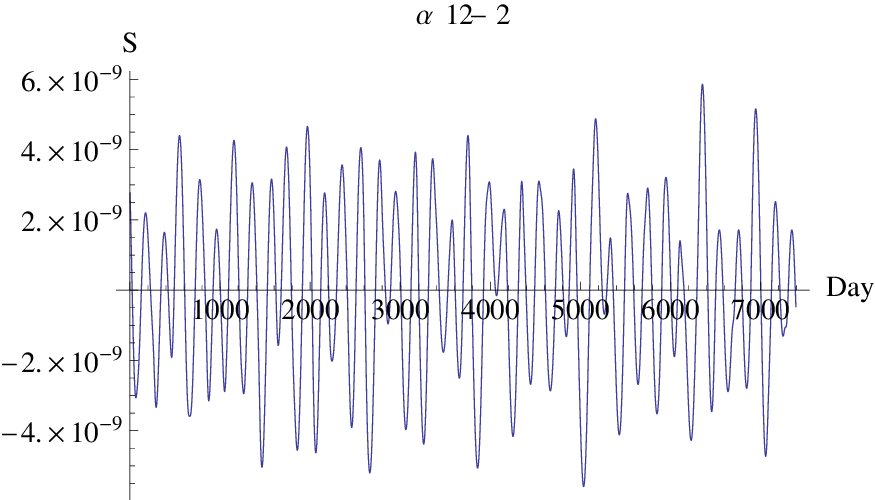}
\includegraphics[width=0.45\textwidth]{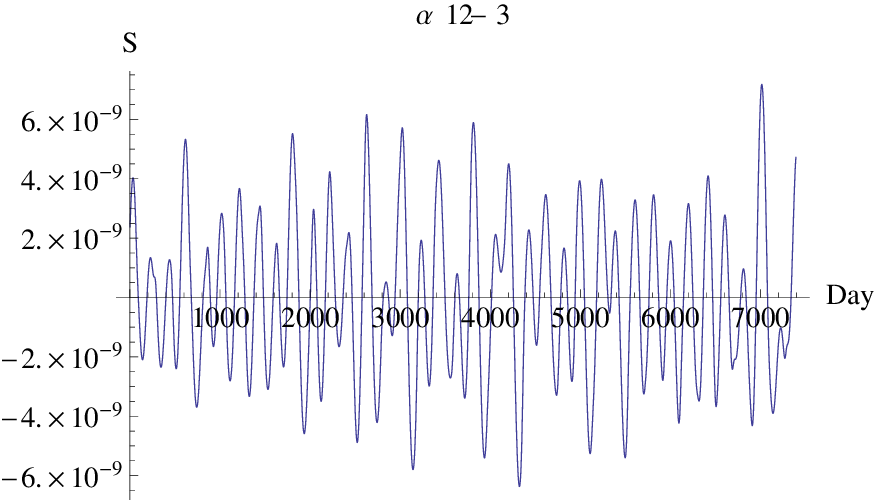}
\caption{\small{The path mismatches in Sagnac ($\alpha$)-type from numerical calculation.}}
\end{figure}

Using this method, based on existing first-generation TDI paths Michelson (X), Relay (U), Beacon (P), and Monitor (E), the second-generation TDI paths are constructed by splicing the original paths with their reversed counterparts. Subsequently, numerical calculations are performed on the resulting second-generation TDI observables.
Michelson (X)-type:
\begin{equation}
\label{equ:BeamX1}
\begin{split} \left.
\begin{split}
  & \overrightarrow{ 3' 3  2  2' } \ \overleftarrow{3  3'  2'  2}  \\
  & \overrightarrow{ 2  2'  3'  3 } \ \overleftarrow{ 2'  2  3  3'} 
  \end{split}   \right\}    \Rightarrow   \left\{
  \begin{split}
  \text{X16-1: } & \overrightarrow{3' 3  2  2' [ 2  2' 3' 3} \ \overleftarrow{2' 2 3 3' ] 3 3' 2' 2} \\
  \text{X16-2: } & \overrightarrow{ 3' 3  2  2' [ 3' 3} \ \overleftarrow{2' 2 3 3'} \ \overrightarrow{2 2'} ] \overleftarrow{3 3' 2' 2} \\
  \text{X16-3: } & \overrightarrow{2 2' 3' 3 [ 2 2'} \ \overleftarrow{3 3' 2' 2} \ \overrightarrow{3' 3} ] \overleftarrow{2' 2 3 3'}
\end{split}  \right.
\end{split}
\end{equation}
\begin{figure}[!ht]
\centering
\includegraphics[width=0.45\textwidth]{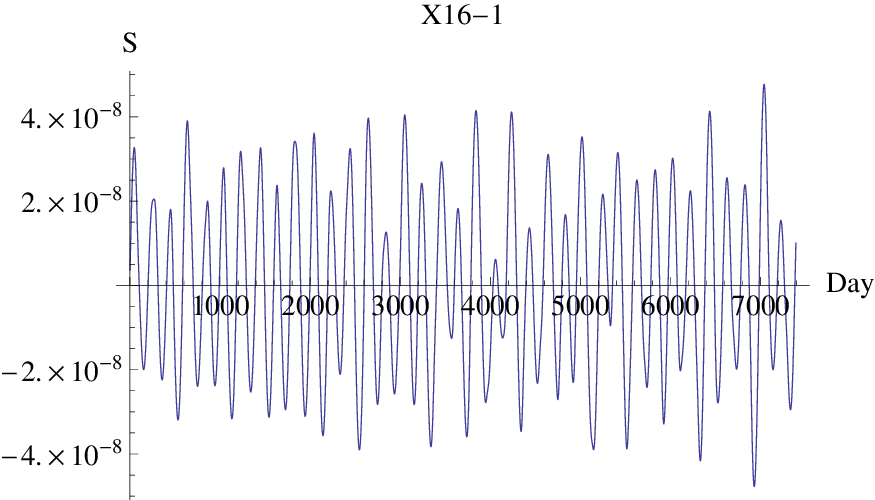}
\includegraphics[width=0.45\textwidth]{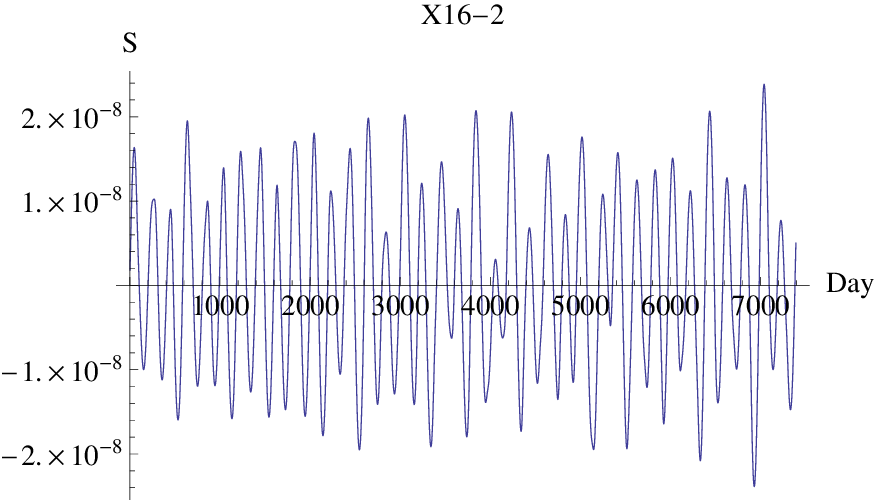}
\includegraphics[width=0.45\textwidth]{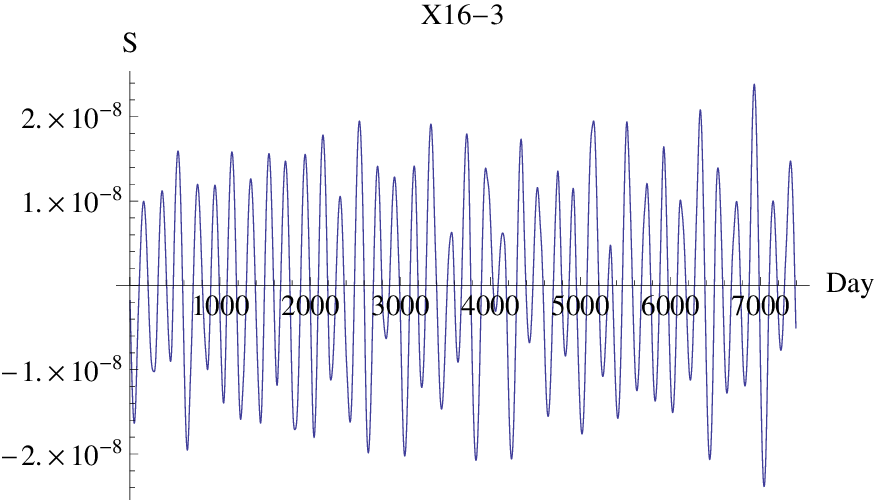}
\caption{\small{The path mismatches in Michelson (X)-type from numerical calculation.}}
\end{figure}

Relay (U)-type: 
\begin{equation}
\label{equ:BeamU1}
\begin{split} \left.
  \begin{split}
  & \overrightarrow{ 2' 3' 1' 1 } \ \overleftarrow{3' 2' 1' 1 } \\
  & \overrightarrow{ 1 1' 2' 3'} \ \overleftarrow{1 1' 3' 2'} 
  \end{split}   \right\}    \Rightarrow   \left\{
  \begin{split}
 \text{U16-1: } & \overrightarrow{ 2' 3' 1' 1 } \ \overleftarrow{3' 2' 1' [ 3' 2'} \ \overrightarrow{ 1 1' 2' 3' } \ \overleftarrow{ 1 1' ] 1} \\
 \text{U16-2: } & \overrightarrow{ 2' 3' 1' 1 } \ \overleftarrow{3' 2'} [ \overrightarrow{ 1 1' 2' 3' } \ \overleftarrow{ 1 1' 3' 2' ] 1' 1} \\
 \text{U16-3: } & \overrightarrow{ 2' 3' 1' 1 [ 1' 2' 3'} \ \overleftarrow{1 1' 3' 2'} \ \overrightarrow{ 1 } ] \overleftarrow{3' 2' 1' 1}
 \end{split} \right.
 \end{split}
\end{equation}
\begin{figure}[!ht]
\centering
\includegraphics[width=0.45\textwidth]{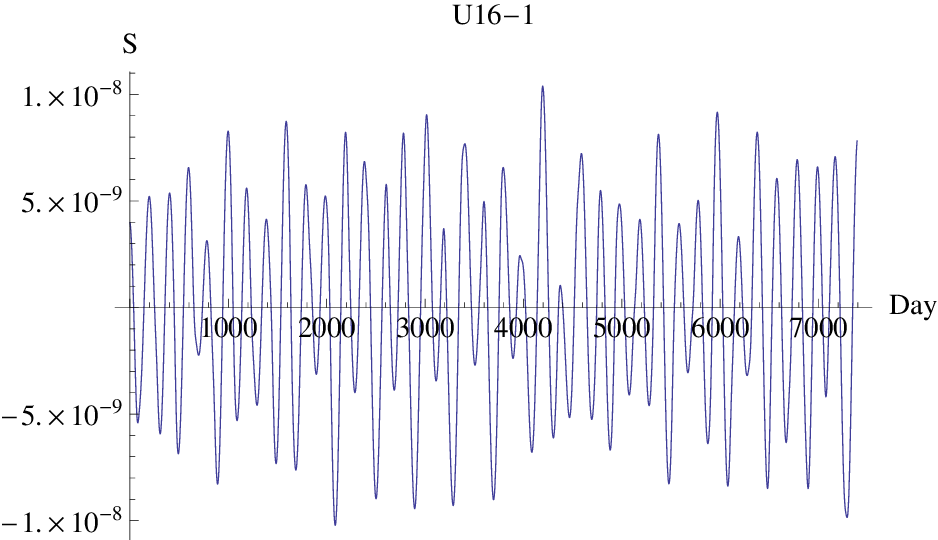}
\includegraphics[width=0.45\textwidth]{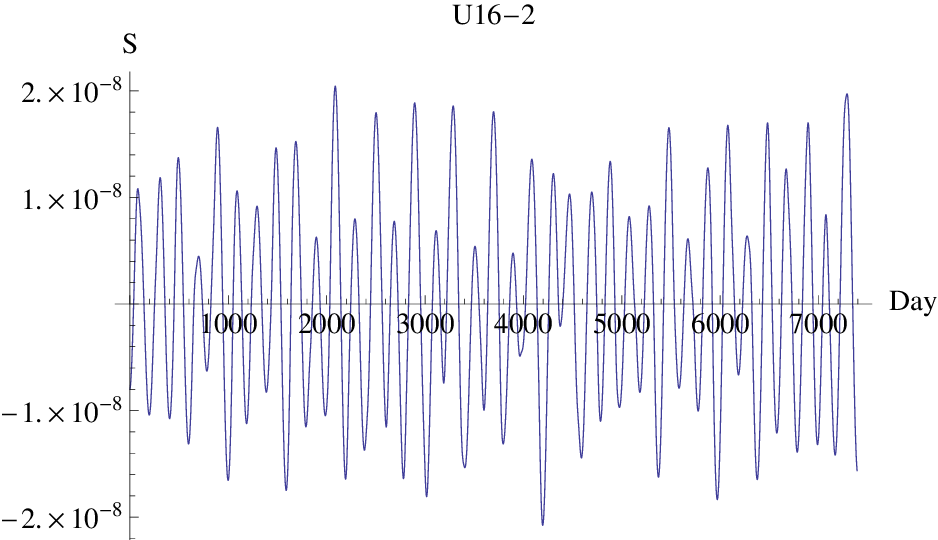}
\includegraphics[width=0.45\textwidth]{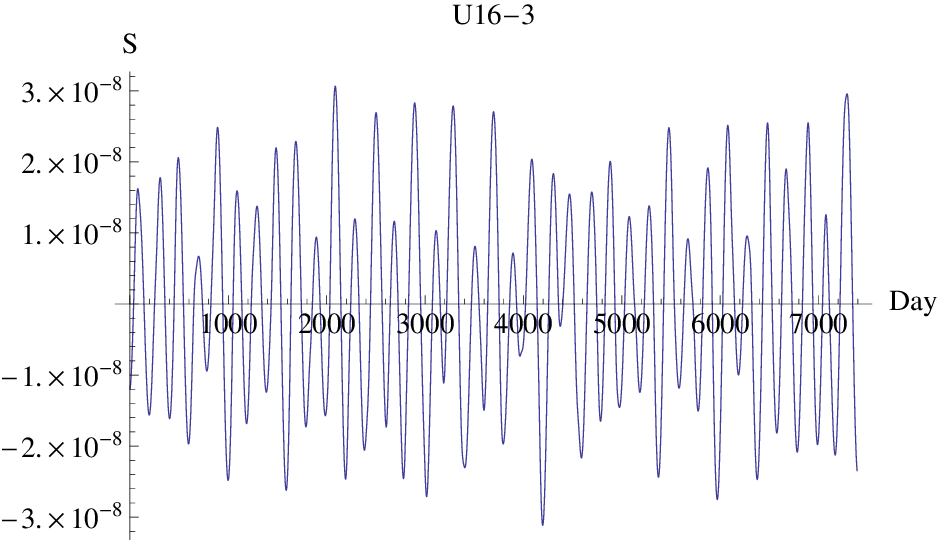}
\caption{\small{The path mismatches in Relay (U)-type from numerical calculation.}}
\end{figure}

Beacon (P)-type:
\begin{equation}
\label{equ:BeamP1}
\begin{split}
   \left.
  \begin{split}  
  & \overrightarrow{ 3' 1' 1 }\ \overleftarrow{3' }\ \overrightarrow{2 }\ \overleftarrow{1' 1 2 }  \\
 &  \overrightarrow{ 2 1 1' }\ \overleftarrow{2 }\ \overrightarrow{3' }\ \overleftarrow{1 1' 3' }  
   \end{split}   \right\}    \Rightarrow   \left\{
  \begin{split}
   \text{P16-1:  } & \overrightarrow{ 3' 1' 1 [1' } \ \overleftarrow{ 2 } \ \overrightarrow{ 3' } \ \overleftarrow{ 1 1' 3' } 
   \ \overrightarrow{ 2 1] } \ \overleftarrow{ 3' } \ \overrightarrow{ 2 } \ \overleftarrow{ 1' 1 2 } \\
   \text{P16-2:  } & \overrightarrow{ 2 1 1' [1 } \ \overleftarrow{ 3' } \ \overrightarrow{ 2 } \ \overleftarrow{ 1' 1 2 } 
   \ \overrightarrow{ 3' 1' ]} \ \overleftarrow{ 2 } \ \overrightarrow{ 3' } \ \overleftarrow{ 1 1' 3' } \\
   \text{P16-3:  } & \overrightarrow{ 2 1 1' } \ \overleftarrow{ 2 } \ \overrightarrow{ [3' 1' 1 } \ \overleftarrow{ 3' }
   \ \overrightarrow{ 2 } \ \overleftarrow{ 1' 1 2] } \ \overrightarrow{ 3' } \ \overleftarrow{ 1 1' 3' }
   \end{split}
   \right.
 \end{split}
\end{equation}
\begin{figure}[!ht]
\centering
\begin{minipage}[t]{0.48\textwidth}
\centering  \includegraphics[width=0.95\textwidth]{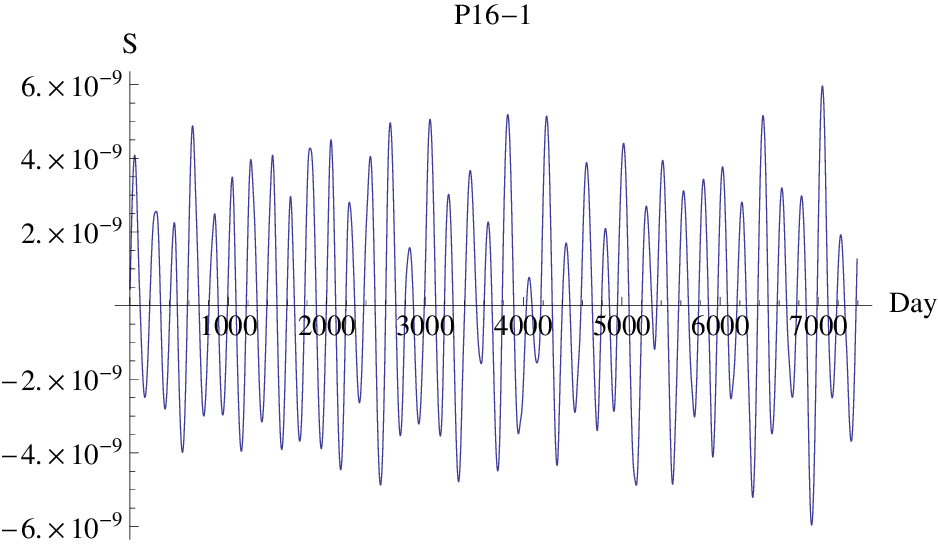} 
   \end{minipage}  
\begin{minipage}[t]{0.48\textwidth}
\centering  \includegraphics[width=0.95\textwidth]{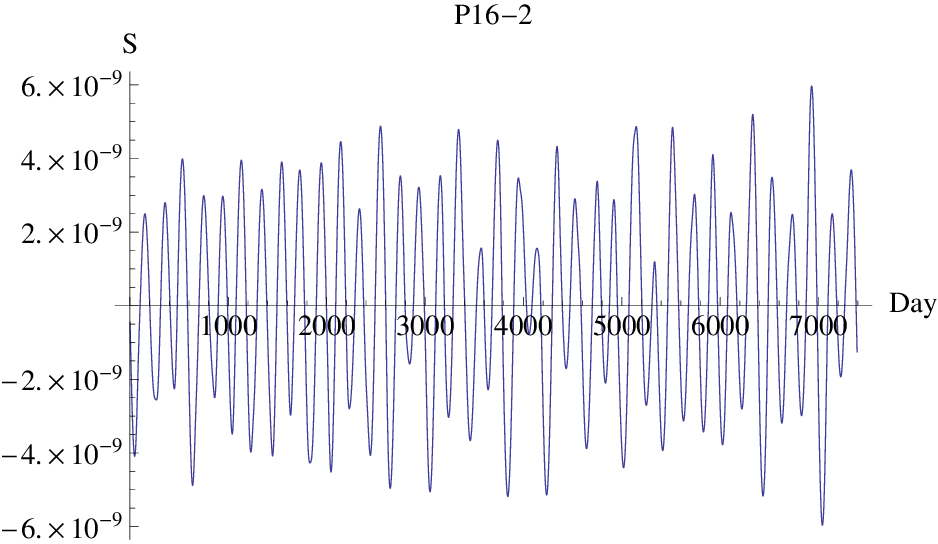} 
   \end{minipage} \\
   \begin{minipage}[t]{0.48\textwidth}
\centering   \includegraphics[width=0.95\textwidth]{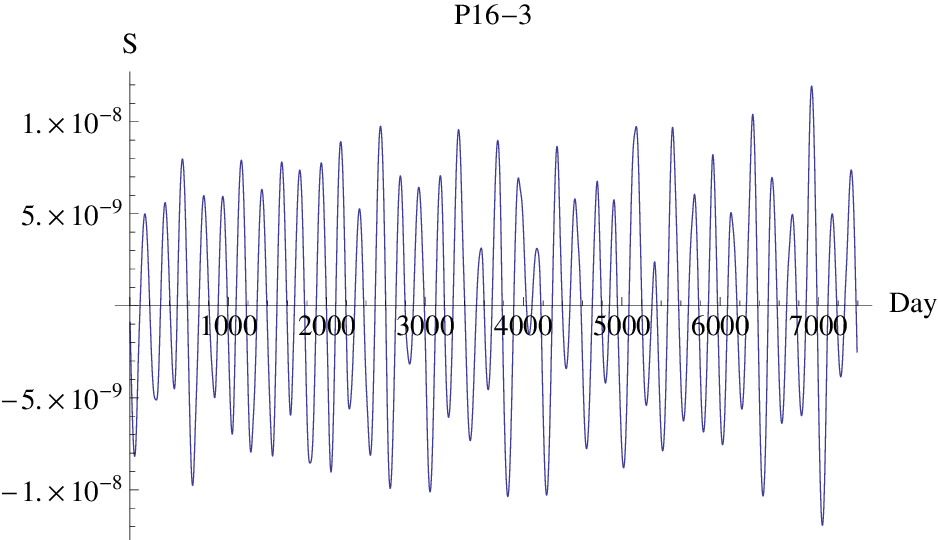} 
   \end{minipage}
   \caption{\small{The path mismatches in Beacon (P)-type from numerical calculation.}}
\end{figure}

Monitor (E)-type: 
\begin{equation}
\label{equ:BeamE1}
\begin{split} \left.
  \begin{split} 
  &  \overrightarrow{1'  1  3 }\ \overleftarrow{2'  1'  1 } \ \overrightarrow{ 2' } \ \overleftarrow{3} \\
  &  \overrightarrow{ 3 } \ \overleftarrow{2'} \ \overrightarrow{1  1'  2'} \ \overleftarrow{3  1  1' } \  \\
 \end{split}  \right\}    \Rightarrow   \left\{
\begin{split}
    \text{E16-1:  } & \overrightarrow{ 1 [1' 1 3 } \ \overleftarrow{ 2' 1' 1 } \ \overrightarrow{ 2' } \ \overleftarrow{ 3] }  
    \ \overrightarrow{ 1' 2' } \ \overleftarrow{ 3 1 1' } \ \overrightarrow{ 3 } \ \overleftarrow{ 2' } \\
    \text{E16-2:  } & \overrightarrow{ 1' [1 1' 2' } \ \overleftarrow{ 3 1 1' } \ \overrightarrow{ 3 } \ \overleftarrow{ 2'] } 
    \ \overrightarrow{ 1 3 } \ \overleftarrow{ 2' 1' 1 } \ \overrightarrow{ 2' } \ \overleftarrow{ 3 } \\
    \text{E16-3:  } & \overrightarrow{ 1 1' 2' } [ \overleftarrow{ 3 } \ \overrightarrow{1' 1 3 } \ \overleftarrow{ 2' 1' 1 } 
    \ \overrightarrow{ 2' } ] \overleftarrow{3 1 1' } \ \overrightarrow{ 3 } \ \overleftarrow{ 2' } 
 \end{split} \right.
 \end{split}
\end{equation}

\begin{figure}[!ht]
\centering
\begin{minipage}[t]{0.48\textwidth}
\centering  \includegraphics[width=0.95\textwidth]{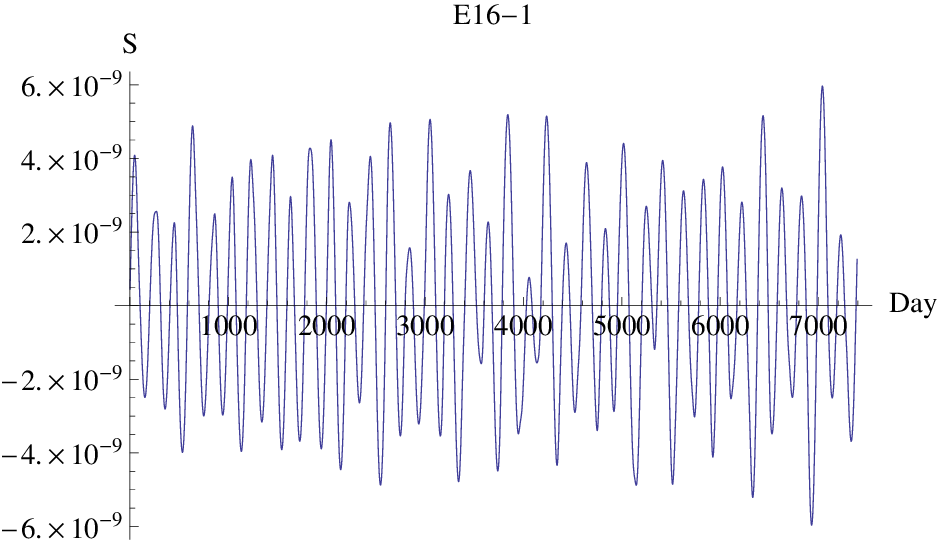} 
   \end{minipage}  
\begin{minipage}[t]{0.48\textwidth}
\centering  \includegraphics[width=0.95\textwidth]{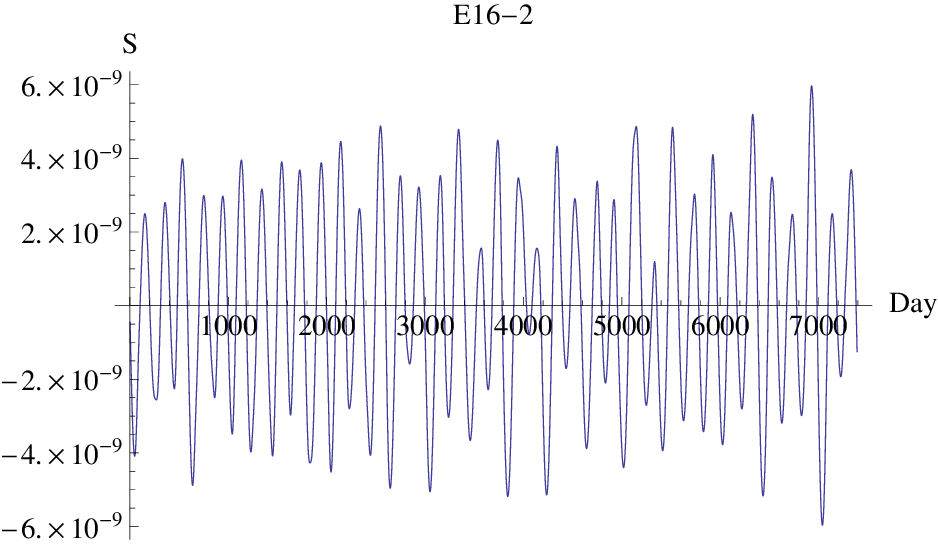} 
   \end{minipage} \\
   \begin{minipage}[t]{0.48\textwidth}
\centering   \includegraphics[width=0.95\textwidth]{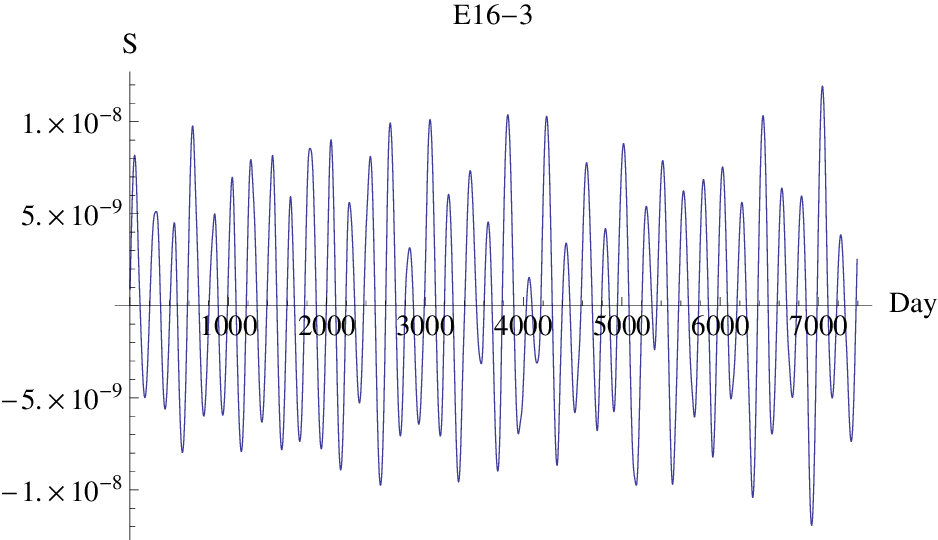} 
   \end{minipage}
   \caption{\small{The path mismatches in Monitor (E)-type from numerical calculation. }}
\end{figure}

\newpage
\newpage

\subsection{Cross-Splicing to Construct Second-Generation TDI Observables}

\subsubsection{Joint Beacon (P) and Monitor (E)}

\begin{figure}[ht]
\begin{minipage}[t]{0.46\textwidth}
\centering
  \includegraphics[width=0.8\textwidth]{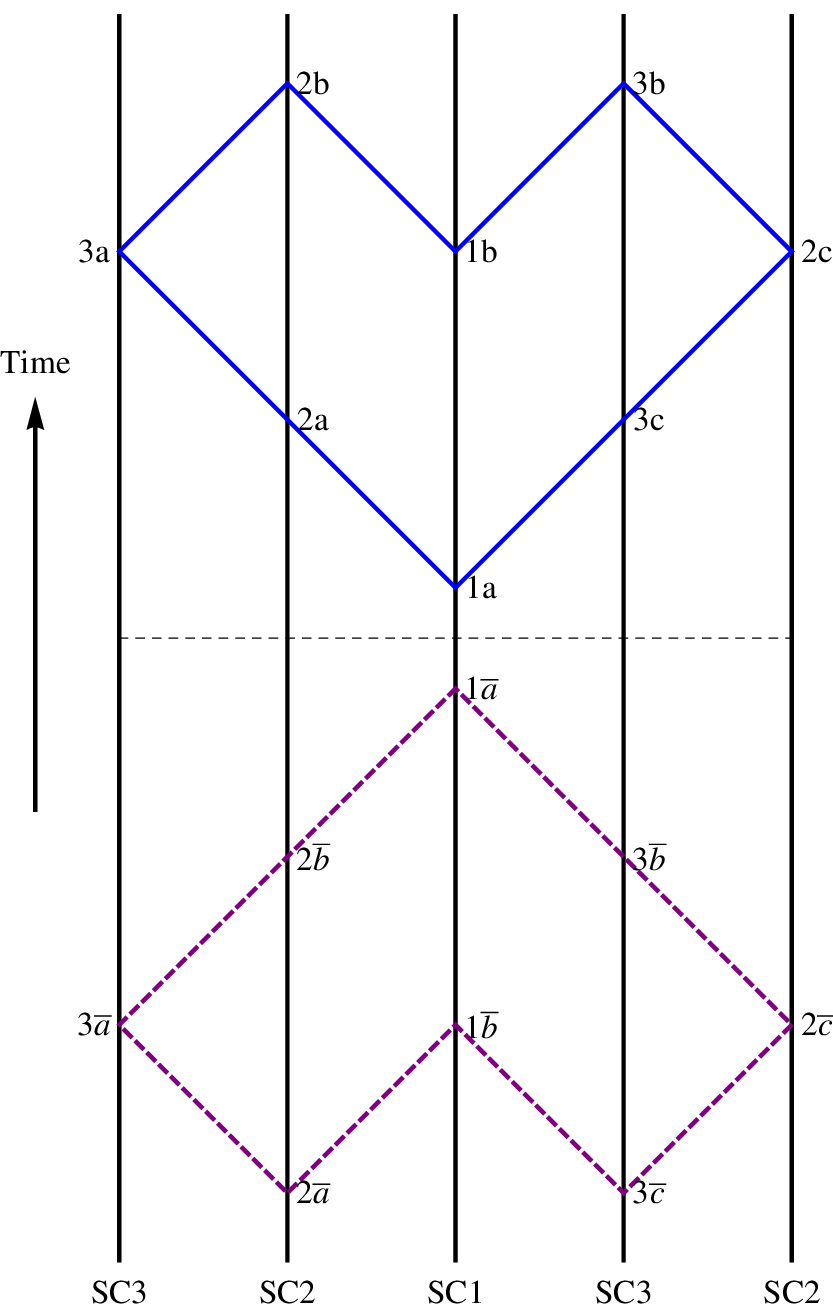} 
   \caption{\small{Geometric diagram of Beacon (P) and $\bar{\text{P}}$.}} \label{fig:PEWL}
   \end{minipage} \hspace{20pt}
\begin{minipage}[t]{0.46\textwidth}
\centering
   \includegraphics[width=0.8\textwidth]{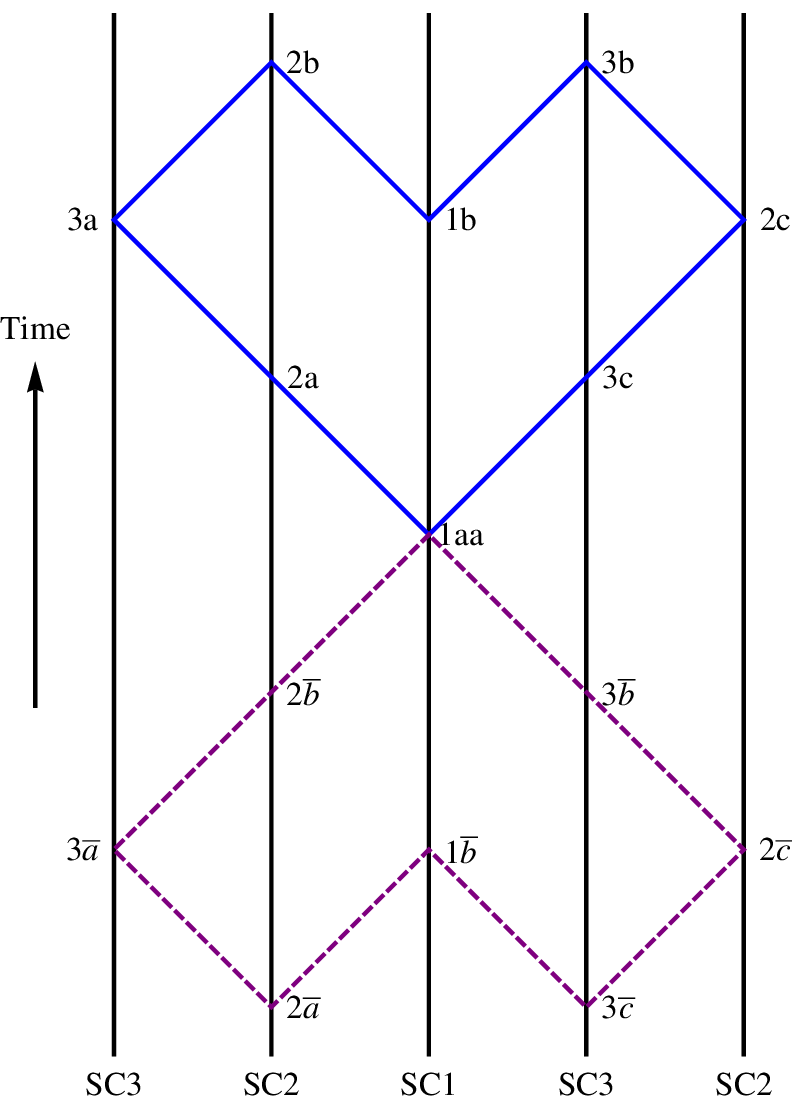}
    \caption{\small{Geometric diagram of PE16-1aa.}} \label{fig:PEWL-1aa}
   \end{minipage}
\end{figure}

As shown in Figure \ref{fig:PEWL}, if we mirror Beacon (P) in the time direction $\bar{\text{P}}$ (shown as dashed lines), we observe that the TDI paths of $\bar{\text{P}}$ aligns with Monitor (E). With first-generation precision, $ dt[\text{Beacon (P)}] = -dt[\text{Monitor (E)}]$, hence we proceed to construct second-generation TDI paths by splicing Beacon (P) and Monitor (E). During computation, we start from the initial spacecraft and proceed clockwise, sequentially numbering spacecraft encountered as primary indices $i (i=1,2,3)$, and using secondary labels $a, b, c$ for each spacecraft visited. In Figure \ref{fig:PEWL}, for clarity, we denote the spacecraft sequence in Monitor (E) as $i \bar{a}, i \bar{b}, i \bar{c} (i=1,2,3)$.

As depicted in Figure \ref{fig:PEWL}, similar to the self-splicing method described earlier, we splice another closed interference path at appropriate spacecraft points. We attempt to splice Monitor (E) into Beacon (P) at suitable spacecraft points. Theoretically, spacecraft $i (i=1,2,3)$ in Monitor (E) can splice into any spacecraft $i$ in Beacon (P). The number of times E passes through SC1, SC2, and SC3 are 2, 3, and 3 respectively, while P passes through SC1, SC2, and SC3 2, 3, and 3 times as well. Therefore, there are 22 (4+9+9) possible connections between them. However, not all configurations are independent. Here, we employ a geometric approach to analyze and select appropriate configurations for computation.

During splicing, we shift Monitor (E)'s path on the world line to align with Beacon (P)'s path at spacecraft $i$ with matching primary indices. This indicates that when traveling along path P to the corresponding spacecraft, we switch to path E, traverse E's closed path, return to spacecraft $i$, and then continue along the remaining path of P. To denote the splicing point, we use the primary index of the splicing spacecraft followed by the secondary indices from P and E's respective splicing points. For instance, $1ab$ signifies splicing at $1a$ on path P into E, then departing from $1 \bar{b}$ on E after encircling it, and returning to path P.
To facilitate further description of the various splicing configurations of Beacon (P) and Monitor (E) resulting in TDI paths, we denote them as PE + actual connection number + splicing points. For example, PE16-1ab indicates the splicing of P and E at SC1a on P and SC1$\bar{\text{b}}$ on E, resulting in 16-link paths.

\begin{figure}[ht]
\begin{minipage}[t]{0.42\textwidth}
  \centering
   \includegraphics[width=0.8\textwidth]{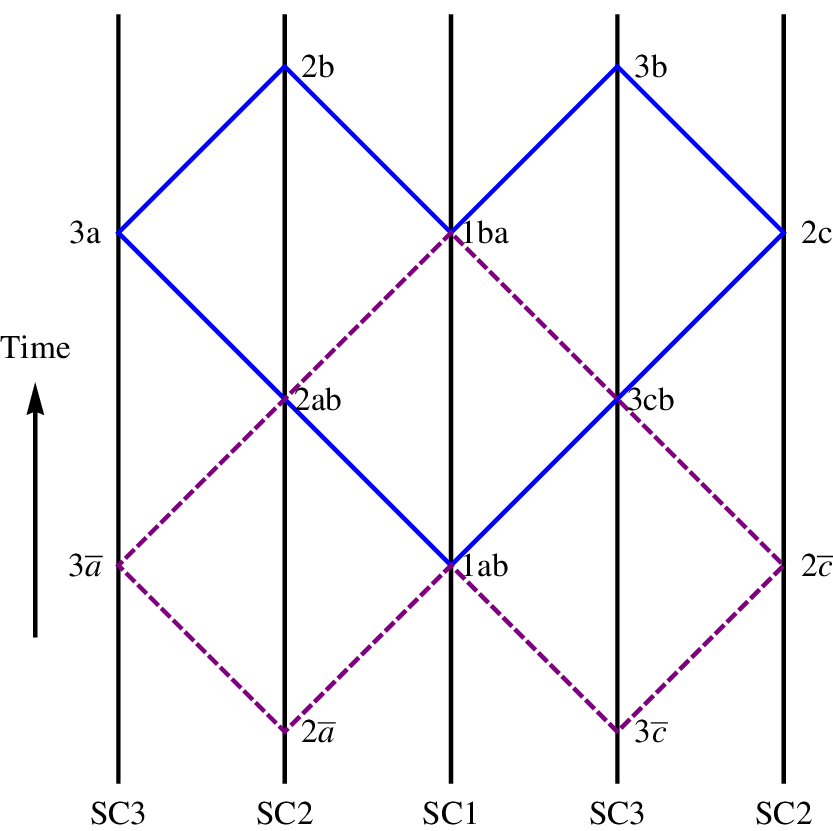}
  \caption{\small{Diagram of PE16-1ab.}}  \label{fig:PEWL-1ab}
  \end{minipage}
  \begin{minipage}[t]{0.55\textwidth}
  \centering
  \includegraphics[width=0.9\textwidth]{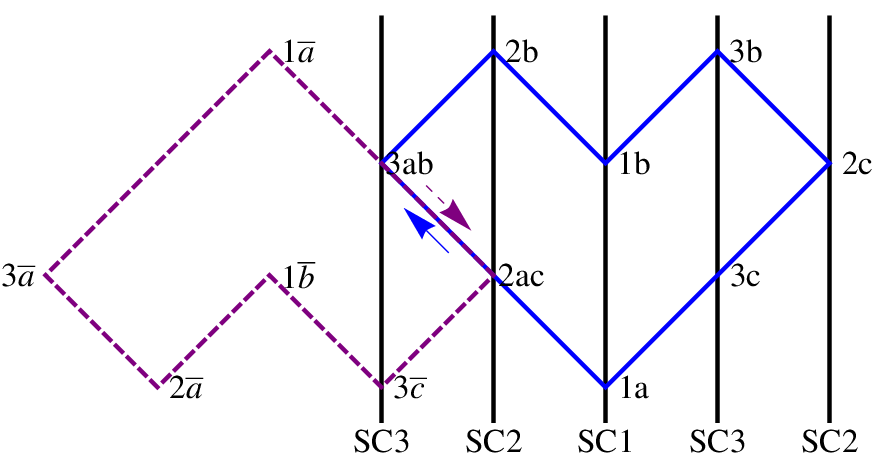}
  \caption{\small{Diagram of PE14-2ac.}} \label{fig:PEWL-2ac}
  \end{minipage}
\end{figure}

\begin{figure}[!ht]
\centering
\includegraphics[width=0.45\textwidth]{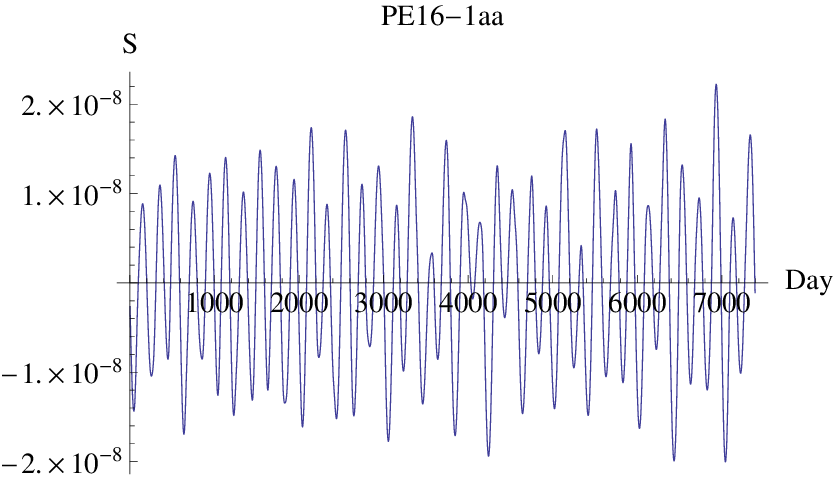}
\includegraphics[width=0.45\textwidth]{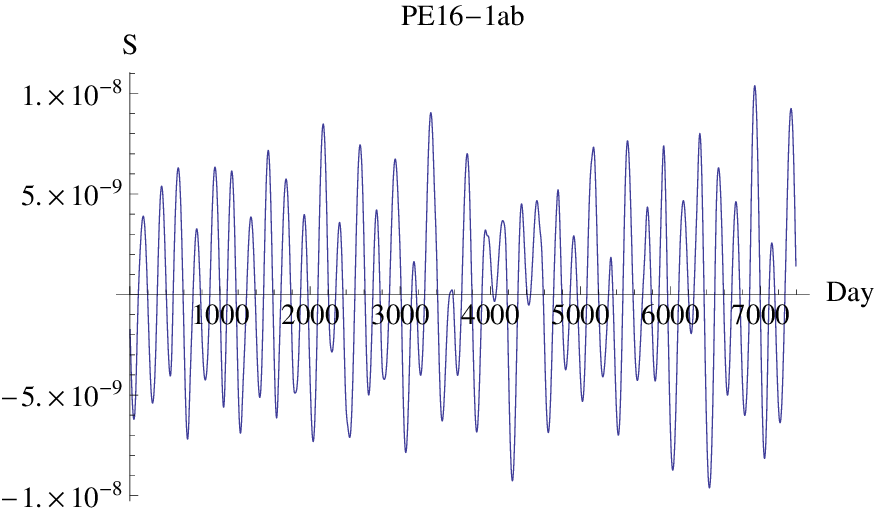} \vspace{10pt} \\
\includegraphics[width=0.45\textwidth]{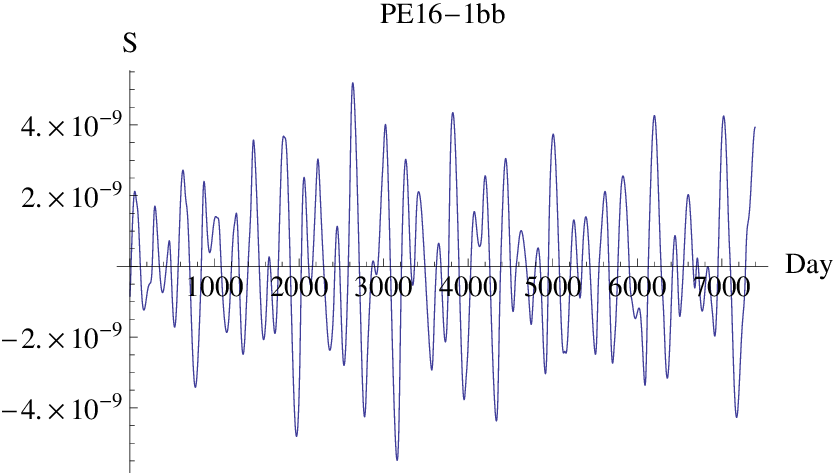}
\includegraphics[width=0.45\textwidth]{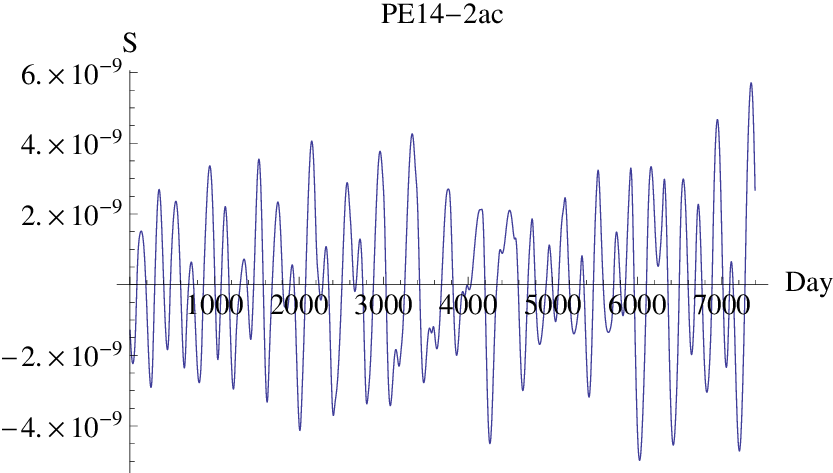} \vspace{10pt} \\
\includegraphics[width=0.45\textwidth]{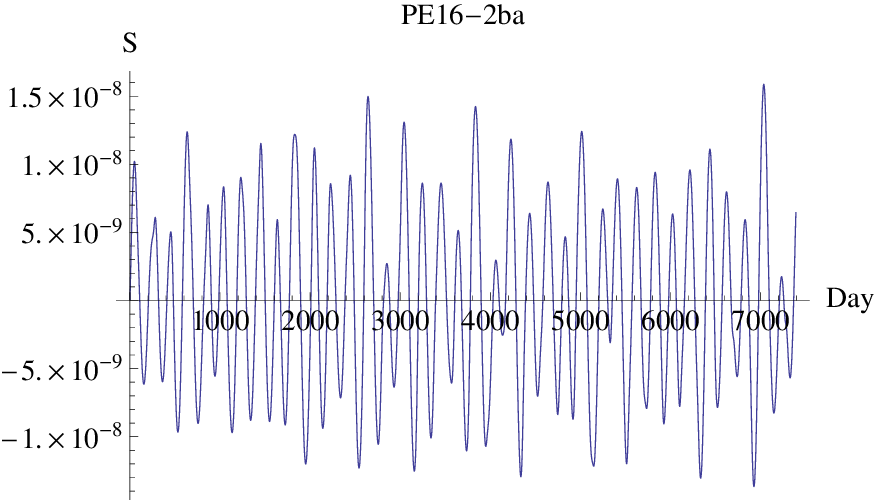}
\includegraphics[width=0.45\textwidth]{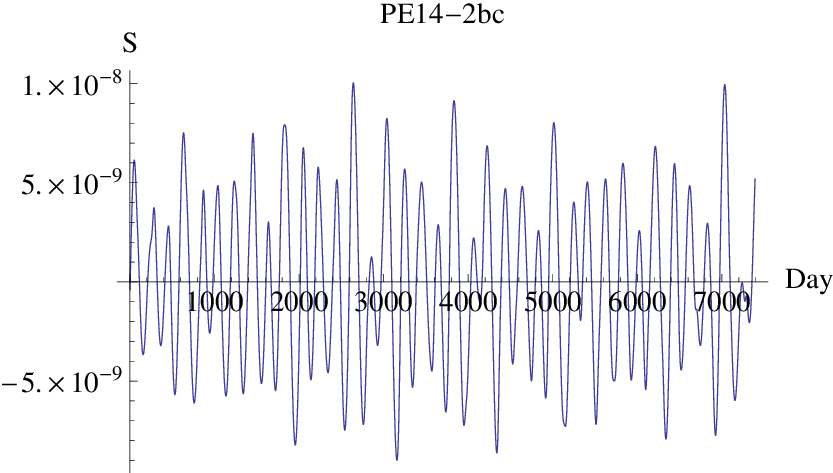}
\caption{\small{The path mismatches in joint Beacon (P) and Monitor (E) type from numerical calculation. }}
\label{fig:ResultPE1}
\end{figure}

When several splicing configurations yield identical path structures, although described differently, their computed results at the second-generation are similar. From Figure \ref{fig:PEWL-1ab}, it can be observed that Beacon (P) and Monitor (E) form the same relative positions on the world line at 1ab, 1ba, 2ab, and 3cc, with differences in computed results being below 1\%. At the second-generation, we consider their interference paths to be inherently identical. Eq. (\ref{equ:PEsame}) lists several overlapping cases in this splicing process, where we select one TDI observable as independent and present it. We identify all independent second-generation TDI paths formed by P and E as shown in (\ref{equ:BeamPE1}), with their respective computed results illustrated in Figure \ref{fig:ResultPE1}.

During the splicing process, we discovered that certain paths cancel out in pairs, reducing the overall number of paths traversed. For instance, paths such as PE14-2ac interference, as depicted in Figure \ref{fig:PEWL-2ac}. When path E splices from P's 2a point into its own $2 \bar{\text{c}}$ point, the final segment of E's path, moving counterclockwise through arm 1 to return to $2 \bar{\text{c}}$ in reverse time direction, immediately proceeds clockwise through arm 1 to 3a or $3\bar{\text{b}}$, these two segments cancel each other completely in numerical computations, and have no physical meaning. Consequently, the actual interference path results in only 14 connections after cancellation.
\begin{equation}
\label{equ:PEsame}
\begin{split}
 & \text{PE16-1aa;} \hspace{90pt}  \text{PE16-1ab: 1ba,2ab,3cb;} \\
 & \text{PE16-1bb: 2aa,2bb,2cc,3aa,3bb,3cc;} \\
 & \text{PE16-2ac: 2cb,3ab,3ca;} \hspace{25pt}  \text{PE16-2ba: 3bc;} \\
 & \text{PE16-2bc: 2ca,3ac,3ba. }
\end{split}
\end{equation}

Beacon (P) and Monitor (E)-type: 
\begin{equation}
\label{equ:BeamPE1}
\begin{split} \left.
  \begin{split} 
  & \overrightarrow{_{1a} 3' _{2a} 1' _{3a} 1 } _{2b} \overleftarrow{3'} _{1b} \overrightarrow{2} _{3b}
   \overleftarrow{1' _{2c} 1 _{3c} 2 _{1a}}  \\
  & \overrightarrow{_{2a} 1' _{3a} 1 _{2b} 3 } _{1a} \overleftarrow{2' _{3b} 1' _{2c} 1} _{3c}
   \overrightarrow{2'} _{1b} \overleftarrow{3 _{2a}} 
 \end{split}  \right\}    \Rightarrow   \left\{
\begin{split}
      \text{PE16-1aa:  } & \overrightarrow{3'1'1} \ \overleftarrow{3'} \ \overrightarrow{2}
    \ \overleftarrow{1'12 [ 2'1'1} \ \overrightarrow{2'} \ \overleftarrow{3} \ \overrightarrow{1'13}] \\
      \text{PE16-1ab:  } & \overrightarrow{3'1'1} \ \overleftarrow{3'} \ \overrightarrow{2}
    \ \overleftarrow{1'12 [ 3}\ \overrightarrow{1'13} \ \overleftarrow{2'1'1} \ \overrightarrow{2'}] \\
      \text{PE16-1bb:  } & \overrightarrow{3'1'1} \ \overleftarrow{3'[3} \ \overrightarrow{1'13}
    \ \overleftarrow{2'1'1} \ \overrightarrow{2']2} \ \overleftarrow{1'12} \\
      \text{PE14-2ac:  } & \overrightarrow{3'} [ \overleftarrow{1} \ \overrightarrow{2'} \ \overleftarrow{3} \ \overrightarrow{1'13}
     \ \overleftarrow{2' \mathbf{1'}} ] \overrightarrow{ \mathbf{1'} 1} \ \overleftarrow{3'} \ \overrightarrow{2} \ \overleftarrow{1'12} \\
     =\ & \overrightarrow{3'} \ \overleftarrow{1} \ \overrightarrow{2'} \ \overleftarrow{3} \ \overrightarrow{1'13}
     \ \overleftarrow{2'} \ \overrightarrow{1} \ \overleftarrow{3'} \ \overrightarrow{2} \ \overleftarrow{1'12} \\
      \text{PE16-2ba:  } & \overrightarrow{3'1'1 [ 1'13} \ \overleftarrow{2'1'1} \ \overrightarrow{2'} 
      \ \overleftarrow{3]3'}   \ \overrightarrow{2} \ \overleftarrow{1'12} \\ 
       \text{PE14-2bc:  } & \overrightarrow{3'1' \mathbf{1}} [ \overleftarrow{\bf{1}} \ \overrightarrow{2'} \ \overleftarrow{3} \ \overrightarrow{1'13}   \ \overleftarrow{2'1'} ] \overleftarrow{3'} \ \overrightarrow{2} \ \overleftarrow{1'12} \\
      =\ & \overrightarrow{3'1'2'} \ \overleftarrow{3} \ \overrightarrow{1'13} \ \overleftarrow{2'1'} \ \overleftarrow{3'} \ \overrightarrow{2} \ \overleftarrow{1'12} \\ 
 \end{split} \right.
 \end{split}
\end{equation}

\newpage

\subsubsection{Joint Relay (U) and $\bar{\text{U}}$ }

\begin{figure}[ht] \centering
\begin{minipage}[t]{0.46\textwidth}
  \centering
  \includegraphics[width=0.8\textwidth]{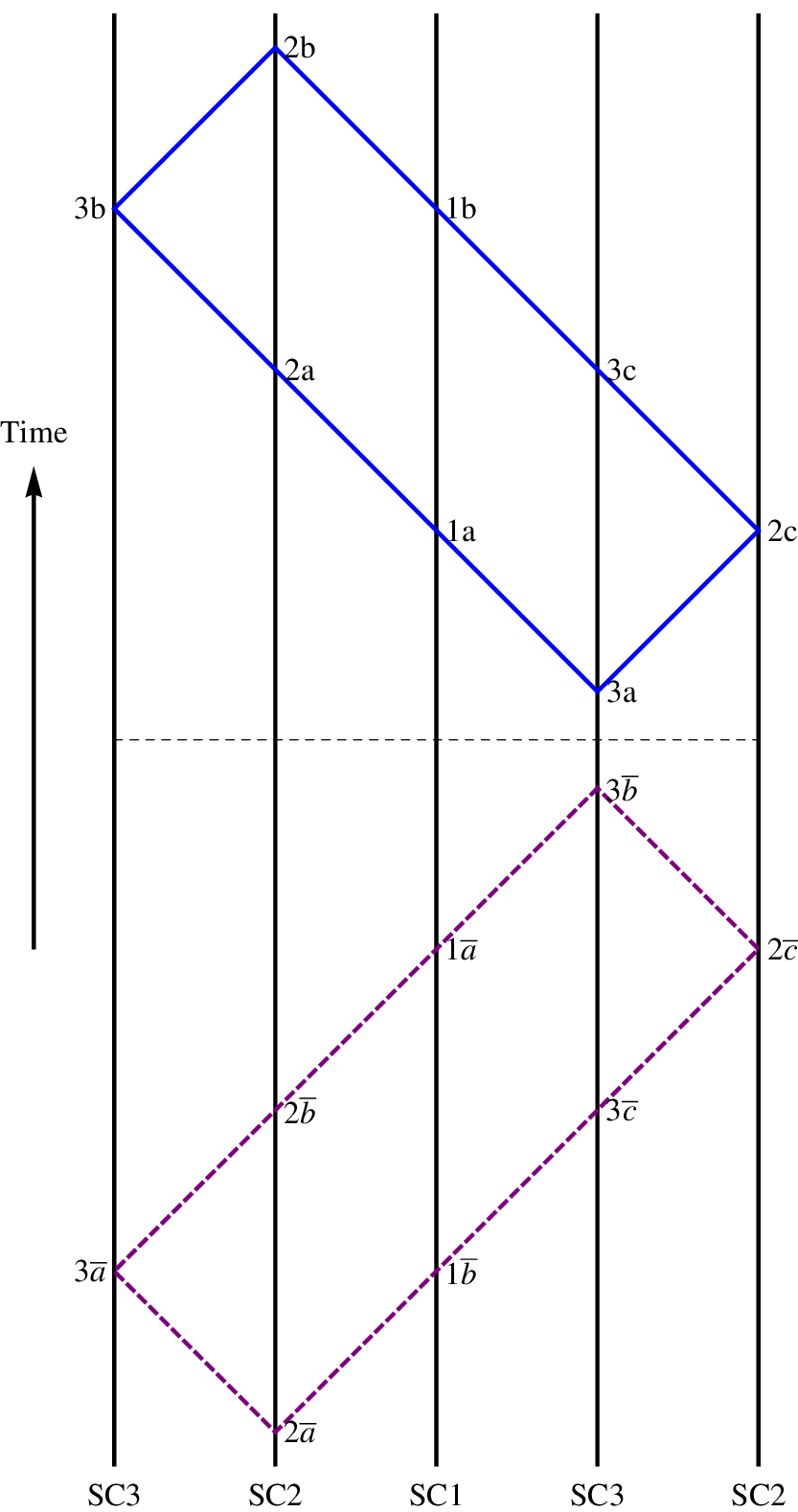}
  \caption{\small{Geometric diagram of Relay (U) and $\bar{\text{U}}$.}} \label{fig:UUWL}
  \hspace{20pt}
  \end{minipage}
  \begin{minipage}[t]{0.46\textwidth}
  \centering
  \includegraphics[width=0.8\textwidth]{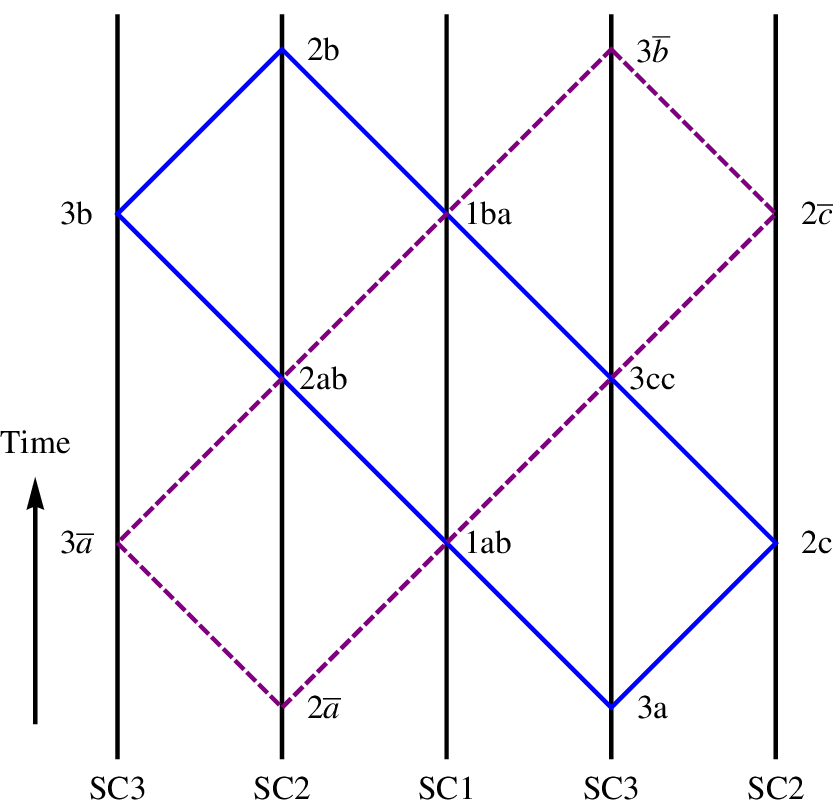}
  \caption{\small{Geometric diagram of UU-1ab.}}  \label{fig:UUWL-1ab}
  \end{minipage}
\end{figure}

Similar to Beacon (P), Relay (U) is mirrored in time to create $\bar{\text{U}}$, as shown in Figure \ref{fig:UUWL}. For first-generation calculations, $dt[\text{U}] = -dt[\bar{\text{U}}]$. By utilizing both paths, we can construct corresponding second-generation TDI paths. We continue to calculate in a clockwise manner for clarity, assigning sequential identifiers to each spacecraft. Relay (U) departs from spacecraft 3a, travels clockwise around U's path, and returns to 3a; $\bar{\text{U}}$ departs from spacecraft $2\bar{\text{a}}$, travels clockwise around $\bar{\text{U}}$'s path, and returns to $2\bar{\text{a}}$.

\begin{figure}[!ht]
\centering
\includegraphics[width=0.45\textwidth]{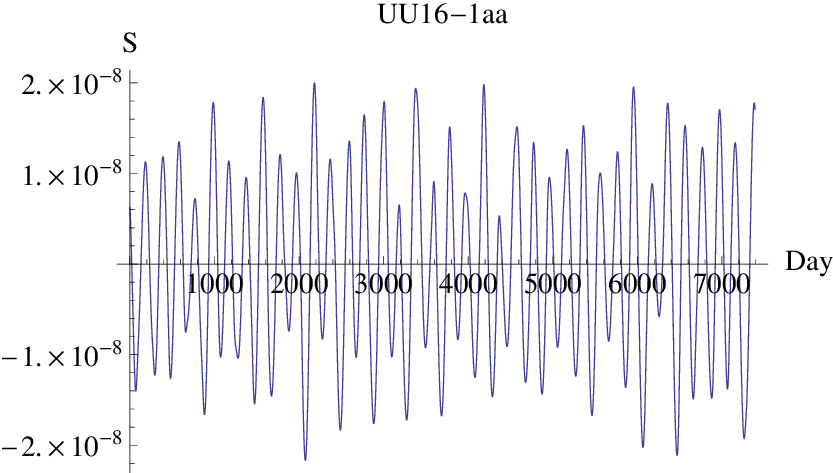}
\includegraphics[width=0.45\textwidth]{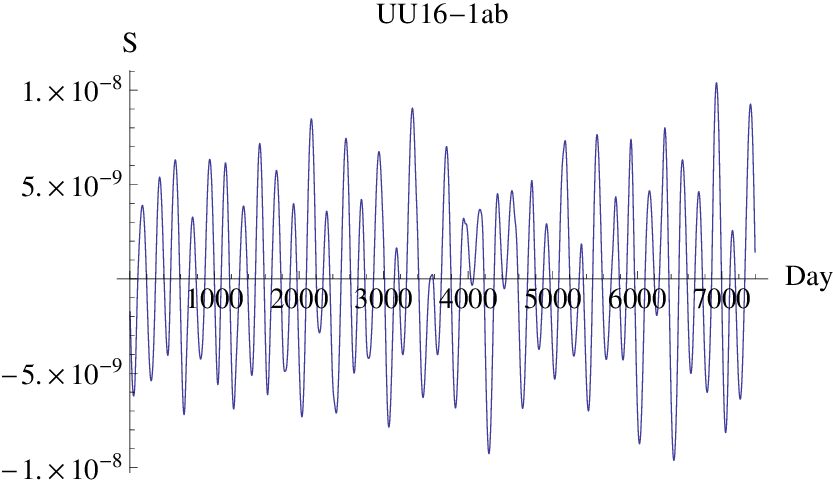} \vspace{10pt} \\
\includegraphics[width=0.45\textwidth]{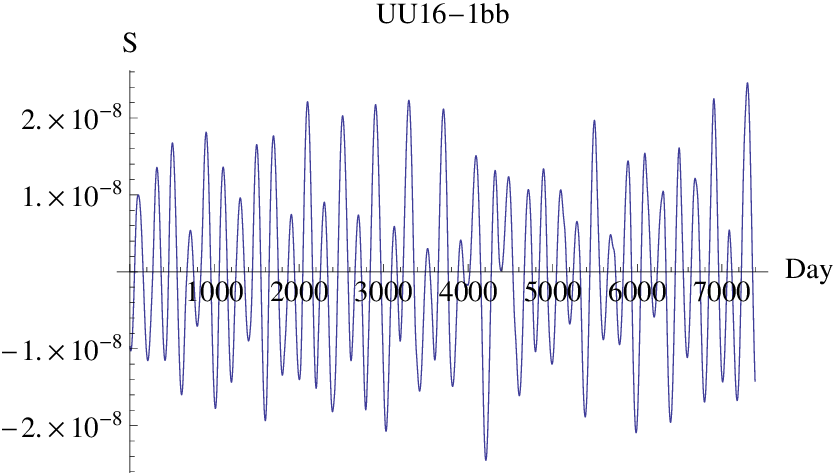}
\includegraphics[width=0.45\textwidth]{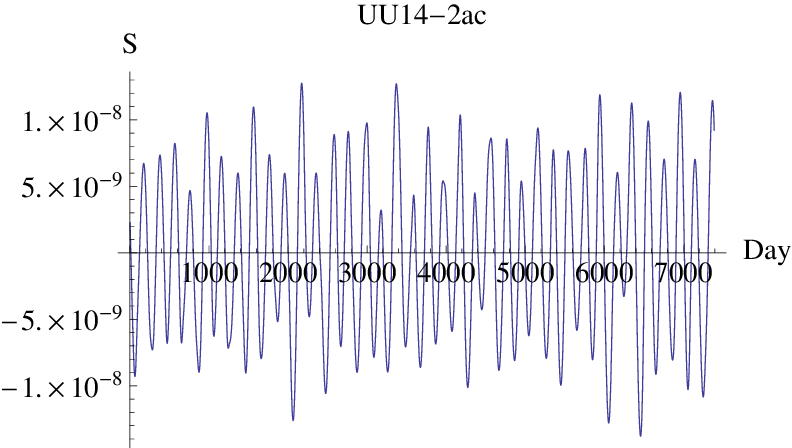} \vspace{10pt} \\
\includegraphics[width=0.45\textwidth]{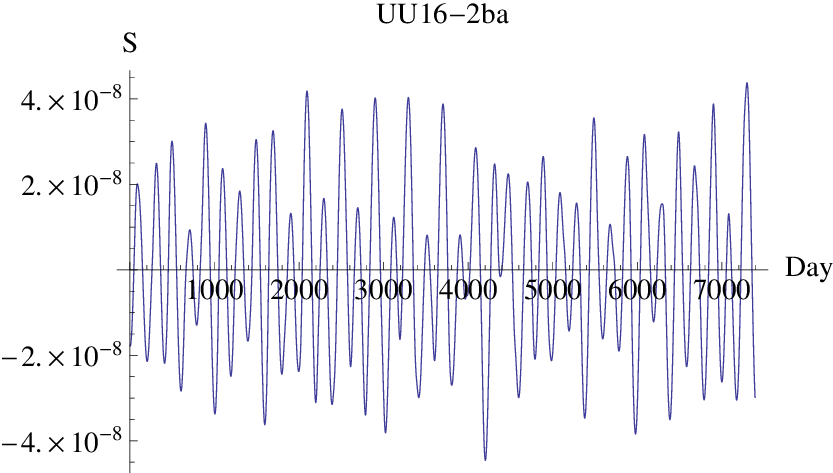}
\includegraphics[width=0.45\textwidth]{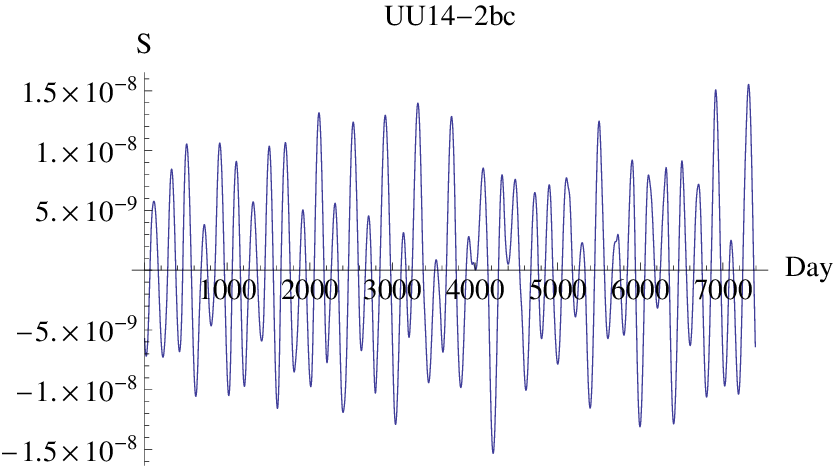} \vspace{10pt} \\
\includegraphics[width=0.45\textwidth]{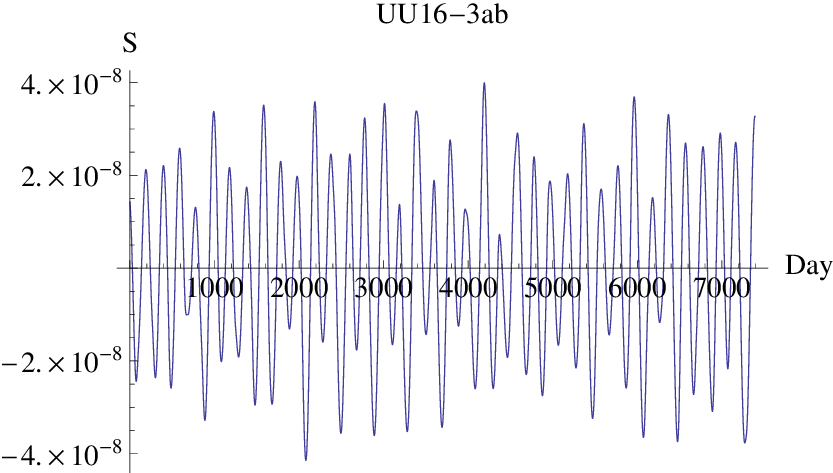}
\includegraphics[width=0.45\textwidth]{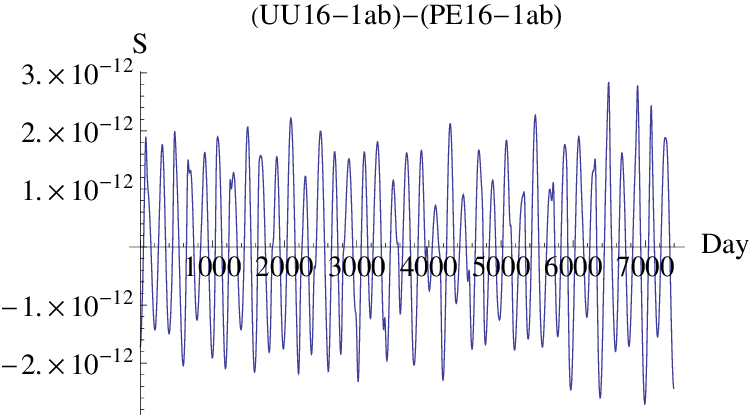}
\caption{\small{The path mismatches in new observables from joint U and $\bar{\text{U}}$.}}
\label{fig:ResultUU}
\end{figure}

According to the previous analysis method, U and $\bar{\text{U}}$ can have 22 (4+9+9) connections, but these 22 connections are not completely independent. The categorization of cases with inherent correlations that we analyzed is shown in Eq. (\ref{equ:UUsame}). It should be noted that in the case of U$\bar{\text{U}}$-1ab, the path formed by U and $\bar{\text{U}}$ is depicted in Figure \ref{fig:UUWL-1ab}. It appears similar to the path discussed earlier in PE-1ab (Figure \ref{fig:PEWL-1ab}), but the actual computation process for the paths is different. However, the difference in the computation results for the second-generation TDI is less than 1\%, as shown in the bottom-right graph of Figure \ref{fig:ResultUU}. There are 7 independent cases in the splicing of U and $\bar{\text{U}}$, and the paths are represented as Eq. (\ref{equ:BeamUU1}). The numerical computation results for each are shown in Figure \ref{fig:ResultUU} (where the labels in the figure use UU instead of U$\bar{\text{U}}$).


\begin{equation}
\label{equ:UUsame}
\begin{split}
 &  \text{U}\bar{\text{U}} \text{16-1aa: 2cc, 3ac, 3cb;} \hspace{25pt}  \text{U}\bar{\text{U}} \text{16-1ab: 1ba,2ab,3cc;} \\
 &  \text{U}\bar{\text{U}} \text{16-1bb: 2aa,2bb,3ba;} \hspace{28pt}  \text{U}\bar{\text{U}} \text{16-2ac: 2cb,3aa,3bb;} \\
 &  \text{U}\bar{\text{U}} \text{16-2ba;}  \hspace{100pt}  \text{U}\bar{\text{U}} \text{16-2bc: 2ca,3bc,3ca;} \\
 &  \text{U}\bar{\text{U}} \text{16-3ab.} 
\end{split}
\end{equation}

Relay (U) and $\bar{\text{U}}$-type: 
\begin{equation}
\label{equ:BeamUU1}
\begin{split} \left.
  \begin{split}
  & \overrightarrow{_{3a} 2' _{1a} 3' _{2a} 1' _{3b} 1} _{2b} \overleftarrow{3' _{1b} 2' _{3c} 1' _{2c} 1 _{3a}} \\
  & \overrightarrow{_{2a} 1' _{3a} 1 _{2b} 3 _{1a} 2} _{3b} \overleftarrow{1' _{2c} 1 _{3c} 2 _{1b} 3 _{2a}} 
  \end{split}   \right\}    \Rightarrow    \left\{ 
  \begin{split} 
  \text{U}\bar{\text{U}}\text{16-1aa: } & \overrightarrow{2' [ 2} \ \overleftarrow{1'123} \ \overrightarrow{1'13 ] 3'1'1} \ \overleftarrow{3'2'1'1} \\
 \text{U}\bar{\text{U}}\text{16-1ab: } & \overrightarrow{2'} [\overleftarrow{3} \ \overrightarrow{1'132} \ \overleftarrow{1'12} ]
   \overrightarrow{3'1'1}\ \overleftarrow{3'2'1'1} \\
 \text{U}\bar{\text{U}}\text{16-1bb: } & \overrightarrow{ 2' 3' 1' 1 } \ \overleftarrow{3'[3} \ \overrightarrow{1'132} \ \overleftarrow{1'12]2'1'1} \\
 \text{U}\bar{\text{U}}\text{14-2ac: } & \overrightarrow{ 2' 3'} [ \overleftarrow{123} \ \overrightarrow{1'132} \ \overleftarrow{\mathbf{1'}} ] \overrightarrow{\mathbf{1'}1} \ \overleftarrow{3'2'1'1}  \\
 =\ & \overrightarrow{2' 3'} \ \overleftarrow{123} \ \overrightarrow{1'1321} \ \overleftarrow{3'2'1'1} \\
 \text{U}\bar{\text{U}}\text{16-2ba: } & \overrightarrow{2' 3' 1' 1 [ 1'132} \ \overleftarrow{1'123 ] 3'2'1'1} \\
 \text{U}\bar{\text{U}}\text{14-2bc: } & \overrightarrow{2' 3'1' \mathbf{1}} [ \overleftarrow{\mathbf{1}23} \ \overrightarrow{1'132} \ \overleftarrow{1']3'2'1'1} \\
 =\ & \overrightarrow{2' 3'1'} \ \overleftarrow{23} \ \overrightarrow{1'132} \ \overleftarrow{1'3'2'1'1} \\
  \text{U}\bar{\text{U}}\text{16-3ab: } & \overrightarrow{2' 3'1'1} \ \overleftarrow{3'2'1'1 [ 1'123} \ \overrightarrow{1'132]} \\
  \end{split} \right. 
 \end{split}
 \end{equation}

The construction of the second-generation TDI paths above involves simply splicing two closed paths. From the splicing process, it can be seen that it essentially involves shifting along the time axis. The number of relatively independent paths obtained from splicing is related to the relative positions of the two closed paths along the time axis. This means that during the splicing process of two paths, one path remains stationary along the time axis while the other path translates along the time direction. Given the conditions for splicing, the number of combinations of relative positions along the time axis determines the number of independent interference paths.
In practical terms, this is easy to understand: the time difference $dt(t)$ after one complete loop is a function of time. Calculation shows that when the same paths are traversed, the more beams that are involved, the smaller the computed time differences.

\newpage

\subsubsection{TDI with more than sixteen links}

For TDI paths with more than 16 connections, their forms are complex and diverse, and the number of paths grows exponentially with the number of connections \( n \). Therefore, it's impractical to enumerate all interference paths, and the previous splicing construction method cannot showcase all of them. In this section, we apply the methods previously used to demonstrate how to construct paths with connection numbers \( n > 16 \).

\begin{figure}[ht]
\centering
 \includegraphics[width=0.6\textwidth]{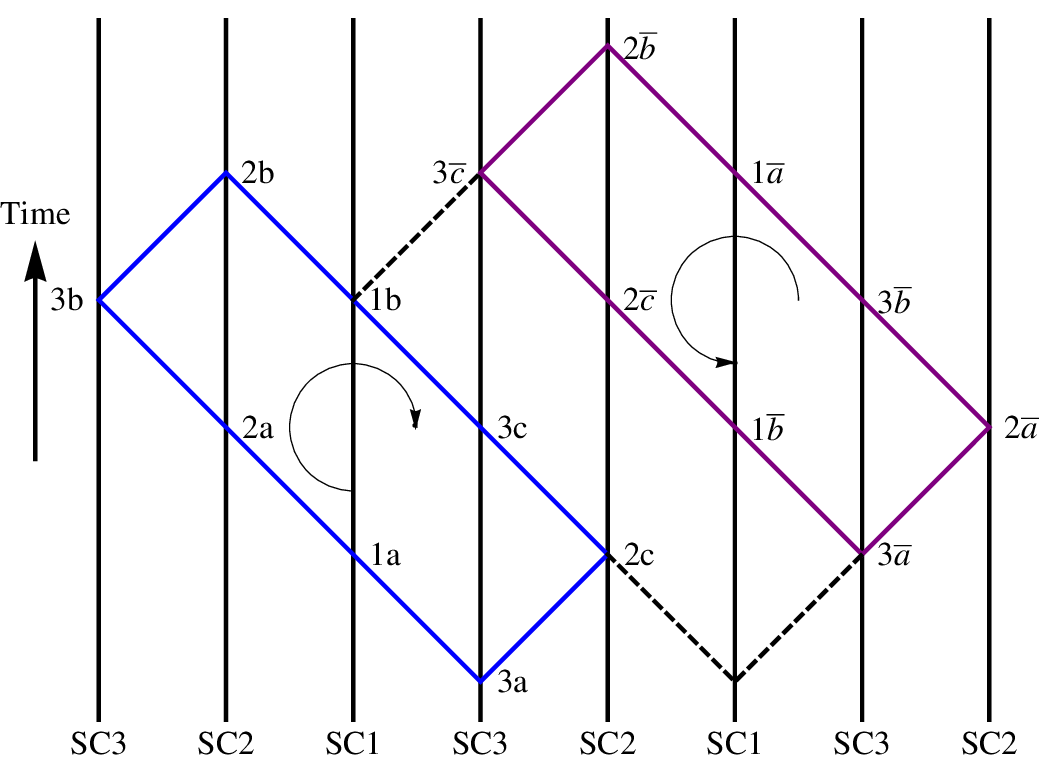}
 \caption{\small{Diagram of U18.}} \label{fig:UWL1820}
\end{figure}

In the case of \( n = 16 \) using self-splicing methods, we use Relay (U) and its reverse calculated path (denoted as \(-U\)) for splicing. As an example, illustrated in Figure \ref{fig:UWL1820}, two interference paths are depicted with solid lines (bottom-left path for U and top-right for \(-U\)). The calculation directions are indicated with arrows, and the labels of spacecraft on both paths are marked. 

To connect U's \( 1b \) and \(-U\)'s \( 3\bar{c} \) using one path, when reaching \( 1b \) on U's path, proceed via a bridging path to \( 3\bar{c} \), circle \(-U\) counterclockwise once, return to \( 3\bar{c} \), then return via the added connection to \( 1b \), and complete the remaining path along U. This process is expressed in Eq. (\ref{equ:U18-20}) as U18-1b3c. Additionally, in Figure \ref{fig:UWL1820}, it's possible to connect U's \( 2c \) and \(-U\)'s \( 3\bar{b} \) using two connections, thereby constructing an interference path with \( n = 20 \), expressed as U20-2c3a in Eq. (\ref{equ:U18-20}). The newly added bridging paths in both interference paths are highlighted in bold in Eq. (\ref{equ:U18-20}).
\begin{equation}
\label{equ:U18-20}
\begin{split}
\text{U18-1b3c:  } & 
\overrightarrow{2'3'1'1} \ \overleftarrow{3'} \ \overrightarrow{\mathbf{2}}[\overleftarrow{1'3'2'} \ \overrightarrow{11'2'3'} \ \overleftarrow{1] \mathbf{2}2'1'1} \\
\text{U20-2c3a:  } &
\overrightarrow{2'3'1'1} \ \overleftarrow{3'2'1' \mathbf{3'}}  \ \overrightarrow{\mathbf{2} [ 11'2'3'}\ \overleftarrow{11'3'2'] \mathbf{2}} \ \overrightarrow{\mathbf{3'}} \ \overleftarrow{1} 
\end{split}
\end{equation}

\begin{figure}[ht]
\centering
 \includegraphics[width=0.6\textwidth]{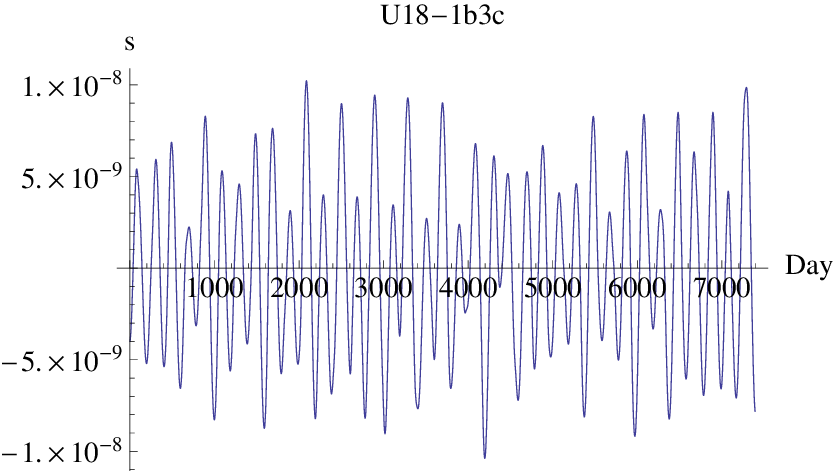}
 \caption{\small{The path mismatches in U18-1b3c from numerical calculation.}} \label{fig:U18}
\end{figure}

Based on our previous analysis, U18-1b3c and U20-2c3a have the same relative positions along the time axis in their closed paths. Therefore, it's inferred that their numerical computation results at the second-generation TDI are the same, which has been verified through numerical calculations. The numerical results for both are shown in Figure \ref{fig:U18}.

\begin{figure}[ht]
\begin{minipage}[t]{0.42\textwidth}
  \centering
  \includegraphics[width=0.9\textwidth]{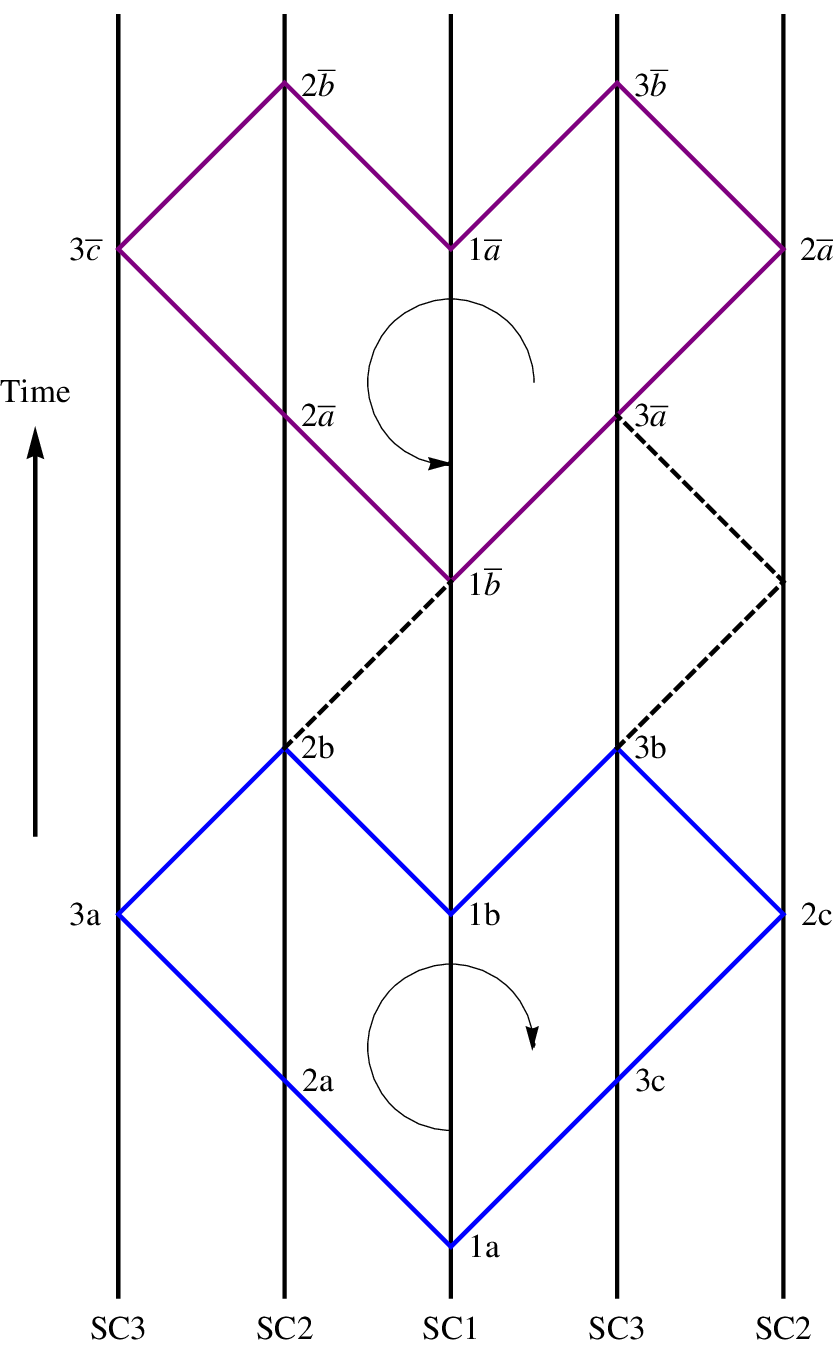}
  \caption{\small{Diagram of P18-2b1b.}} \label{fig:PWL18-20}
  \end{minipage}
  \begin{minipage}[t]{0.55\textwidth}
  \centering
 \includegraphics[width=0.9\textwidth]{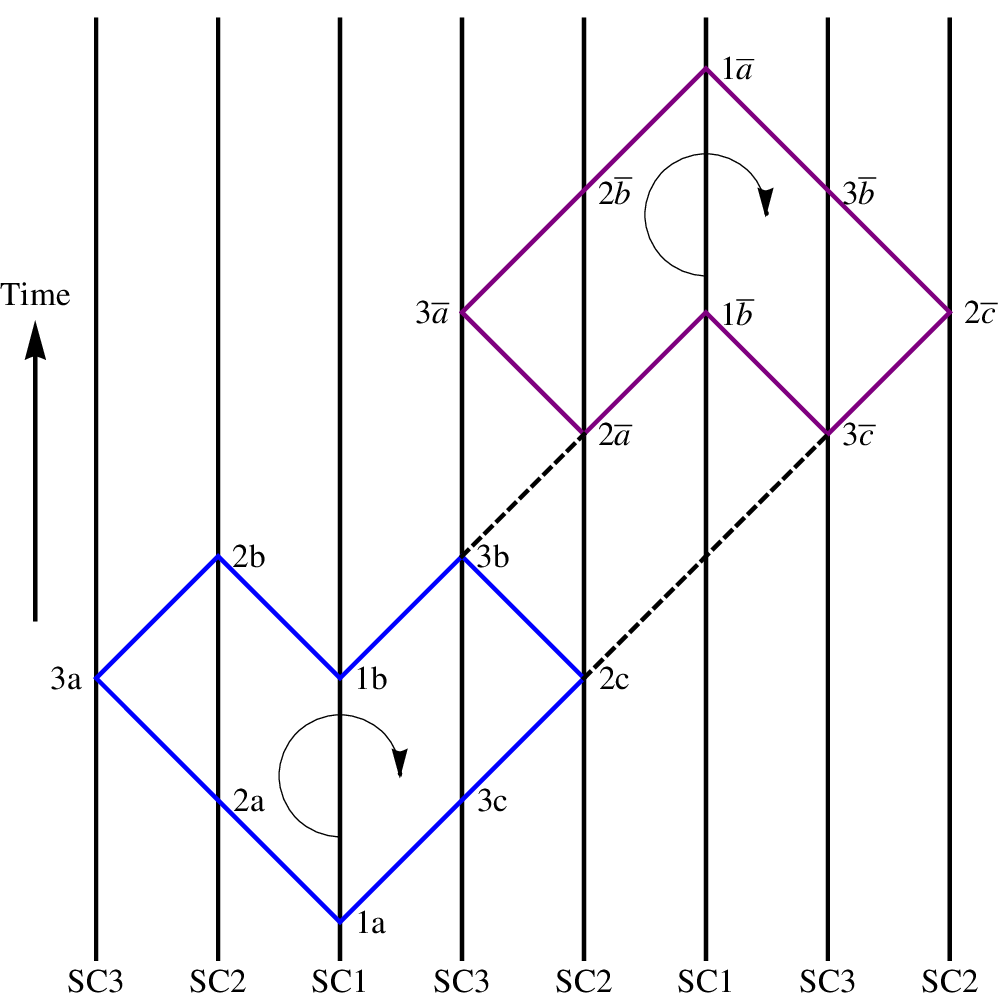}
  \caption{\small{Diagram of PE18-3b2a.}}  \label{fig:PEWL18-20}
  \end{minipage}
\end{figure}

\begin{figure}[!ht]
\centering
 \includegraphics[width=0.48\textwidth]{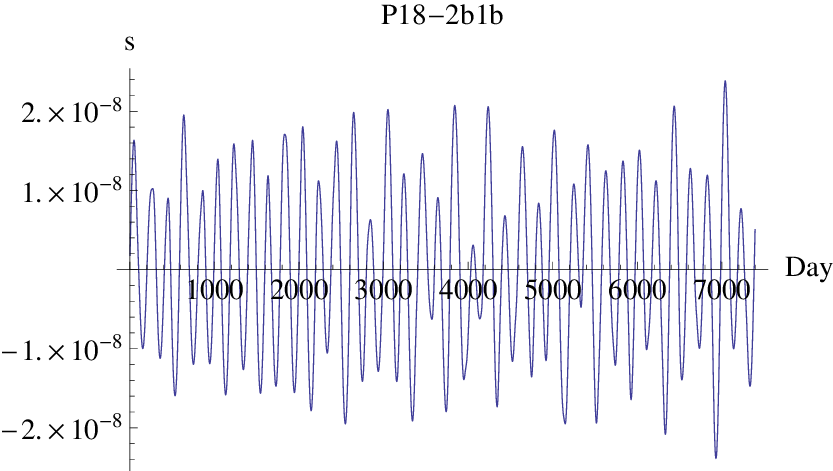}
 \includegraphics[width=0.48\textwidth]{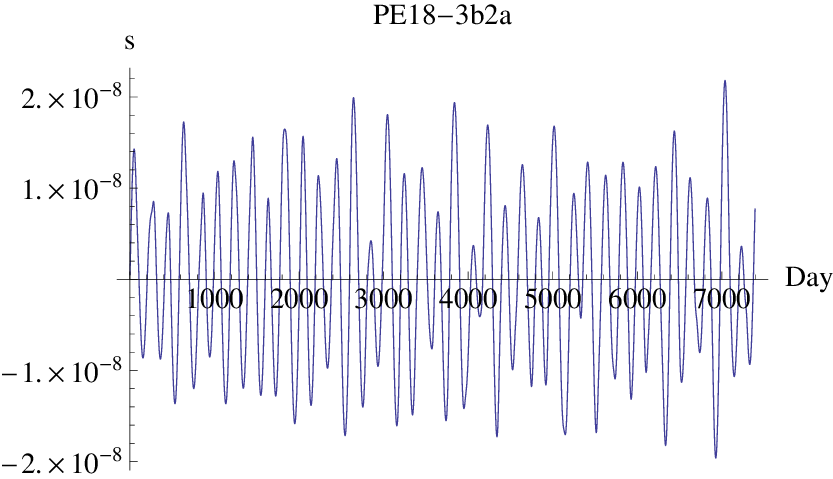}
 \caption{\small{The path mismatches for $n$=18 from numerical calculation.}} \label{fig:R18-20}
\end{figure}

According to the above method, examples of interference paths constructed using Beacon (P) and \(-P\) for \( n = 18 \) and \( n = 20 \) are shown in Figure \ref{fig:PWL18-20}. The path expressions are given in Eq. (\ref{equ:P18-20}), and the computation results are illustrated in Figure \ref{fig:R18-20}.
\begin{equation} 
\label{equ:P18-20}
\begin{split}
\text{P18-2b1b:  } &
\overrightarrow{3'1'1 \mathbf{3} [ 211'} \ \overleftarrow{2} \ \overrightarrow{3} \ \overleftarrow{11'3'] \mathbf{3} 3'} \ \overrightarrow{2} \ \overleftarrow{1'12} \\
\text{P20-3b3a:  } &
\overrightarrow{3'1'1} \ \overleftarrow{3'} \ \overrightarrow{2 \mathbf{11'} [11'} \ \overleftarrow{2} \ \overrightarrow{3'} \ \overleftarrow{11'3'} \ \overrightarrow{2} ] \overleftarrow{\mathbf{1'1} 1'12} \\
\end{split}
\end{equation}

Examples of interference paths constructed using Beacon (P) and Monitor (E) for \( n = 18 \) and \( n = 20 \) are shown in Figure \ref{fig:PEWL18-20}. The path expressions are given in Eq. (\ref{equ:PE18-20}). The computation results are illustrated in Figure \ref{fig:R18-20}.
\begin{equation}
\label{equ:PE18-20}
\begin{split}
\text{PE18-3b2a:  } &
\overrightarrow{3'1'1} \ \overleftarrow{3'} \ \overrightarrow{2 \mathbf{1} [1'13} \ \overleftarrow{2'1'1} \ \overrightarrow{2'} \ \overleftarrow{3] \mathbf{1} 1'12} \\
\text{PE20-2c3c:  } &
\overrightarrow{3'1'1} \ \overleftarrow{3'} \ \overrightarrow{2} \ \overleftarrow{1'} \ \overrightarrow{\mathbf{32} [ 2'} \ \overleftarrow{3} \ \overrightarrow{1'13} \ \overleftarrow{2'1'1 ] \mathbf{23} 12} \\
\end{split}
\end{equation}

\begin{figure}[ht]
\centering
 \includegraphics[width=0.45\textwidth]{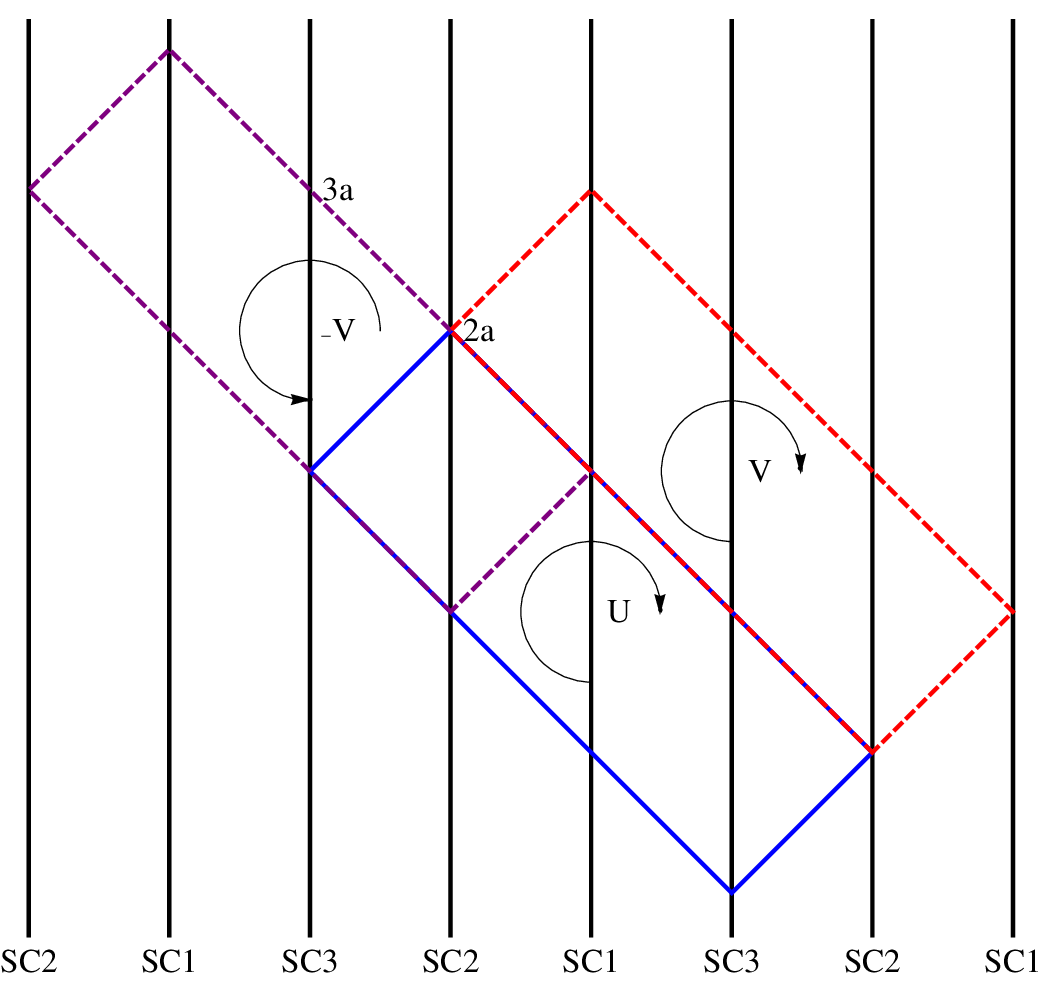} \hspace{20pt}
 \includegraphics[width=0.45\textwidth]{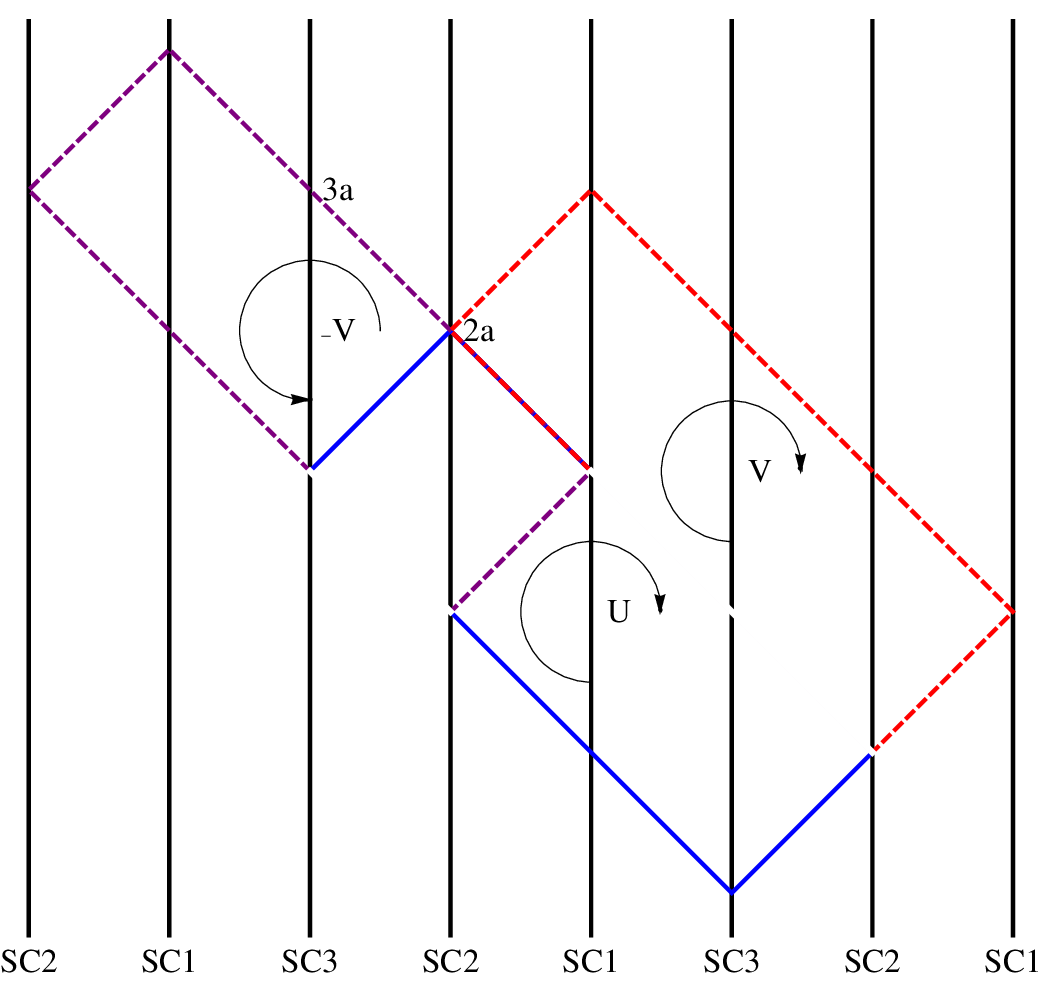}
 \caption{\small{Diagrams of joint U, V and -V.}} \label{fig:UVWL22}
\end{figure}

\begin{figure}[ht]
\begin{minipage}[t]{0.48\textwidth}
  \centering
 \includegraphics[width=0.95\textwidth]{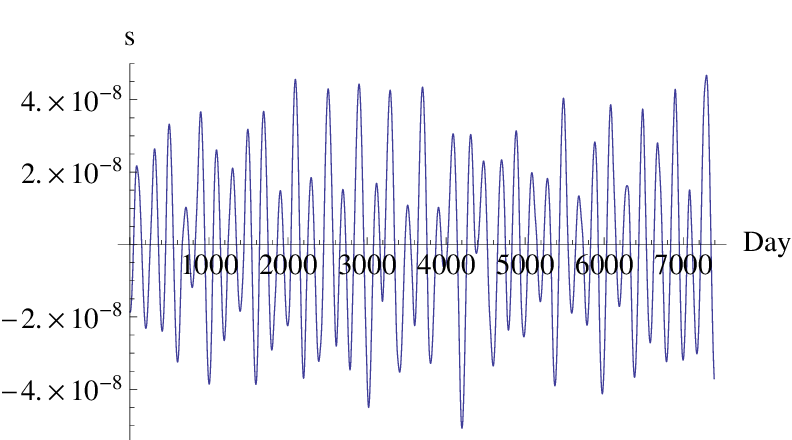}
  \caption{\small{The path mismatches in UV22 from numerical calculation.}} \label{fig:UV22R}
  \end{minipage}
  \begin{minipage}[t]{0.48\textwidth}
  \centering
 \includegraphics[width=0.95\textwidth]{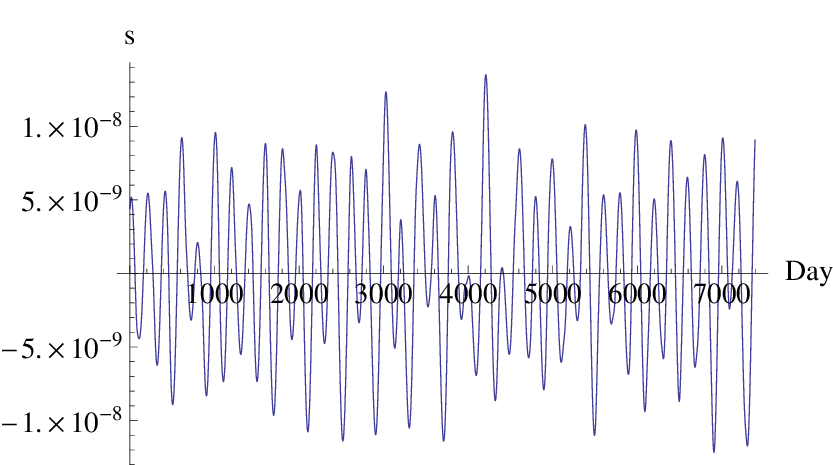}
  \caption{\small{The path mismatches in XW18 from numerical calculation.}} \label{fig:XW18R}
  \end{minipage}
\end{figure}

When the connection number \( n \geqslant 18 \), it is possible to have combinations of two pairs of second-generation TDI paths. Some paths are combined in a way that results in paired cancellation, reducing the total number of connections \( n \) below the sum of the original connection numbers. Below are two examples. Eq. (\ref{equ:UV22}) shows the connection of four first-generation interference paths using Relay types U, -U, V, and -V. 
\begin{equation}
\label{equ:UV22}
\begin{split} 
\left. \begin{split}
 \left. \begin{split}  
 \text{V:   } & \overrightarrow{ 1' 2' 3' 3} \ \overleftarrow{2' 1' 3' 3} \\
 \text{-V:   } & \overrightarrow{ 3 3' 1' 2'} \ \overleftarrow{3 3' 2' 1'} \\
 \end{split} \right\} \overset{\text{\tiny{splicing}}}{\Rightarrow} \text{2nd-generation TDI} &\\
 \left. \begin{split}  
 \text{U:   } & \overrightarrow{ 2' 3' 1' 1 } \ \overleftarrow{3' 2' 1' 1} \\
 \text{-U:   } & \overrightarrow{ 1 1' 2' 3'} \ \overleftarrow{1 1' 3' 2'} 
 \end{split} \right\}  \overset{\text{\tiny{splicing}}}{\Rightarrow} \text{2nd-generation TDI} & \\
 \end{split} \right\} \overset{\text{\tiny{splicing}}}{\Rightarrow} \text{2nd-generation TDI} \\
\end{split} 
\end{equation}
The connections of U and -U form a second-generation TDI path, and similarly, V and -V form another second-generation TDI path. Connecting these two second-generation TDI paths results in at least a second-generation TDI path (potentially achieving higher-order TDI paths). The construction process is shown in Eq. (\ref{equ:UV22R}). 
\begin{equation}
\label{equ:UV22R}
\begin{split}  
  \left.   \begin{split}  
 \text{V:  } & \overrightarrow{ 1' 2' 3' 3} \ \overleftarrow{2' 1' 3' 3} \\
 \text{U:  } & \overrightarrow{ 2' 3' 1' 1 } \ \overleftarrow{3' 2' 1' 1} 
 \end{split} \right\}  \Rightarrow \  &
 \overrightarrow{2'3'1'1[3} \ \overleftarrow{2'1'3'3} \ \overrightarrow{\mathbf{1'2'3'}} ] \overleftarrow{\mathbf{3'2'1'}1} \\
  = \ & \overrightarrow{2'3'1'13} \ \overleftarrow{2'1'3'31}  \\
  \left.   \begin{split}  
 \Rightarrow \ & \overrightarrow{2'3'1'13} \ \overleftarrow{2'1'3'31} \\
 \text{-V:  } & \overrightarrow{ 3 3' 1' 2'} \ \overleftarrow{3 3' 2' 1'} \\
  \end{split} \right\}  \Rightarrow \ &
   \overrightarrow{2'3' [3 3' 1' 2'}\ \overleftarrow{3 3' 2' \mathbf{1'}} ] \overrightarrow{\mathbf{1'}13} \ \overleftarrow{2'1'3'31} \\
  = \ & \overrightarrow{2'3' 3 3' 1' 2'}\ \overleftarrow{3 3' 2'} \ \overrightarrow{13} \ \overleftarrow{2'1'3'31} \\
   \left.   \begin{split}  
 \Rightarrow  & \overrightarrow{2'3' 3 3' 1' 2'}\ \overleftarrow{3 3' 2'} \ \overrightarrow{13} \ \overleftarrow{2'1'3'31} \\
\text{-U:  } & \overrightarrow{ 1 1' 2' 3'} \ \overleftarrow{1 1' 3' 2'} \\
\end{split} \right\}  \Rightarrow \ &
  \overrightarrow{2'3' 3 3' 1'[1 1' 2' 3'}  \ \overleftarrow{1 1' 3' \mathbf{2'}}] \overrightarrow{\mathbf{2'}}\ \overleftarrow{3 3' 2'} \ \overrightarrow{13} \ \overleftarrow{2'1'3'31} \\
  = \ & \overrightarrow{2'3' 3 3' 1'1 1' 2' 3'}  \ \overleftarrow{1 1' 3' 3 3' 2'}\ \overrightarrow{13} \ \overleftarrow{2'1'3'31}
\end{split}
\end{equation}
Connections that cancel each other are highlighted in bold. Finally, we obtain a interference path with \( n = 22 \). The numerical computation results are shown in Figure \ref{fig:UV22R}. The process of connection on the spacecraft's world lines is illustrated more clearly in Figure \ref{fig:UVWL22}, which depicts the cancellation process of arms in U, V, and -V connections: the left panel shows the initial situation before arm cancellation, while the right panel shows the situation after arm cancellation.

For the construction of four different types of first-generation TDI paths, such as the Michelson type X and -X, and the Relay type $\bar{\text{W}}$ and -$\bar{\text{W}}$, the connection process is illustrated in Eq. (\ref{equ:XW18}). The construction process is shown in Eq. (\ref{equ:XW18R}). Finally, we obtain an interference path with \( n = 18 \). The numerical results are shown in Figure \ref{fig:XW18R}.
\begin{equation}
\label{equ:XW18}
\begin{split} 
 \left. \begin{split}
 \left. \begin{split}  
 \bar{\text{W}}:\  & \overrightarrow{ 2' 2 1 3} \ \overleftarrow{2' 2 3 1} \\
-\bar{\text{W}}:\  & \overrightarrow{1 3 2 2'} \ \overleftarrow{3 1 2 2'} \\
 \end{split} \right\} \overset{\text{\tiny{splicing}}}{\Rightarrow} \text{2nd-generation TDI} &\\
 \left. \begin{split}  
 \text{X}:\ & \overrightarrow{ 3' 3 2 2' } \ \overleftarrow{3 3' 2' 2} \\
 -\text{X}:\  & \overrightarrow{2 2' 3' 3} \ \overleftarrow{2' 2 3 3'} 
 \end{split} \right\}  \overset{\text{\tiny{splicing}}}{\Rightarrow} \text{2nd-generation TDI} & \\
 \end{split} \right\} \overset{\text{\tiny{splicing}}}{\Rightarrow} \text{2nd-generation TDI} \\
\end{split} 
\end{equation}

\begin{equation}
\label{equ:XW18R}
\begin{split}  
  \left.   \begin{split}  
 \bar{\text{W}}: \ & \overrightarrow{2' 2 1 3} \ \overleftarrow{2' 2 3 1} \\
 \text{X: } & \overrightarrow{ 3' 3 2 2' } \ \overleftarrow{3 3' 2' 2} 
 \end{split} \right\}  \Rightarrow \  &
 \overrightarrow{2'21 \mathbf{3}} [ \overleftarrow{\mathbf{3} 3'2'2} \ \overrightarrow{3' \mathbf{322'}} ] \overleftarrow{\mathbf{2'23}1} \\
  = \ & \overrightarrow{2'21} \ \overleftarrow{3'2'2} \ \overrightarrow{3'} \ \overleftarrow{1}  \\
  \left.   \begin{split}  
 \Rightarrow \ & \overrightarrow{2'21} \ \overleftarrow{3'2'2} \ \overrightarrow{3'} \ \overleftarrow{1}  \\
 -\bar{\text{W}}:\ & \overrightarrow{1 3 2 2'} \ \overleftarrow{3 1 2 2'}  \\
  \end{split} \right\}  \Rightarrow \ &
   \overrightarrow{2'21}\ \overleftarrow{3' [ 3122' } \ \overrightarrow{13 \mathbf{22'}} ] \overleftarrow{\mathbf{2'2}} \ \overrightarrow{3'} \ \overleftarrow{1}  \\
  = \ & \overrightarrow{2'21}\ \overleftarrow{3' 3122' } \ \overrightarrow{13 3'} \ \overleftarrow{1} \\
   \left.   \begin{split}  
 \Rightarrow  & \overrightarrow{2'21}\ \overleftarrow{3' 3122' } \ \overrightarrow{13 3'} \ \overleftarrow{1} \\
 -\text{X}:\  & \overrightarrow{2 2' 3' 3} \ \overleftarrow{2' 2 3 3'} \\
\end{split} \right\}  \Rightarrow \ &
  \overrightarrow{2'21}\ \overleftarrow{3' 31 \mathbf{2}} [\overrightarrow{\mathbf{2} 2' 3' 3} \ \overleftarrow{2' 2 3 3' ]2' } \ \overrightarrow{13 3'} \ \overleftarrow{1} \\
  = \ &  \overrightarrow{2'21}\ \overleftarrow{3' 31} \ \overrightarrow{2' 3' 3} \ \overleftarrow{2' 2 3 3' 2' } \ \overrightarrow{13 3'} \ \overleftarrow{1}
\end{split} 
\end{equation}

\chapter{Conclusions and Outlooks}

In the work, we discussed the principles of time-delay interferometry, the CGC2.7 planetary ephemeris framework, and the orbit selection and optimization for the ASTROD-GW mission. We focused on the numerical simulation for the TDI in space-based GW detection. We covered the paths of first-generation TDI and the process of constructing second-generation TDI paths using geometric methods based on existing first-generation paths. Based on these constructed interference paths, extensive numerical calculations were performed for TDI in the ASTROD-GW mission. Additionally, building upon previous work, a detailed geometric analysis of TDI paths was conducted.

TDI is a crucial component in space-based GW detection, directly impacting the sensitivity of GW detection. The numerical results obtained from our studies represent a preliminary results in this research. In future work, we can further enhance our understanding by simulating various GW signals and evaluating the sensitivity of different TDI paths for GW detection. This research can be complemented by experimental efforts to identify interferometry methods that best suit practical needs.
\appendix
\backmatter

\renewcommand{\bibname}{References}
\newcommand{\kai}{\CJKfamily{gbsn}}
\small{

}


\end{spacing}
\end{CJK}
\end{document}